\documentclass[useAMS,usenatbib]{mn2e}
\usepackage{graphicx}
\usepackage{natbib}
\usepackage{lscape}
\usepackage{longtable}
\usepackage{subfig}
\usepackage{amssymb,amsmath}
\usepackage[T1]{fontenc}
\usepackage{hyperref}
\usepackage{amsmath}
\usepackage{breakurl}
\usepackage{times}
\usepackage{morefloats}
\usepackage{bm}
\usepackage{flafter}

\newcommand{\wise}{\textit{WISE}}

\newcommand{\galex}{\textit{GALEX}}

\newcommand{\CIV}{C\,{\sc iv}}
\newcommand{\MgII}{Mg\,{\sc ii}}

\newcommand{\CIII}{C\,{\sc iii}]}

\newcommand\ion[2]{#1$\;${\small\rmfamily\@Roman{#2}}\relax}%

\def\lsim{\lower0.3em\hbox{$\,\buildrel <\over\sim\,$}}
\def\gsim{\lower0.3em\hbox{$\,\buildrel >\over\sim\,$}}

\title[\emph{XDQSOz} with WISE]{Quasar Probabilities and Redshifts from \textit{WISE} mid-IR through \textit{GALEX} UV Photometry}
\author[DiPompeo et al.]{M.A. DiPompeo$^1$, J. Bovy$^2$\thanks{Hubble Fellow, John N. Bahcall Fellow}, A.D. Myers$^1$, D. Lang$^3$\\
$^1$ University of Wyoming, Dept. of Physics and Astronomy 3905, 1000 E. University, Laramie, WY 82071, USA \\
$^2$ Institute for Advanced Study, Einstein Drive, Princeton, NJ 08450, USA \\
$^3$ McWilliams Center for Cosmology, Department of Physics, Carnegie Mellon University, 5000 Forbes Avenue, Pittsburgh, PA 15213, USA}

\begin{document}
\date{Accepted 2015 July 10; Received 2015 July 9; in original form 2014 December 02}

\pagerange{\pageref{firstpage}--\pageref{lastpage}} \pubyear{2014}

\maketitle

\label{firstpage}

\begin{abstract}
\textit{Extreme deconvolution (XD)} of broad-band photometric data can both separate stars from quasars and generate probability density functions for quasar redshifts, while incorporating flux uncertainties and missing data.  Mid-infrared photometric colors are now widely used to identify hot dust intrinsic to quasars, and the release of all-sky \wise\ data has led to a dramatic increase in the number of IR-selected quasars.  Using forced-photometry on public \wise\ data at the locations of SDSS point sources, we incorporate this all-sky data into the training of the \emph{XDQSOz} models originally developed to select quasars from optical photometry.  The combination of \wise\ and SDSS information is far more powerful than SDSS alone, particularly at $z>2$.  The use of SDSS$+$\wise\ photometry is comparable to the use of SDSS$+$ultraviolet$+$near-IR data.  We release a new public catalogue of 5,537,436 (total; 3,874,639 weighted by probability) potential quasars with probability $P_{\textrm{QSO}} > 0.2$.  The catalogue includes redshift probabilities for all objects.  We also release an updated version of the publicly available set of codes to calculate quasar and redshift probabilities for various combinations of data.  Finally, we demonstrate that this method of selecting quasars using \wise\ data is both more complete and efficient than simple \wise\ color-cuts, especially at high redshift.  Our fits verify that above $z \sim 3$ \wise\ colors become bluer than the standard cuts applied to select quasars.  Currently, the analysis is limited to quasars with optical counterparts, and thus cannot be used to find highly obscured quasars that \wise\ color-cuts identify in significant numbers.
\end{abstract}


\section{INTRODUCTION}
As newer and larger imaging surveys are conducted over more area and frequency ranges, photometric classification of quasars using as much available information as possible is becoming increasingly important.  Spectroscopic follow-up of complete samples will pose greater and greater technical challenges as the depth of imaging surveys increases.  Studies of photometric quasar samples have already shed light on the full quasar population  \citep{2005ApJ...633..589S, 2006PhRvD..74f3520G, 2008PhRvD..77l3520G, 2006ApJ...638..622M, 2007ApJ...658...85M, 2007ApJ...658...99M, 2011ApJ...731..117H, 2014ApJ...789...44D, 2014MNRAS.442.3443D}.  Not only is classification important, but photometric redshift probability density functions (PDFs) are also useful for many applications \citep[e.g.][]{2006ApJ...638..622M, 2007ApJ...658...85M}. The ability to generate photo-$z$s for quasars has improved with deeper and more precise multi-filter photometry \citep[][]{2001AJ....121.2308R, 2001AJ....122.1151R, 2001AJ....122.1163B, 2012ApJ...749...41B}.

The problem of photometric quasar classification is well studied, and often involves breaking up samples into redshift bins that allow the classification schemes to also serve as broad redshift indicators \citep[e.g.][]{2004ApJS..155..257R, 2009AJ....137.3884R, 2009ApJS..180...67R, 2005AJ....130.2439S, 2007ApJ...663..774B, 2008ApJ...683...12B}.  \citet{2011ApJ...729..141B} developed one of the most successful current quasar classification techniques (\emph{XDQSO}), by using a large number of Gaussians to model the flux space of quasars and stars/unresolved galaxies  \citep[incorporating missing and/or noisy data using \textit{extreme deconvolution, XD;}][]{Bovy:2011bt} to assign quasar probabilities in broad redshift bins.  This method was successfully applied in selecting quasars (primarily at $z > 2$) for the Baryon Oscillation Spectroscopic Survey \citep[BOSS;][]{2012ApJS..199....3R, 2013AJ....145...10D}.  \citet{2012ApJ...749...41B} extended this method to incorporate redshift information into the model (\emph{XDQSOz}) such that it can be integrated analytically to provide quasar probabilities over arbitrary redshift ranges and generate photometric redshift PDFs.  \citet{2012ApJ...749...41B} also incorporated ultraviolet (UV) and near-infrared (NIR) forced-photometry at SDSS source positions into the model, which improved quasar classification and essentially broke all redshift degeneracies evident when using photometry from the Sloan Digital Sky Survey \citep[SDSS;][]{2000AJ....120.1579Y} $u$$g$$r$$i$$z$ filters \citep[][]{1996AJ....111.1748F} alone.

While \citet{2012ApJ...749...41B} illustrated the power of NIR photometry in the \emph{XDQSOz} method, its use is limited because of the shallow depths and relatively smaller areas of current NIR surveys (see section 2.2.4).  It is now well established that the mid-IR is an efficient identifier of quasars \citep{Stern:2005p2563, 2007ApJ...671.1365H}, and the public all-sky data from the \textit{Wide-Field Infrared Survey Explorer} \citep[\wise;][]{2010AJ....140.1868W} allows the use of this method to identify very large samples.  With no additional information, over redshifts from $0< z < 3$, \wise\ colors can select luminous quasars at completeness levels near 90\% \citep[at conservative flux limits of $m \lesssim 15$;][]{2012ApJ...753...30S, 2013ApJ...772...26A}.  These techniques have been used to classify the first large samples of obscured quasars \citep[][]{2012MNRAS.426.3271M, 2013MNRAS.434..941M} and begin to statistically compare obscured and unobscured quasar populations \citep{2014ApJ...789...44D, 2014MNRAS.442.3443D}.  

In this paper, we incorporate the mid-IR data of \wise\ in the \emph{XDQSOz} model of the relative-flux-redshift density, and show that these data dramatically increase the power for \emph{XDQSOz} to identify quasars, especially at high redshift, as well as improve photometric redshift estimation.  We also present a new catalogue of quasar probabilities for point sources in SDSS DR8, including a potential quasar catalogue for objects with probabilities above 0.2 that includes redshift PDFs.  We caution that this catalogue is probabilistic in nature, and not suitable as a statistical sample (at least not in its entirety --- statistical subsamples can be selected from the full catalogue).  Section 3.3 discusses this in more detail.

\section{METHODS \& DATA}
\subsection{Photometric classification and redshift estimation}
Details of the method used to calculate quasar/star probabilities for objects and redshift PDFs are presented fully in \citet{2011ApJ...729..141B} and \citet{2012ApJ...749...41B}.  We refer the reader there for complete details, with a brief summary of the general considerations provided here.

The important factor for photometrically classifying quasars and estimating redshifts is the joint probability of an object's fluxes, redshift, and the possibility that it is a quasar: $p$(flux, $z$, quasar).  This can be written in many different ways depending on the approach, but \emph{XDQSOz} is based on:
\begin{equation}
p(\textrm{fluxes, } z\textrm{, quasar}) = p(\textrm{fluxes, } z | \textrm{quasar}) P (\textrm{quasar}).
\end{equation}
Under the assumption that an object is a quasar, photometric redshifts are given by:
\begin{equation}
p(z|\textrm{fluxes, quasar}) = \frac{p(\textrm{fluxes, } z \textrm{, quasar})}{p(\textrm{fluxes, quasar})}.
\end{equation}
To determine if an object is a quasar in a given redshift range, we integrate the joint probability over redshift:
\begin{equation}
P(\textrm{quasar in } \Delta z | \textrm{fluxes}) = \int_{\Delta z} dz\ p(\textrm{quasar, } z| \textrm{fluxes})
\end{equation}
\begin{equation}
= \int_{\Delta z} dz\ \frac{p(\textrm{quasar, } z\textrm{, fluxes})}{p(\textrm{fluxes})}
\end{equation}
where
\begin{equation}
p(\textrm{fluxes}) = p(\textrm{fluxes, quasar}) + p(\textrm{fluxes, not a quasar})
\end{equation}
and $p(\textrm{fluxes, not a quasar})$ is found by modeling the fluxes of non-quasars (the stellar sample described below).  The probability that an object is a quasar at any $z$ is calculated with $\Delta z = [0, \infty]$.  Given that the quasar training set is in the range $0.3 < z < 5.5$, in practice these are the redshifts that should be considered and are the limits applied when we refer to a quasar at ``any redshift''.


\subsection{Training data}

The training sets used are essentially identical to those of \citet{2011ApJ...729..141B} and \citet{2012ApJ...749...41B}, with the addition of \wise\ data described in section 2.2.5, and so we only provide a brief summary of these, referring the reader to the original papers for full details.

\subsubsection{Sloan Digital Sky Survey optical data}
The SDSS imaged $\sim$10,000 deg$^2$ of the northern and southern Galactic sky in $u$, $g$, $r$, $i$, and $z$ filters.  SDSS-III extended this by $\sim$2500 deg$^2$ in the southern Galactic cap \citep[][]{2011AJ....142...72E}.  In addition, the SDSS has taken follow-up spectroscopy of millions of sources, which have generated several catalogues of various source subclasses.  All of the optical data used to train the \emph{XDQSOz} algorithms are based on SDSS imaging and spectroscopy.

\subsubsection{Quasars and stars}
Our goal is to classify point-like objects as either stars or quasars.  This of course assumes that all resolved sources are galaxies, automatically removing them from the samples, or conversely that all quasars are unresolved.  While not strictly true, this is largely handled by limiting the redshifts probed to $z \ge 0.3$.  Limiting to unresolved sources will discard a significant fraction of optically faint obscured (``type 2'') quasars, but should not strongly impact optically bright unobscured (``type 1'') sources.

The quasar training data consists of 103,601 spectroscopically confirmed quasars from the SDSS data release 7 (DR7) quasar catalogue \citep{2010AJ....139.2360S} with $z \ge 0.3$.  While the training is performed in bins of SDSS $i$-band magnitude, all of the quasars are included in each bin in order to have large enough numbers to properly constrain the fits.  Thus the quasar flux in each filter is rescaled based on its $i$-band magnitude and redshift, using an apparent-magnitude-dependent redshift prior.  The prior is obtained by integrating a model of the quasar luminosity function \citep{2007ApJ...654..731H}\footnote{Preliminary work with BOSS quasars suggests that this luminosity function may be incorrect above $z \sim 2.5$ (Ross, private communication).  At present, this is still the best choice, but it is possible future updates may be needed.}  over the appropriate apparent magnitude bin \citep[see Figure 1 of][and section 2.2.2 for more details]{2012ApJ...749...41B}.  This is used to re-weight the quasars in each $i$-band bin so that the weighted histogram of the quasars reproduces the prediction of the luminosity function.  This method also works under the assumption that quasar colors are independent of their absolute magnitude.  While there are known correlations between quasar properties and luminosity \citep[e.g.][]{1977ApJ...214..679B, 2004AJ....128.2603Y}, these effects are small compared to the intrinsic scatter and are generally washed out in broadband colors.  

To model the flux space of stars, we use a large set of point sources from the co-added photometry of SDSS Stripe 82 \citep[][]{2009ApJS..182..543A} that are selected to have low variability, because essentially all quasars are variable and most stars are not \citep[][]{2011ApJ...743..125K}.  This identifies a stellar training set of 701,215 objects.  Of those with available spectra (23,540), only 221 are quasars, indicating that the contamination is extremely low.  Additionally, as unresolved galaxies are also not highly variable, these objects are included as part of the stellar training set and thus accounted for when the quasar probability is calculated.

\subsubsection{\textit{Galaxy Evolution Explorer} UV data}
We utilize data from the all-sky \textit{Galaxy Evolution Explorer} \citep[\galex;][]{2005ApJ...619L...1M} in the near- and far-UV (NUV and FUV).  Rather than using \galex\ catalogue fluxes, we use values force-photometered from \galex\ images at SDSS source positions \citep{2011ApJS..193...29A}.  This allows us to obtain low signal-to-noise PSF fluxes of objects that are not detected by \galex, and thus not included in the \galex\ catalogues.  One update in this work as compared to \citet{2012ApJ...749...41B} is that the \galex\ forced photometry was not complete at the time of the original analysis, and so UV data for more objects are included here --- 81,011 quasars have NUV data, and 68,523 have FUV data, with the numbers increasing slightly for those that only have upper limits.  This additional UV data in the training set improves \emph{XDQSOz} slightly on its own (see section 3.1).  The distributions of the UV signal-to-noise ratio (SNR) of these sources are shown in the middle panel of Figure~\ref{fig:snr}.

\subsubsection{UKIRT Infrared Deep Sky Survey NIR data}
The UKIRT Infrared Deep Sky Survey (UKIDSS) covers $\sim$4000 deg$^2$ of the SDSS footprint in the $Y$, $J$, $H$, and $K$ NIR filters.  We make use of this imaging, again force-photometered at SDSS source positions.  We find that $\sim 26,000$ objects have complementary fluxes in all of these filters, again with the number increasing when only those with upper limits are considered.  The numbers of objects with flux measurements per band are 26,487 ($Y$); 26,450 ($J$); 26,486 ($H$); 26,561 ($K$).  Again, these numbers are slightly updated from \citet{2012ApJ...749...41B}, though the differences are quite small.  The distributions of SNR values for the UKIDSS imaging are shown in the bottom panel of Figure~\ref{fig:snr}.

\subsubsection{\textit{Wide-Field Infrared Survey Explorer} MIR data}
\wise\ has mapped the entire sky multiple times, in four bands centered at 3.4, 4.6, 12, and 22 $\mu$m ($W1$, $W2$, $W3$, and $W4$ respectively).  The 5$\sigma$ limit in each band is at least 0.08, 0.11, 1, and 6 mJy, respectively, and improves toward the ecliptic poles where the observing strategy leads to deeper imaging.  The angular resolutions in each band are 6.1, 6.4, 6.5, and 12 arc sec, respectively.  \wise\ has released two full-sky source catalogs --- however, again in order to use the full power of the \textit{XD} method and incorporate information on sources below the flux limit of the survey, we make use of forced photometry of the \wise\ All-Sky Release imaging at SDSS positions \citep[to be included in a future SDSS release; see][]{2014AJ....147..108L, 2014arXiv1410.7397L}.

\begin{figure}
\centering
\hspace{0cm}
   \includegraphics[width=7cm]{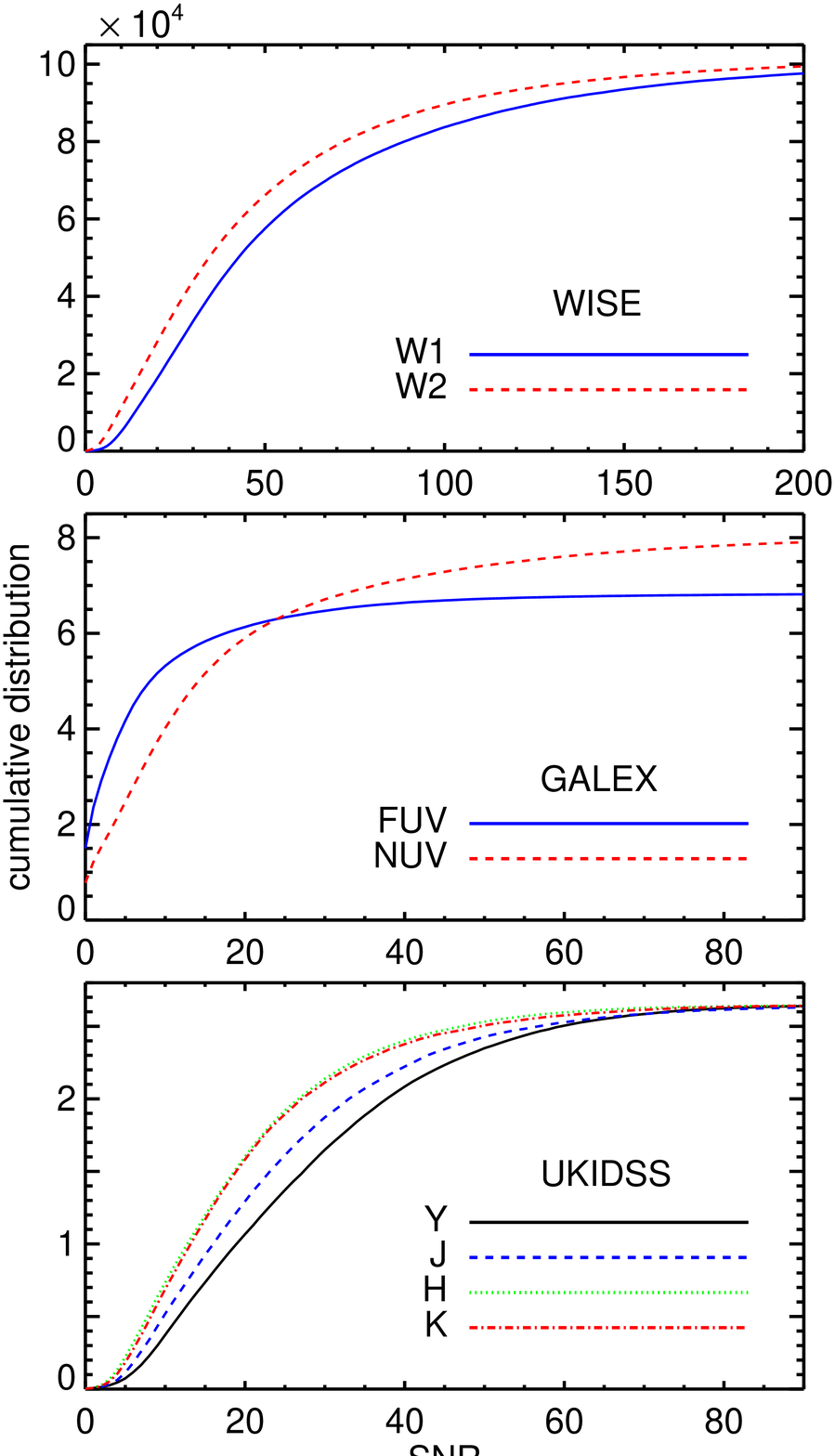}
    \vspace{0.4cm}
  \caption{Cumulative distributions of the SNR for \wise, \galex, and UKIDSS photometric data of spectroscopically confirmed quasars.  \wise\ has data available for more sources, and is also deeper than the other surveys.\label{fig:snr}}
\end{figure}

We utilize the two most sensitive bands in \wise, $W1$ and $W2$, also the two bands most efficient for selecting quasars via their IR colors.  We find 103,050 quasars with $W1$ fluxes and 103,019 with $W2$ fluxes.  The SNR distributions for these sources are shown in Figure~\ref{fig:snr}. Not only does \wise\ have the most data for the training set, it is also the deepest.

One possible complication with using \wise\ forced-photometry is that it is performed using source positions from DR12 imaging while this work uses DR8 imaging.  In addition to the known astrometry error that was corrected with DR9 \citep[especially above Dec $\sim$ 41$^{\circ}$;][]{2012ApJS..203...21A}, some objects have updated ``primary'' imaging from DR8 to DR12, causing them to shift positions or change primary run number.  Our catalogue is built by running \emph{XDQSOz} on all available data for an object, matching objects in the forced-photometry catalogues by run.  If an object's primary SDSS observation changes runs, we may not find its corresponding \wise\ data.  However, testing reveals that these issues affect a very small fraction of objects --- over 99.9\% of DR8 primary point sources match to force-photometered \wise\ sources in the same runs in DR12.  Of the unmatched objects, $\sim$85\% are below the SDSS $i$-band flux limit, and so are more likely to be spurious sources.  The remaining missing sources are likely due to astrometry changes, but these are an essentially negligible fraction in our final training set and catalogue (section 3.3).

The different epochs of the surveys utilized mean that observations span over 10 years in the observed frame (over 3 years at $z =2$).  Typical UV/optical variability over these rest-frame timescales is 0.1-0.2 mag, depending on luminosity \citep{1994MNRAS.268..305H}.  Variability in the IR over these timescales is also $\sim0.1$ mag \citep[$J$-band;][]{2010ApJ...716..530K}.  These variability levels are less than than the average errors for individual filters, and definitely less than the range in colors, and are therefore largely unimportant for this work.

\subsection{Relative-flux-redshift density model}
\emph{XDQSOz} models the relative-flux-redshift  distribution with a large number of Gaussians by deconvolving the distribution for the training sets described above with the $XD$ technique \citep[][]{Bovy:2011bt}.  This method is uniquely suited to the data available for modeling quasar flux space, because not all objects have all fluxes available and the uncertainties are heterogeneous.  $XD$ assumes that the flux uncertainties are Gaussian \citep[as they are for the SDSS;][]{2003MmSAI..74..978I, 2007AJ....134..973I}.  The spectroscopic redshifts are assumed to have null uncertainties because the typical values are on the order of $10^{-3}$ \citep[][]{2010AJ....139.2360S}, far smaller than the errors on photometric redshifts.  

As in \citet{2012ApJ...749...41B}, we use a sum of 60 Gaussians to model the deconvolved flux-redshift density.  All fits are performed in 0.2 mag wide bins of $i$-band magnitude, starting at $i=17.7$ up to $i=22.5$ (for 47 overlapping bins centered 0.1 mag apart).  All of the quasars are used in the fit to each bin, by rescaling the fluxes according to the luminosity function as described in section 2.2.1.  The model contains a total of $47 \times (60 \times[1+ d +d(d+1)/2]-1)$ parameters --- for five SDSS fluxes, two \galex\ fluxes, four UKIDSS fluxes, and two \wise\ fluxes, this is a total of 296,053 parameters (plus an additional 85,493 from the stellar fits from \emph{XDQSO} with all of the available fluxes).

\begin{figure}
\centering
\vspace{0.3cm}
\hspace{0cm}
   \includegraphics[width=4cm]{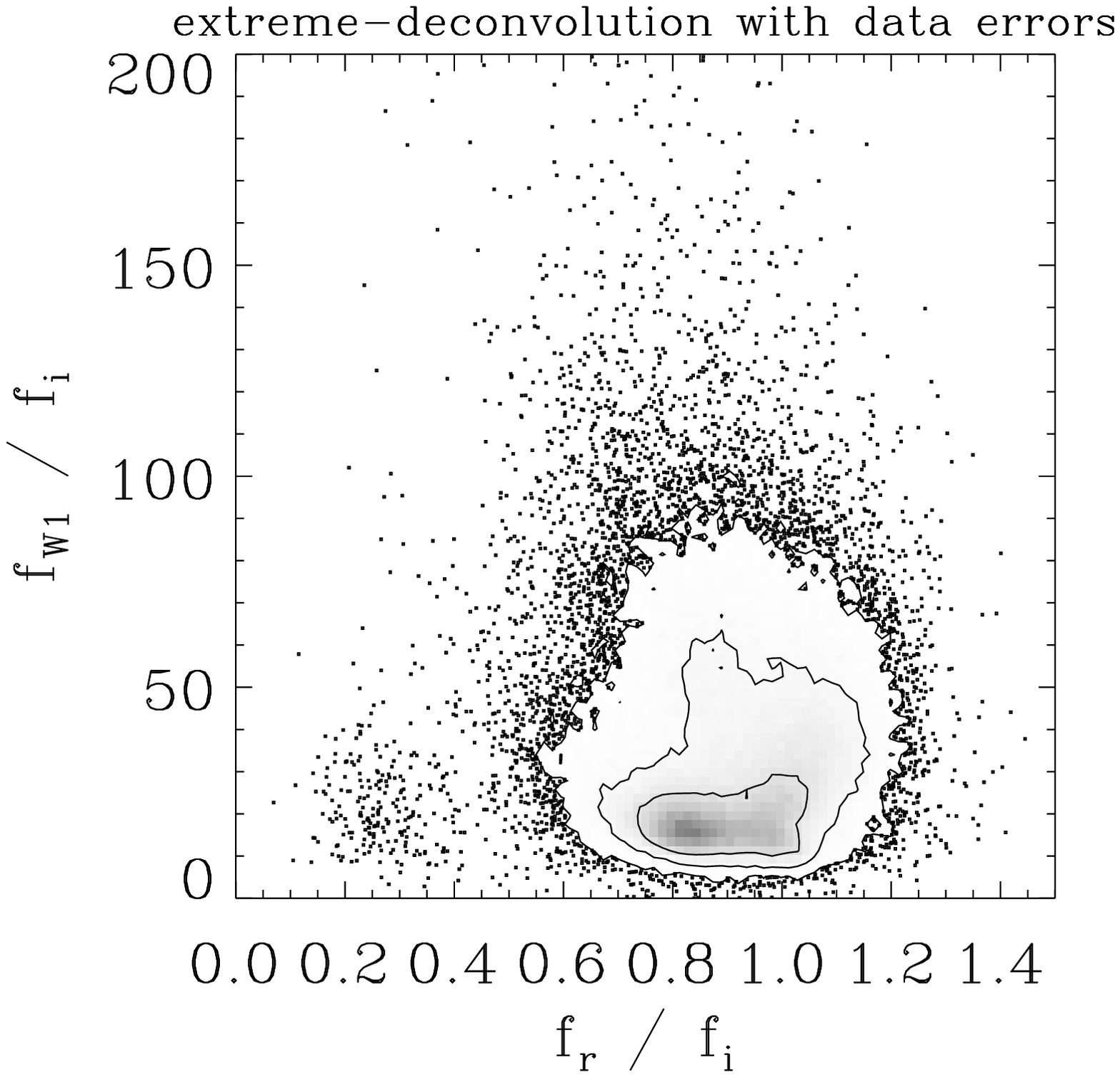}
   \includegraphics[width=4cm]{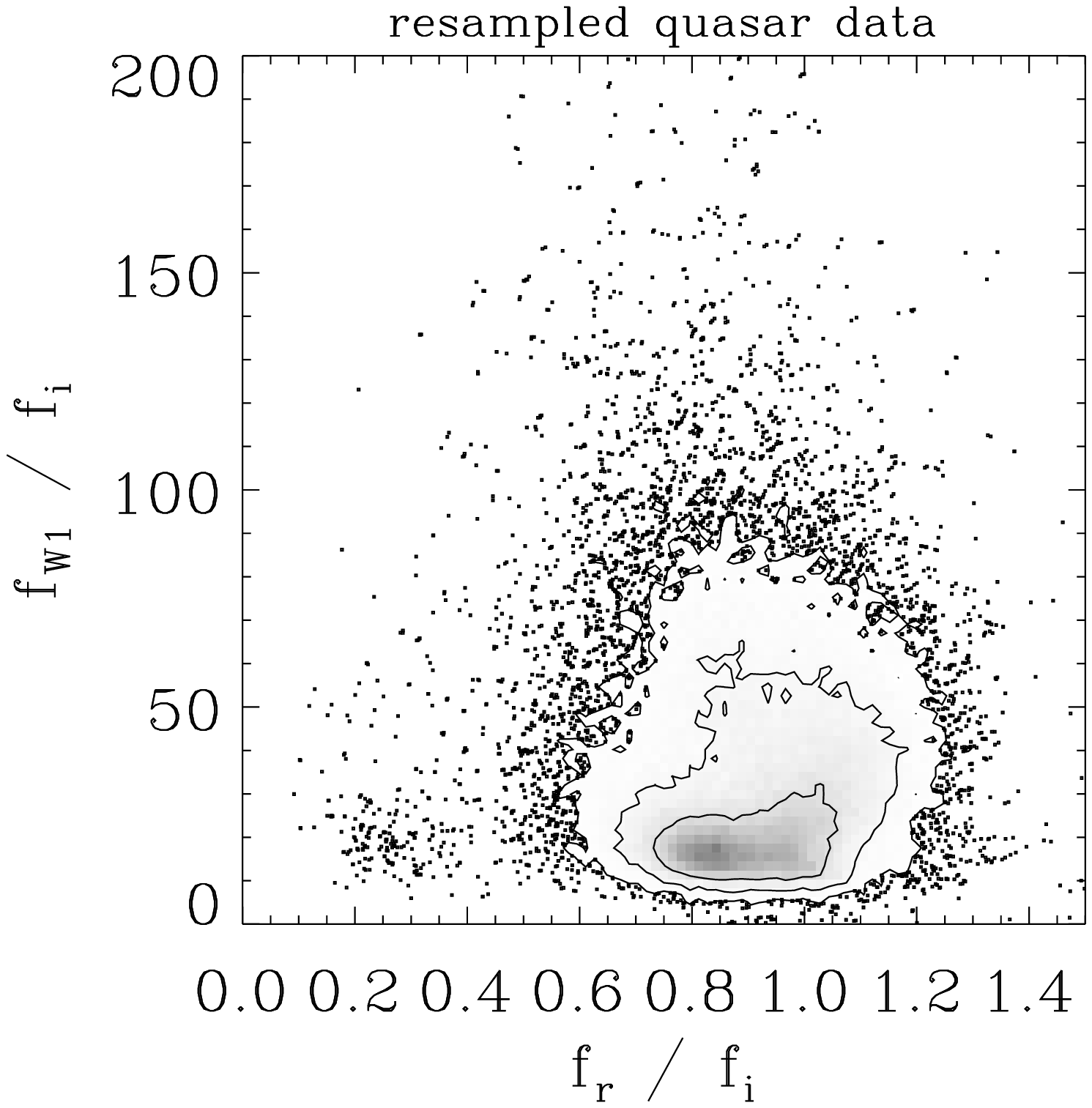}

   \vspace{0.4cm}

   \includegraphics[width=4cm]{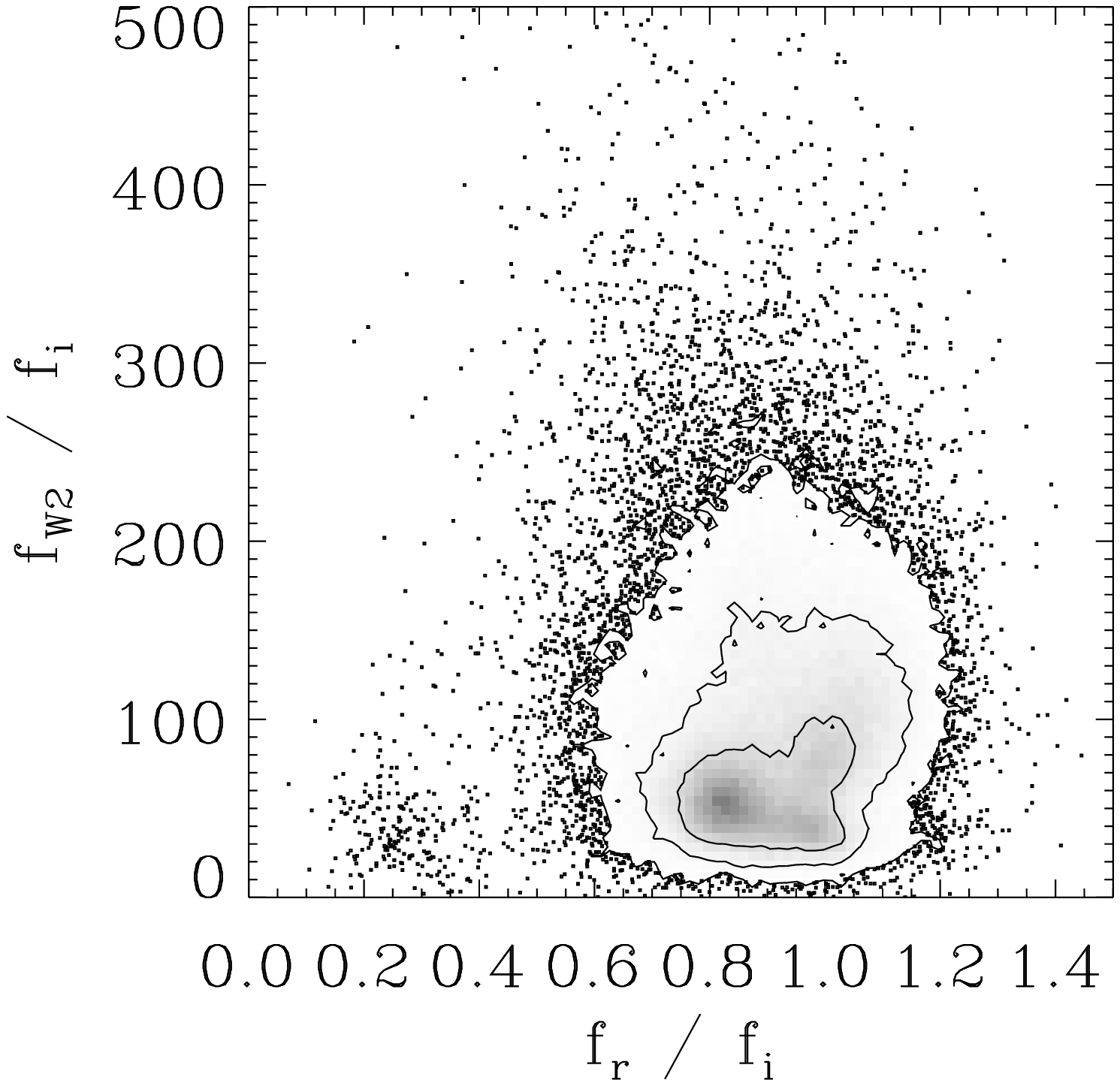}
   \includegraphics[width=4cm]{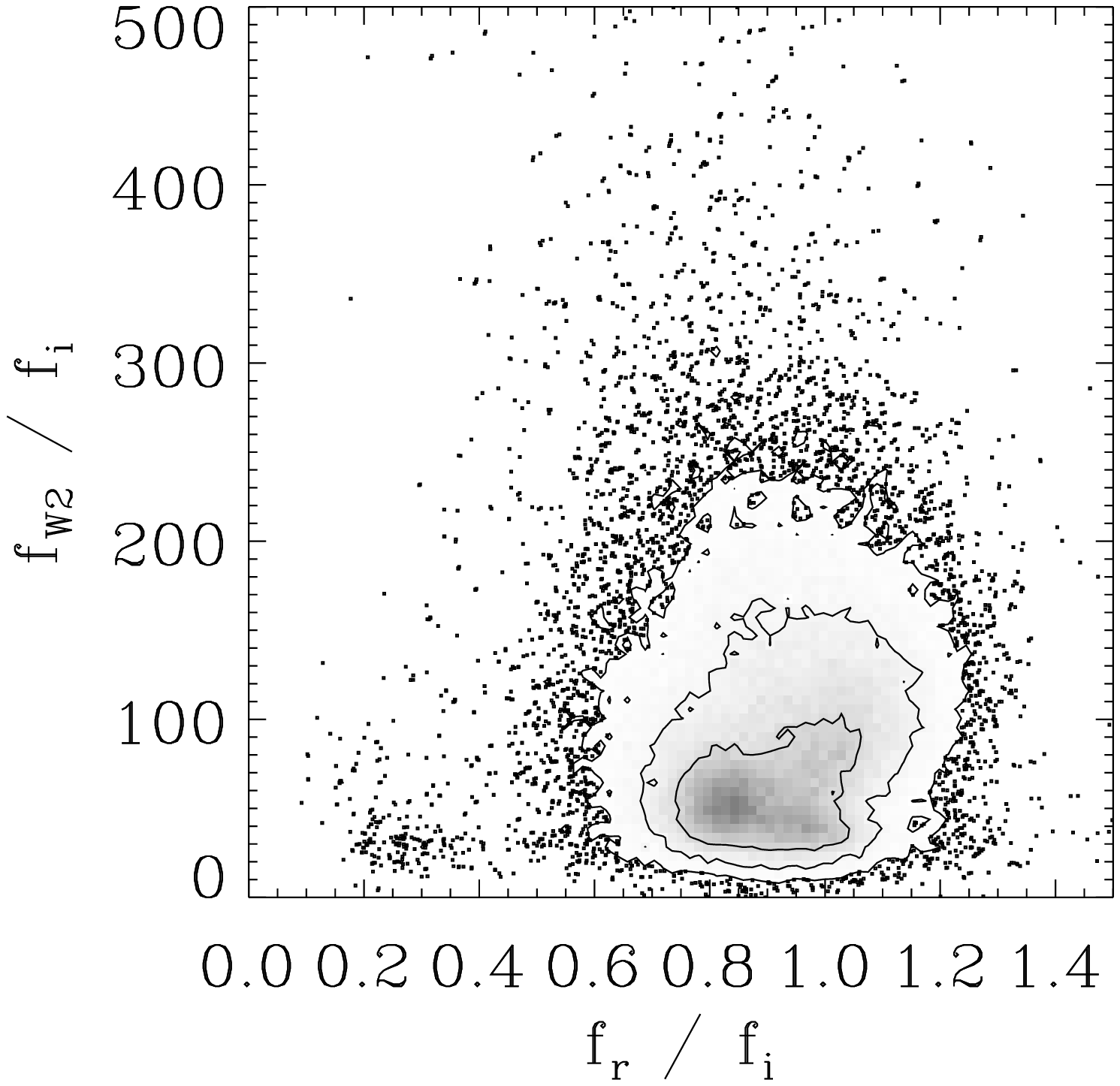}

    \vspace{0cm}
  \caption{$W1$ (top) and $W2$ (bottom) versus SDSS $r$ (all normalized by the $i$-band flux), illustrating a projection of the space in which the fits are performed. The left columns show a random sampling from the \emph{XD} fits with errors from the real data added in, and the right columns show the real quasar data resampled according to the quasar luminosity function (see sections 2.2.1 and 2.3).  Note that the \wise\ fluxes retain their native Vega zero-point, while the SDSS fluxes are in AB, hence the large relative \wise-SDSS fluxes.\label{fig:wise_vs_r}}
\vspace{0.2cm}
\end{figure}

\begin{figure}
\centering
\vspace{0.3cm}
\hspace{0cm}
   \includegraphics[width=4cm]{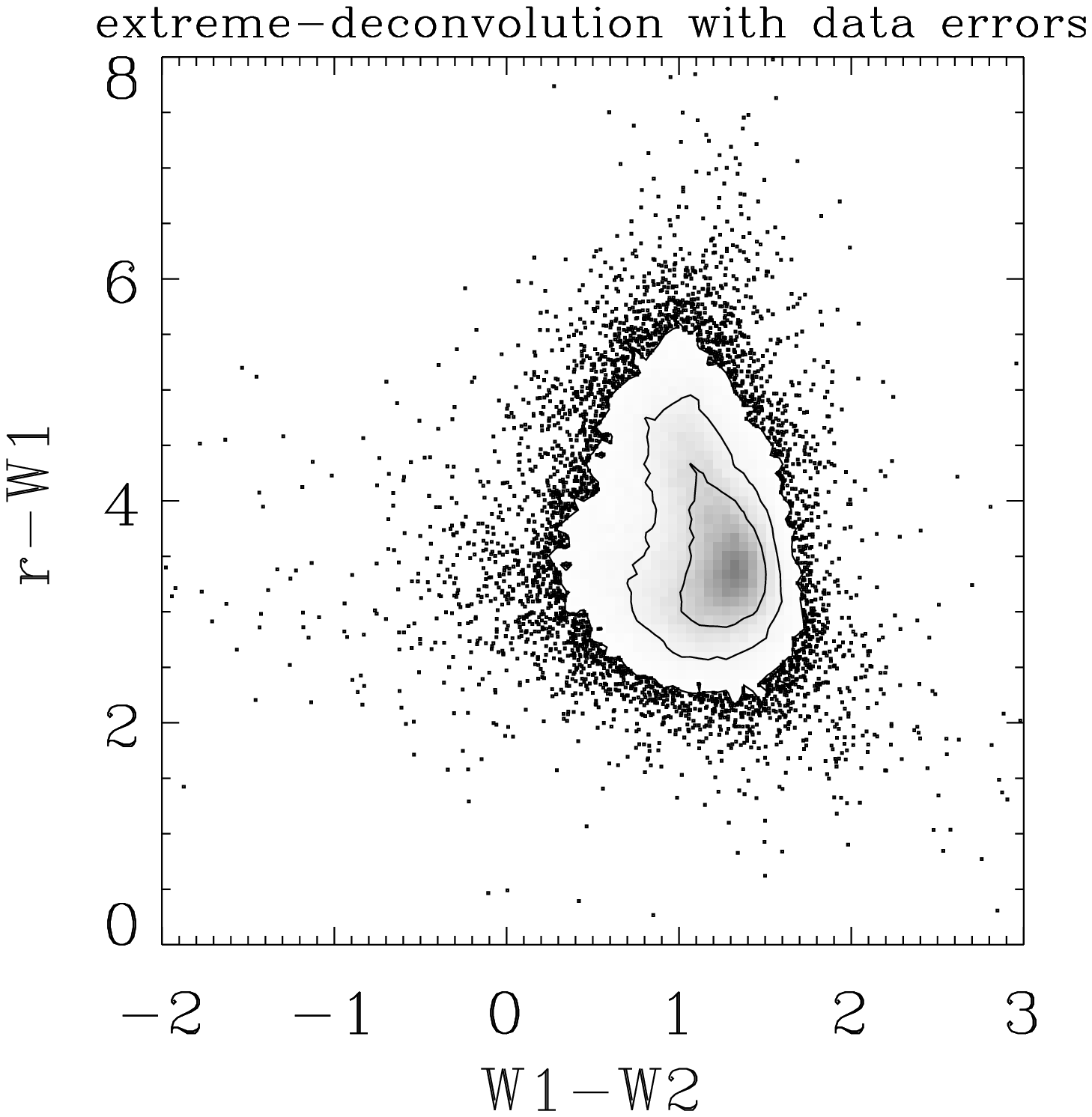}
   \includegraphics[width=4cm]{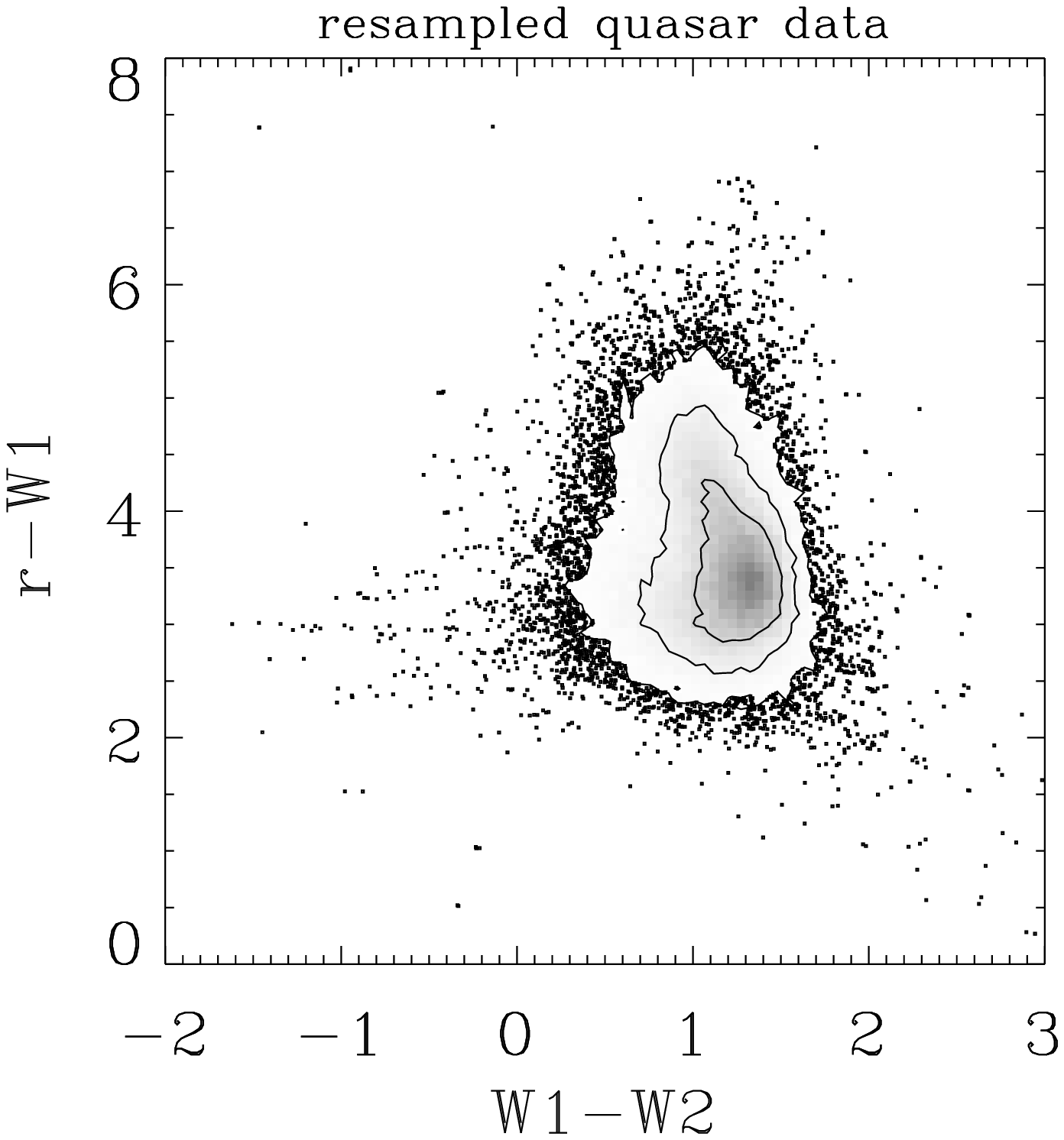}
     \vspace{0cm}
  \caption{The same as Figure~\ref{fig:wise_vs_r}, but for $r-W1$ versus $W1-W2$ colors.\label{fig:color_color}}
\end{figure}

\begin{figure}
\centering
\vspace{0.3cm}
\hspace{0cm}
   \includegraphics[width=4cm]{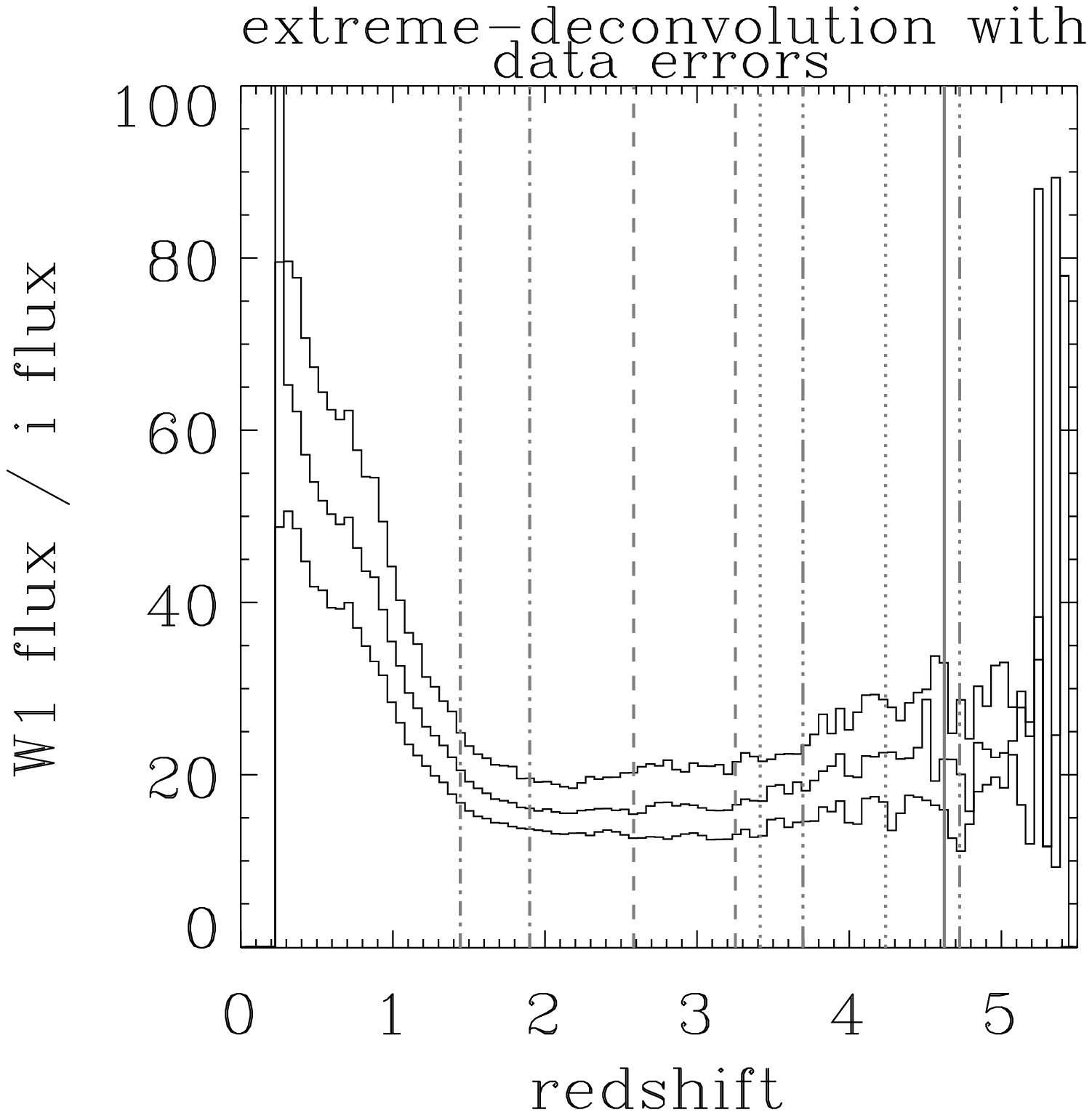}
   \includegraphics[width=4cm]{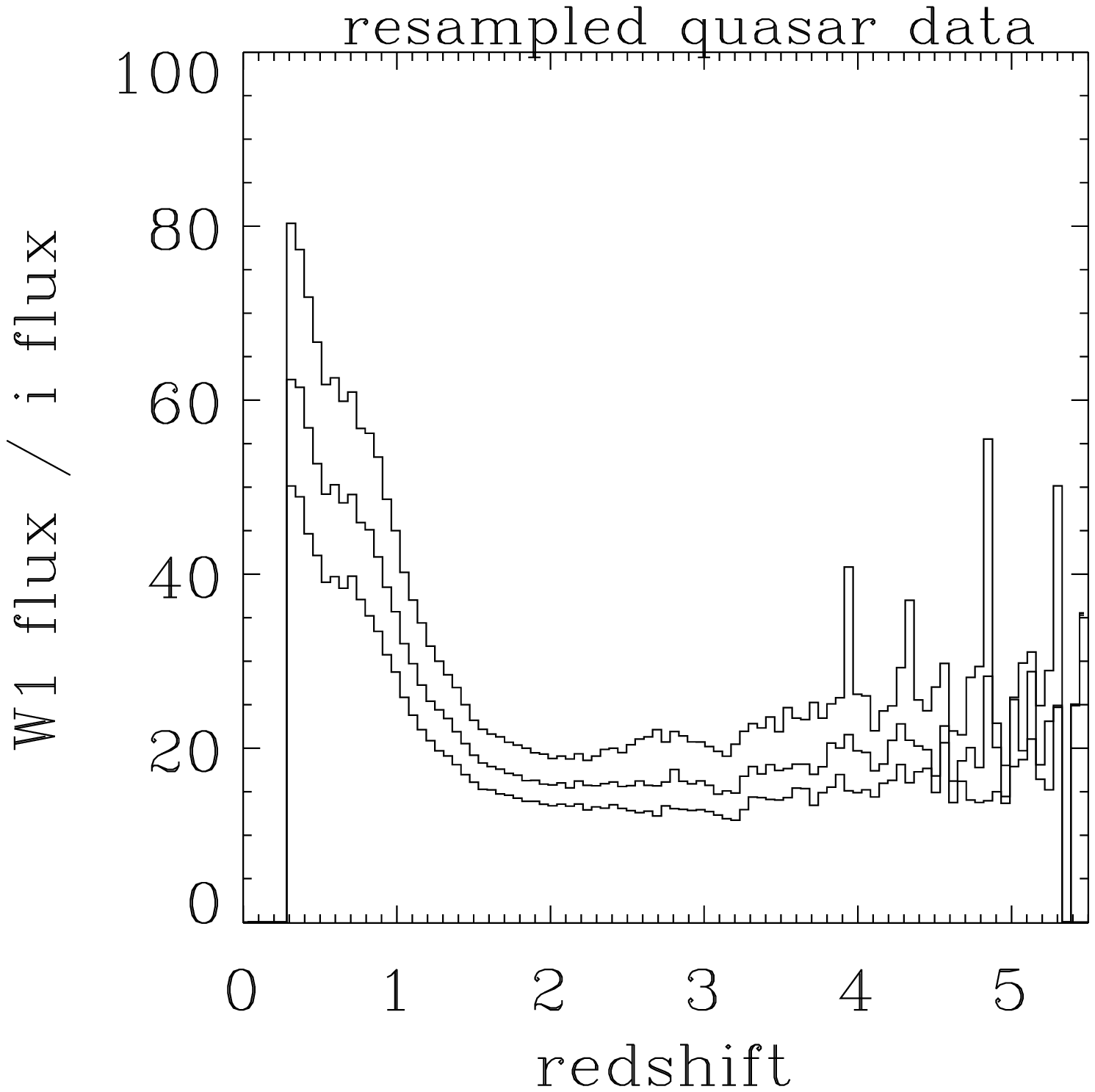}

   \vspace{0.4cm}

   \includegraphics[width=4cm]{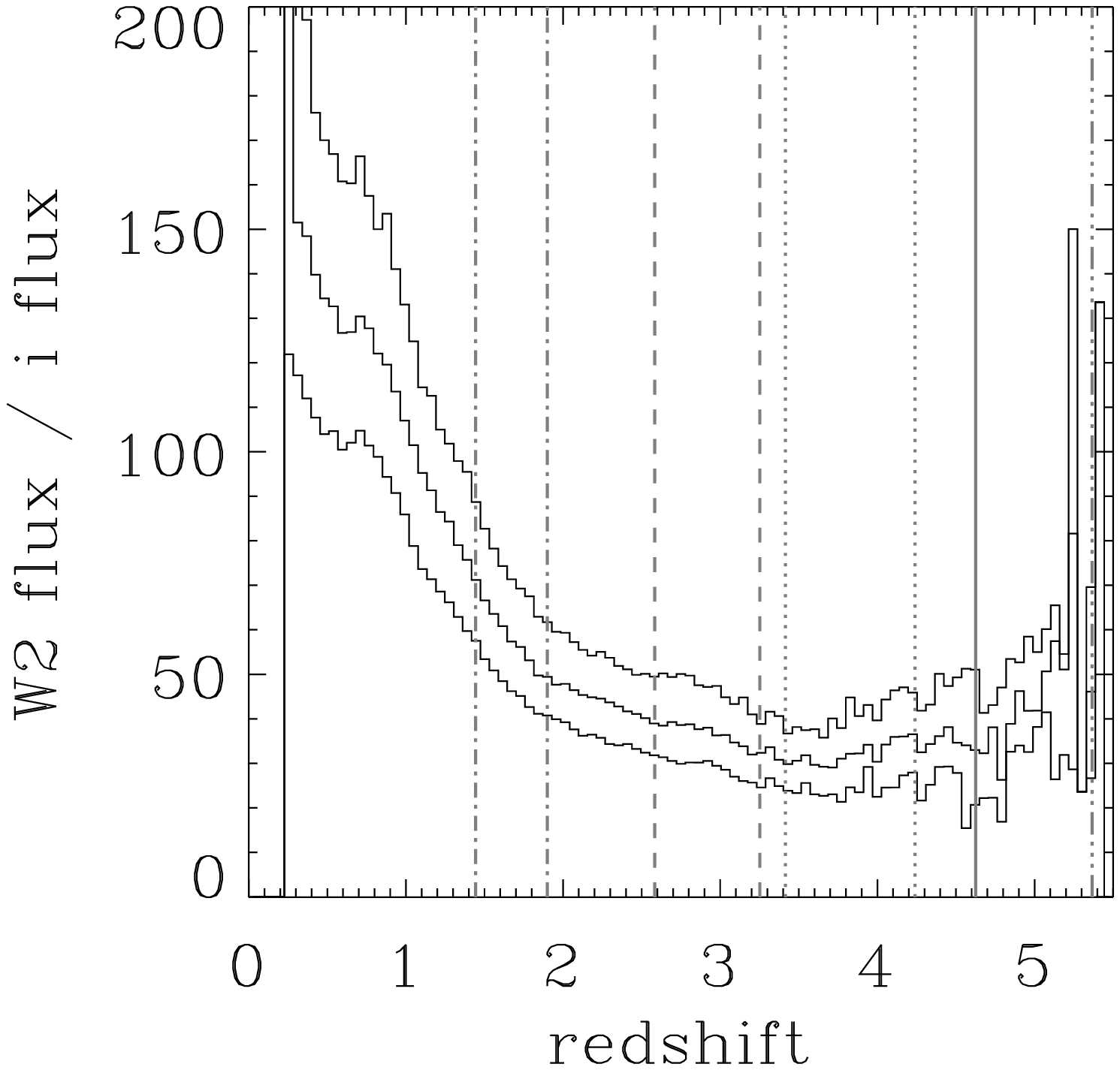}
   \includegraphics[width=4cm]{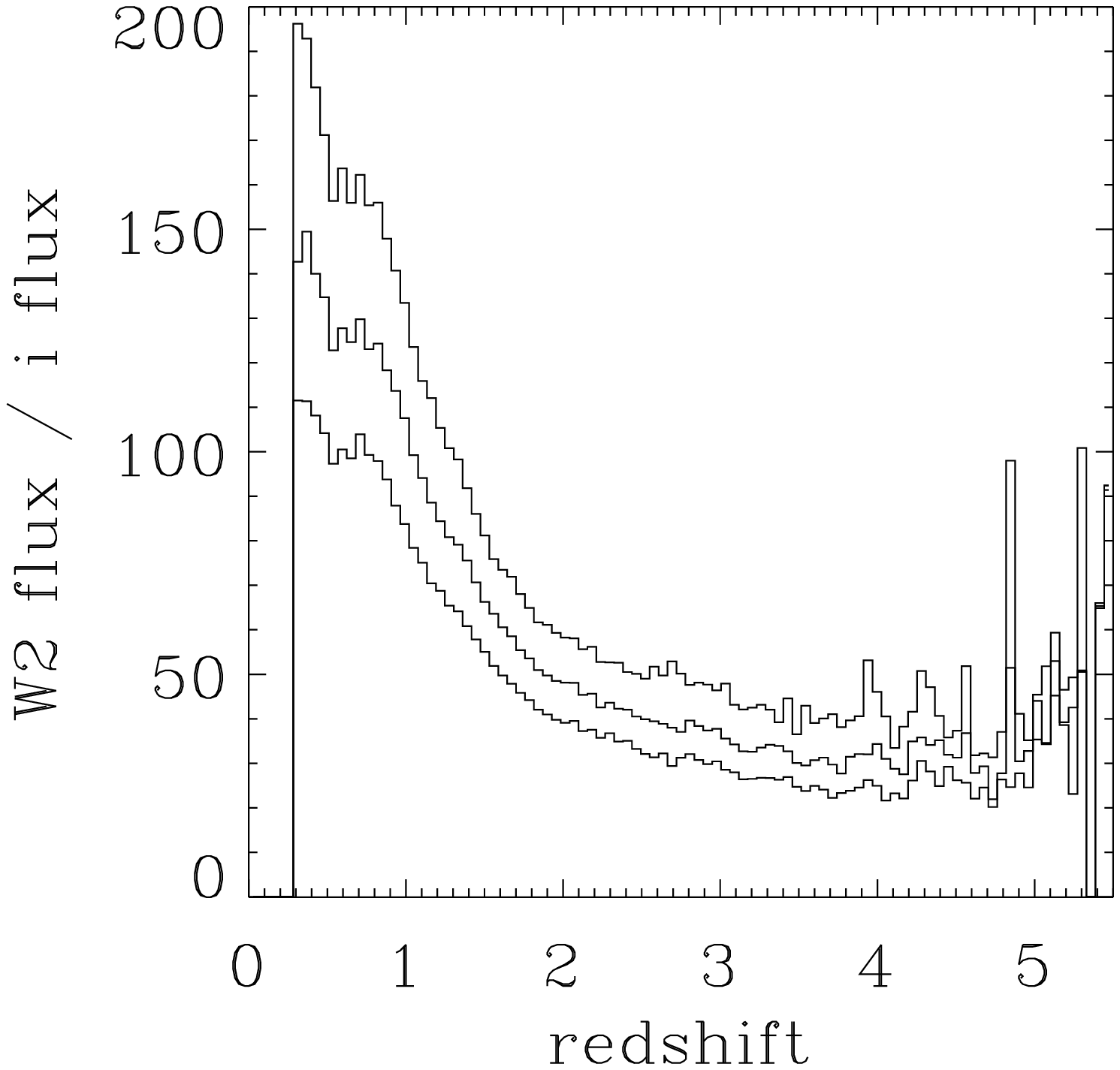}
    \vspace{0cm}
  \caption{The same as Figure~\ref{fig:wise_vs_r}, but for $W1$ and $W2$ versus $z$, again showing a projection of the space in which fits are performed.  The \wise\ fluxes retain their native Vega zero-point, hence the large relative \wise-SDSS fluxes.  The vertical lines indicate the redshifts at which various strong quasar emission lines fall within any of the relevant filters ($i$, $W1$, or $W2$): solid is Ly$\alpha$, dotted is \CIV, dashed is \CIII, dash-dotted is \MgII, dash-dot-dot-dotted is H$\alpha$.  None of these lines have a strong effect on the behavior of the normalized $W1$ or $W2$ fluxes as a function of redshift.\label{fig:flux_vs_z}}
 \vspace{0.5cm}
\end{figure}

\begin{figure}
\centering
\vspace{0.3cm}
\hspace{0cm}
   \includegraphics[width=4cm]{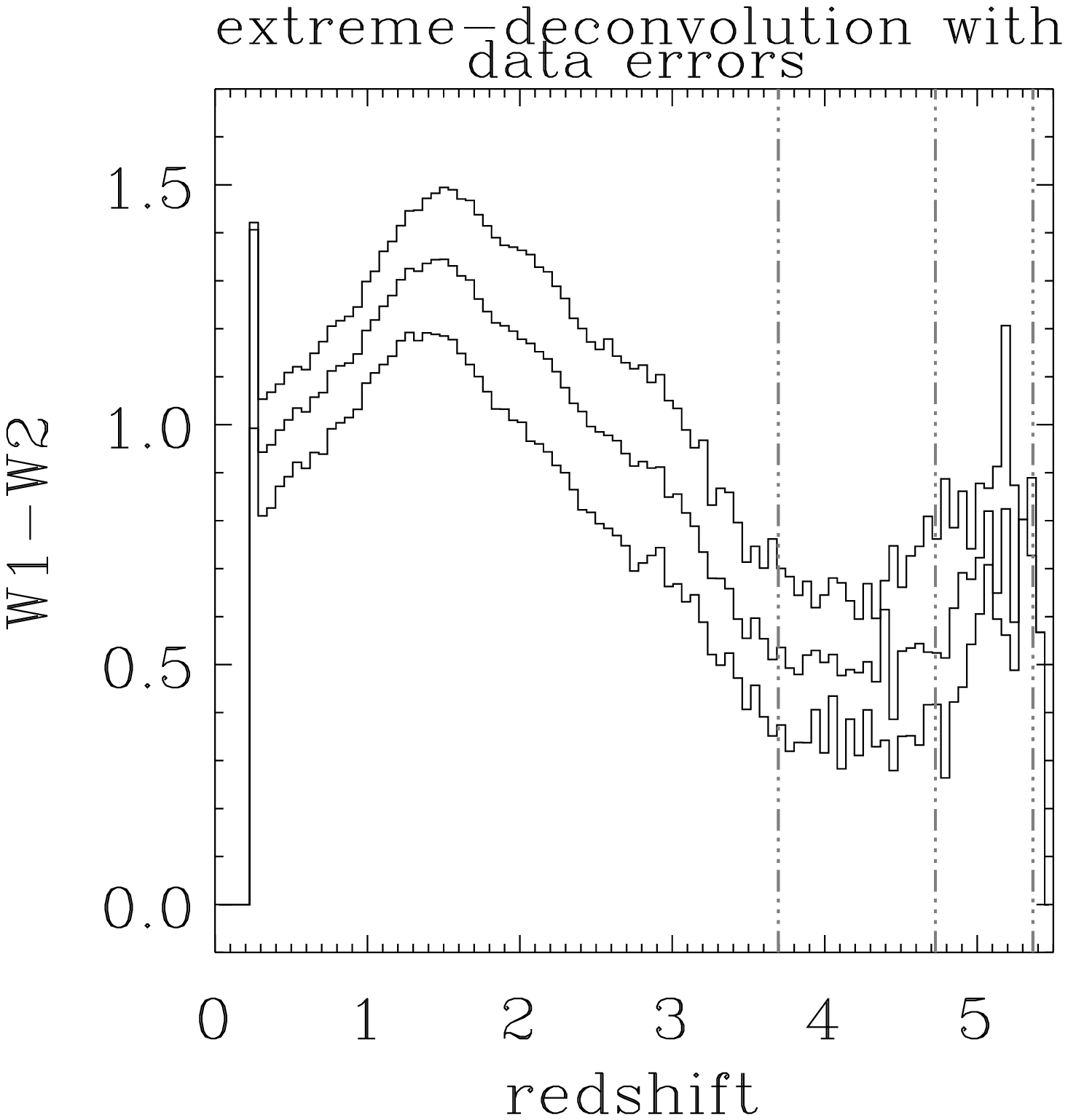}
   \includegraphics[width=4cm]{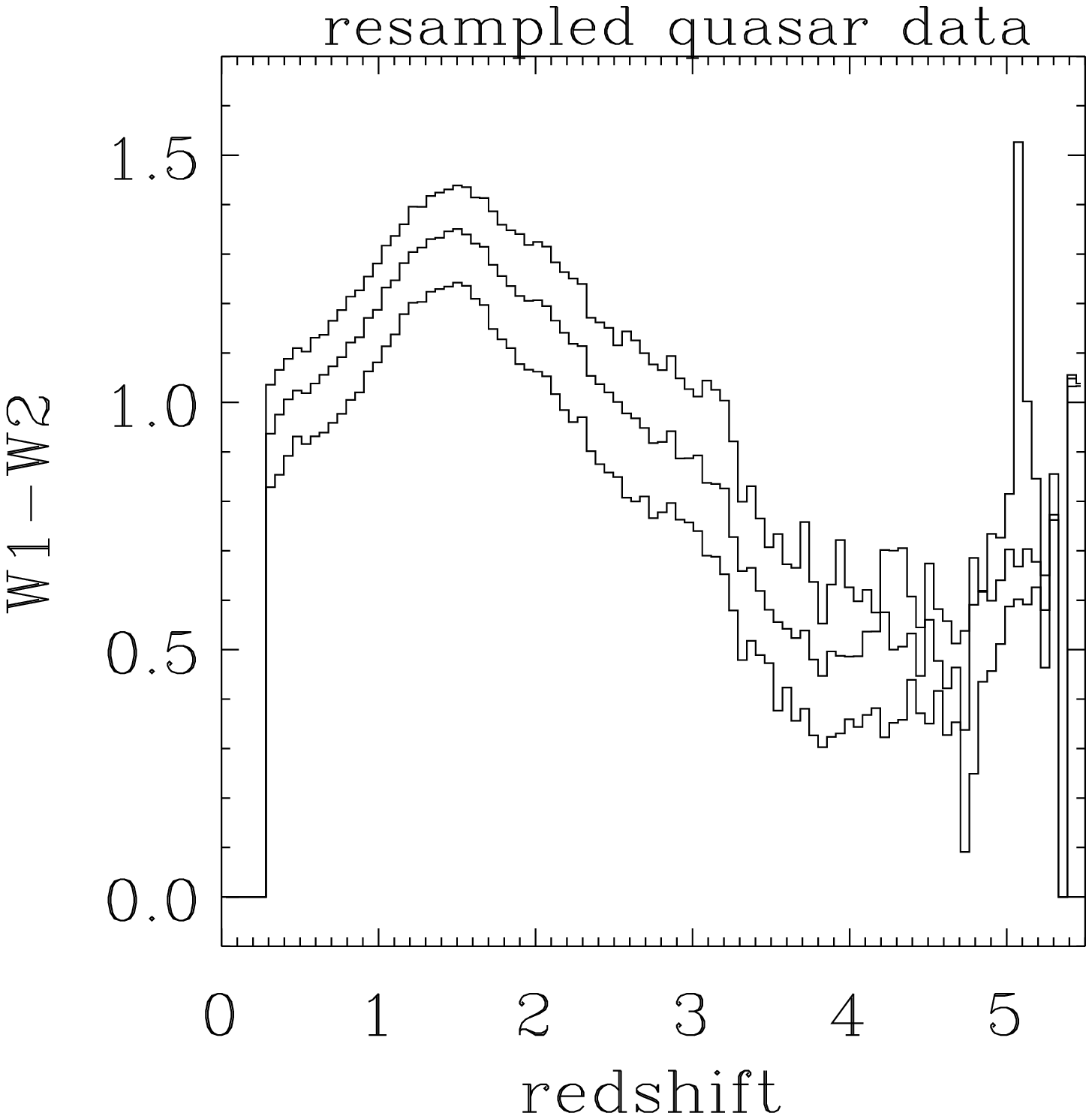}
     \vspace{0cm}
  \caption{The same as Figure~\ref{fig:flux_vs_z}, but for $W1-W2$ colors instead of fluxes.  The vertical lines indicate the range of redshifts for which H$\alpha$ lies in the $W1$ filter bandpass.\label{fig:color_vs_z}}
\end{figure}

Figures~\ref{fig:wise_vs_r} and~\ref{fig:color_color} show the flux-flux ($W1$ and $W2$ versus SDSS $r$) and color-color ($r-W1$ versus $W1-W2$) plots in one $i$-band bin ($18.6 \le i < 18.8$) for a random re-sampling from the best-fit $XD$ model (convolved with actual data errors, left columns) and the data these fits are based on, resampled according to the quasar luminosity function (right columns).  Note that all of the \wise\ data are kept in the native Vega magnitude system.  Figures~\ref{fig:flux_vs_z} and~\ref{fig:color_vs_z} show similar plots comparing samplings of the redshift fits to the redshifts of the real data.  The resampled error-convolved fits to the data are nearly identical to the real data in all cases, illustrating the accuracy of the fits.  We also point out in Figure~\ref{fig:color_vs_z} that above $z \sim 3$ the colors fall below $W1-W2 = 0.8$, a common cut used to select quasars from \wise\ data alone \citep[e.g.][see section 3.2]{2012ApJ...753...30S}

\section{RESULTS \& DISCUSSION}
\subsection{Performance}

\begin{figure*}
\centering
\hspace{0cm}
   \includegraphics[width=5.8cm]{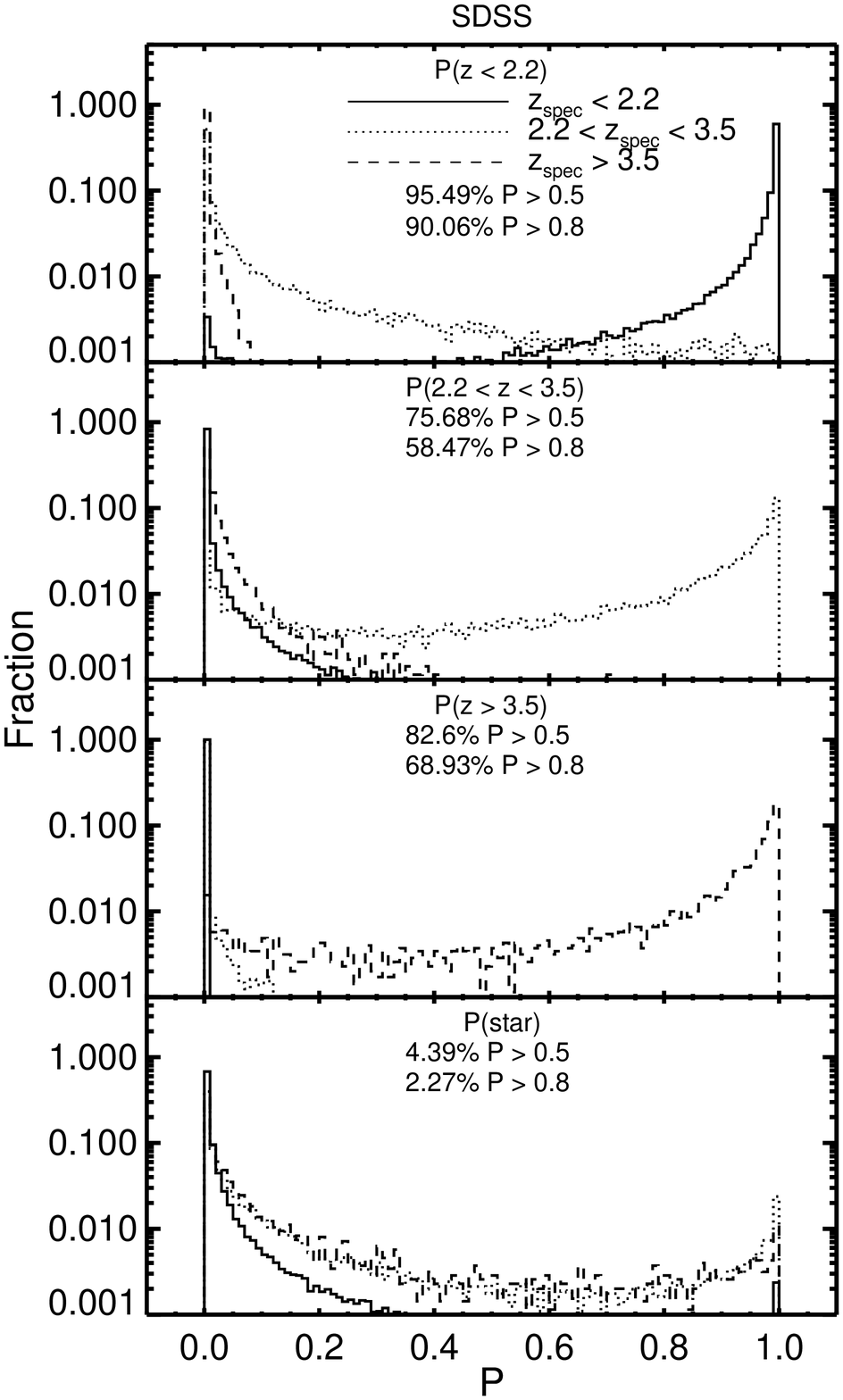}
   \includegraphics[width=5.8cm]{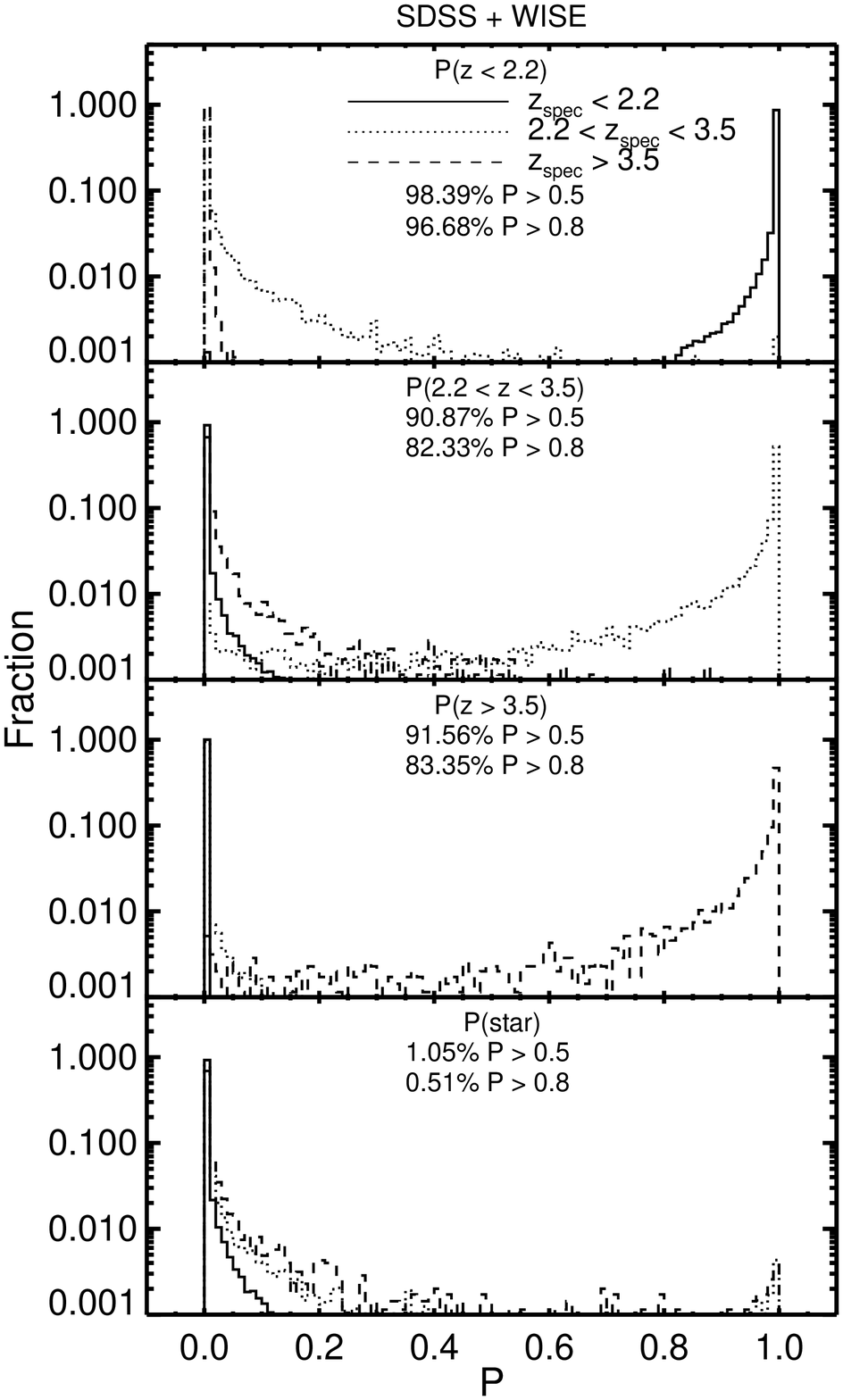}
   \includegraphics[width=5.8cm]{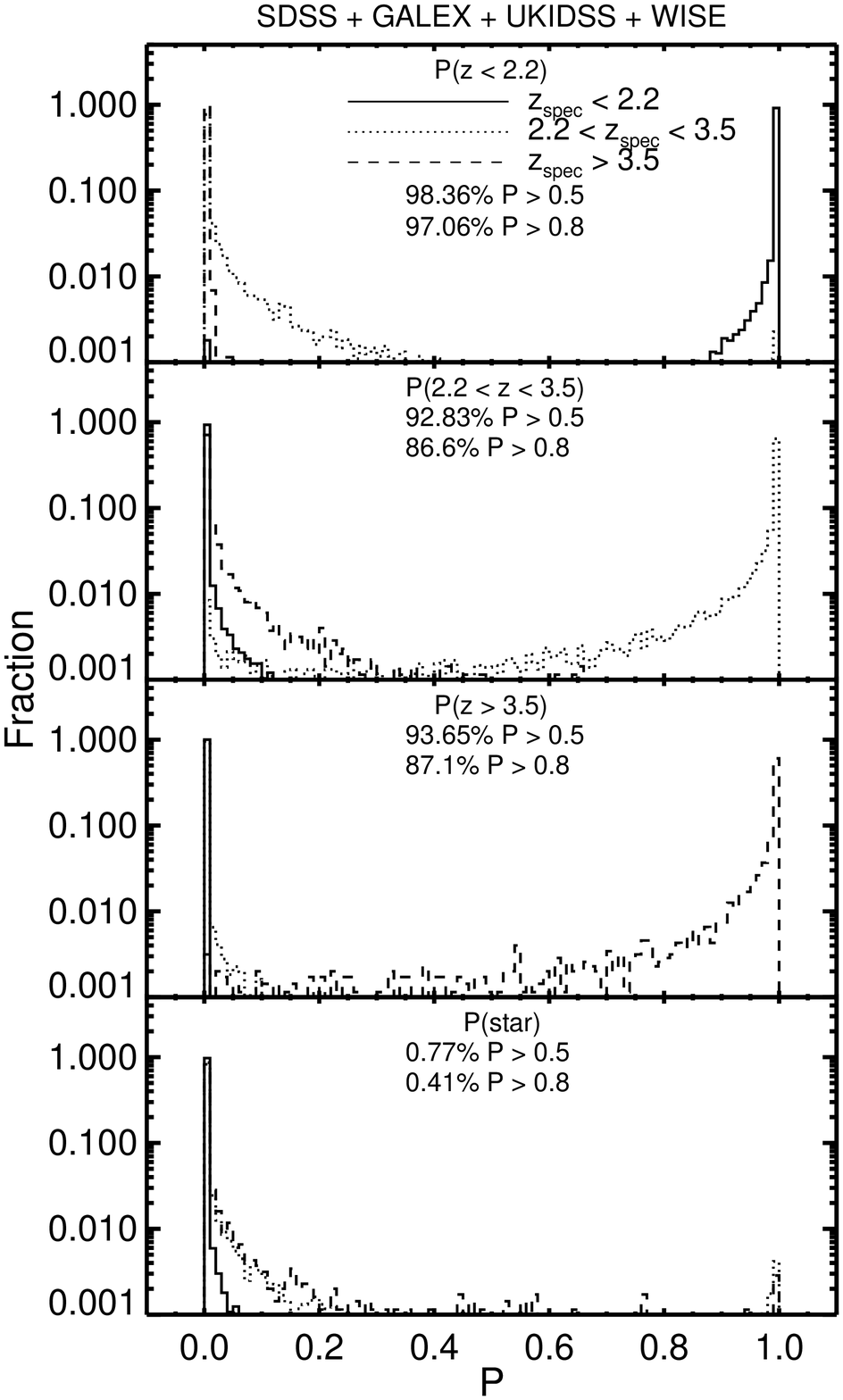}
   \vspace{0cm}
  \caption{The \emph{XDQSOz} probability that a spectroscopically confirmed quasar is a quasar in broad redshift bins, using just SDSS photometry (left), SDSS$+$\wise\ photometry (center), and SDSS$+$\galex$+$UKIDSS$+$\wise\ photometry (right).  Within each panel are the distributions of the probabilities that objects are low-redshift quasars ($z<2.2$, first panel), medium-redshift quasars ($2.2 < z < 3.5$, second panel), high-redshift quasars ($z > 3.5$, third panel), or stars (bottom panel).  Line styles indicate the spectroscopic classification of each subset.  The percentage of objects that belong to each redshift bin of interest and meet the thresholds $P_{\textrm{QSO}} > 0.5$ and $P_{\textrm{QSO}} > 0.8$ are listed in each panel.  In the case of $P_{\textrm{star}}$, these percentages are for quasars at all redshifts.  Similar figures showing additional combinations of data are provided in the appendix.\label{fig:quasars}}
\end{figure*}

\begin{figure*}
\centering
\hspace{0cm}
   \includegraphics[width=5.8cm]{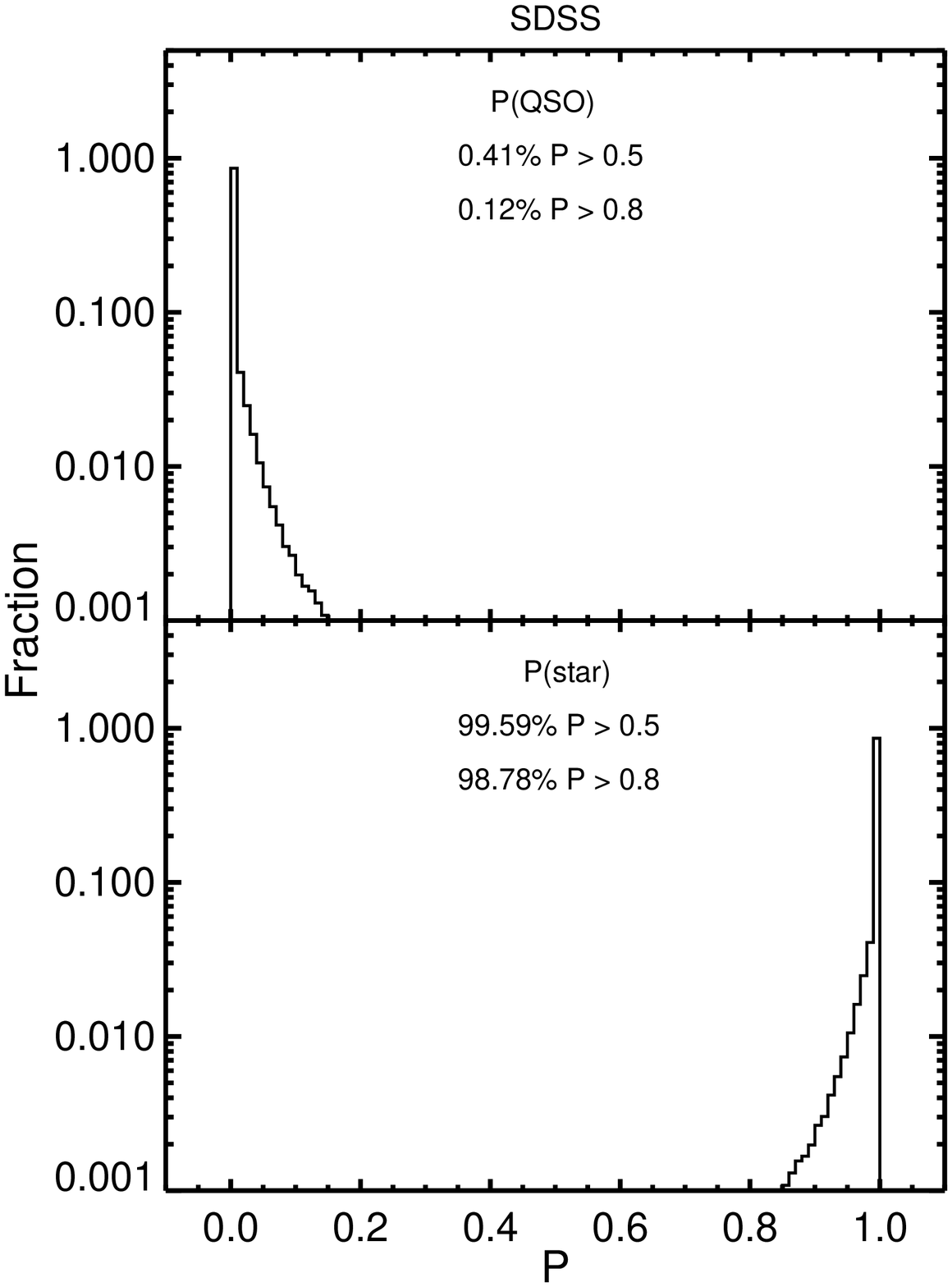}
   \includegraphics[width=5.8cm]{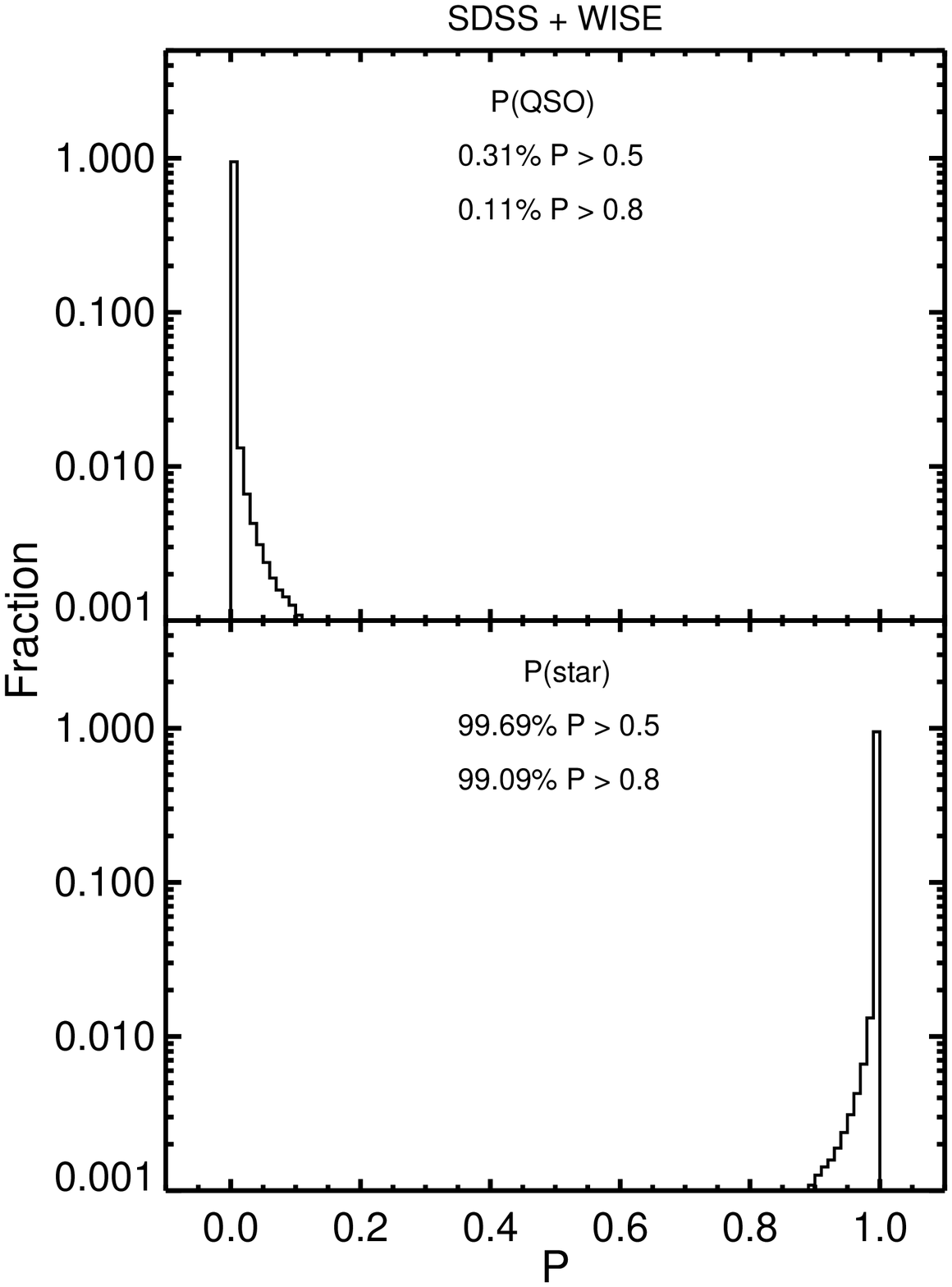}
   \includegraphics[width=5.8cm]{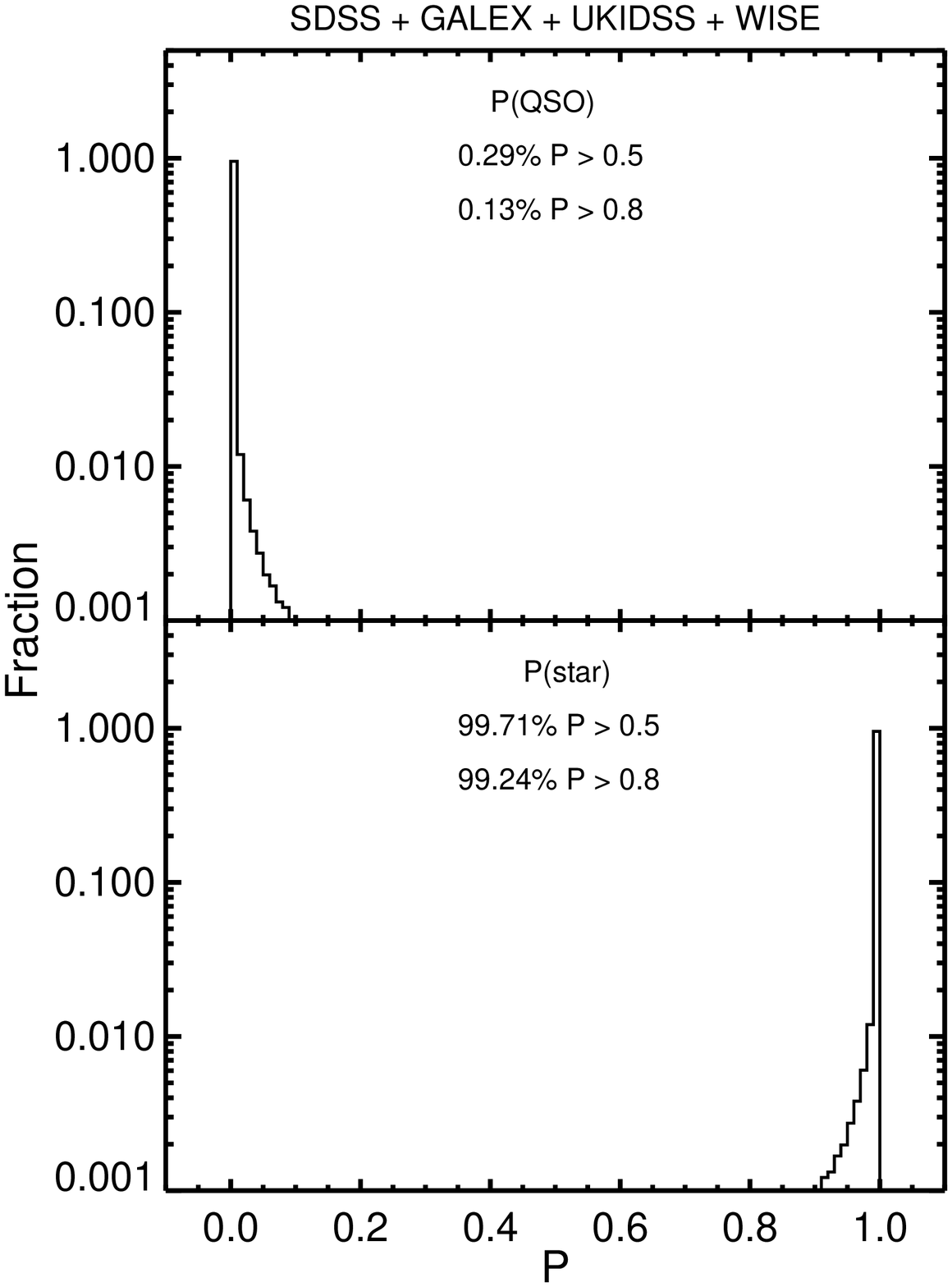}
    \vspace{0cm}
  \caption{The \emph{XDQSOz} probabilities for the stellar training set, using various combinations of photometric data.  The top panels shows the probability that an object is a quasar at any redshift, and the bottom shows the probability that it is a star.  Note that there is some contamination from quasars in the stellar training set (on the order of $\sim$1\%).  Similar figures showing additional combinations of data are provided in the appendix.\label{fig:stars}}
\end{figure*}

Sections 5 and 6 of \citet{2012ApJ...749...41B} already demonstrated that \emph{XDQSOz} outperforms many other methods when calculating quasar probabilities and photometric redshifts.  Here, we simply present the improvements with the addition of \wise\ flux information.  In order to assess the performance of \emph{XDQSOz}, we calculate the quasar probability $P_{QSO}$ for the known SDSS DR7 spectroscopic quasars as well as the stars in our  training set.  We use the entire samples for all of the below analysis, including additional fluxes where they are available (and the analysis requires them).  

First, we present the results using the broad redshift bins of the original \emph{XDQSO} ($z < 2.2$, $2.2 < z < 3.5$, and $z > 3.5$), shown in Figures~\ref{fig:quasars} (quasars) and~\ref{fig:stars} (stars).  These figures show a comparison between using only SDSS fluxes, SDSS$+$\wise, and SDSS$+$\galex$+$UKIDSS$+$\wise.  The percentages of the subsamples belonging to the redshift bins of interest that are identified as quasars at $P > 0.5$ and $P > 0.8$ are given in each panel for easy comparison.

The method does an excellent job identifying quasars in the appropriate redshift bins, and not as stars, even with only the SDSS fluxes.  However, there is significant improvement with the addition of \wise\ information, especially at $z > 2.2$.  At low redshift, gains are on the order of $\sim$5\%. In the mid-redshift range, the improvement is $\sim$15-20\%.  At high redshifts, the improvement is $\sim$10-15\%.  The addition of \galex\ and UKIDSS data adds improvements of a few percent at most over the SDSS$+$\wise\ information --- adding \wise\ fluxes has the single greatest effect on the performance of \emph{XDQSOz} when classifying quasars.  Similar results are found when testing the stellar training set (Figure~\ref{fig:stars}).  Very few stars are identified as quasars, with a significant improvement when \wise\ data are added and only a marginal further increase with \galex\ and UKIDSS information.  Similar plots for other combinations of data are provided in the appendix for comparison.

The top panels of Figure~\ref{fig:xdqsoz} show a similar analysis, using probabilities that an object is a quasar at \textit{any} redshift.   With just the SDSS data alone, $\sim$90\% of the quasars are recovered with $P_{\textrm{QSO}} > 0.8$.  The addition of \wise\ photometry improves this \textit{dramatically}, increasing the fraction with $P_{\textrm{QSO}} > 0.8$ to $\sim$97\%.  The further addition of \galex\ and UKIDSS data only increases the performance by another $\sim$1\%.  Without the use of \wise\ these percentages only reach $\sim$93\% for any other combinations of data (see the appendices).

Because our training set has additional \galex\ and UKIDSS data not available in the training set of \citet{2012ApJ...749...41B}, we explicitly compare the percentage of known quasars identified at $P_{\textrm{QSO}} > 0.8$ using our new fits using SDSS$+$\galex\ data and SDSS$+$UKIDSS data and the same fits of \citet{2012ApJ...749...41B} available from the previous \emph{XDQSOz} release.  The new data do not greatly alter the fits, but the changes are large enough such that the new fits improve the ability of \emph{XDQSOz} to identify quasars by $\sim$0.3\% for SDSS$+$\galex, and $\sim$0.1\% for SDSS$+$UKIDSS.  

The bottom panels of Figure~\ref{fig:xdqsoz} show the comparison of the peaks of the redshift PDFs versus the spectroscopic redshifts.  These panels include all objects in all of the panels, regardless of which data are available for a given source (e.g.\ an object is included in the panel using fits to all of the fluxes even if it doesn't have all of the fluxes available) or if its redshift PDF has multiple peaks (see below).  There is a clear improvement over just SDSS data with the addition of \wise, and again only a marginal further improvement by adding \galex\ and UKIDSS data.  The groups of points that are far off-axis and most strongly present in the SDSS-only panel are largely due to strong quasar emission lines (e.g.\ \MgII\ or \CIV) moving between filters \citep[see e.g.\ Figure 12 of][]{2008ApJ...683...12B}, and additional data at other wavelengths helps to reduce these affects.

To better illustrate the benefit of \wise, Figure~\ref{fig:z_comp} shows the difference between the peak photometric redshifts and the spectroscopic redshifts for several combinations of data.  The use of \wise\ data clearly has the single largest effect, with smaller additional gains from including \galex\ and UKIDSS.  In the case of SDSS$+$\wise, $\sim$95\% of the objects have $\Delta z = | z_{\textrm{spec}} - z_{\textrm{phot}} | < 0.3$ and $\sim$76\% have $\Delta z < 0.1$.  The large bumps around $ |\Delta z| =1.5$ in most combinations of data without \wise\ represent the off-axis clumps discussed above, and are largely removed with the additional \wise\ data because only one emission line (H$\alpha$) affects the \wise\ bandpasses, and only at higher redshifts ($z \gtrsim 3.5$; see e.g.\ Figure~\ref{fig:flux_vs_z}),.

\begin{figure*}
\centering
\hspace{0cm}
   \includegraphics[width=5.8cm]{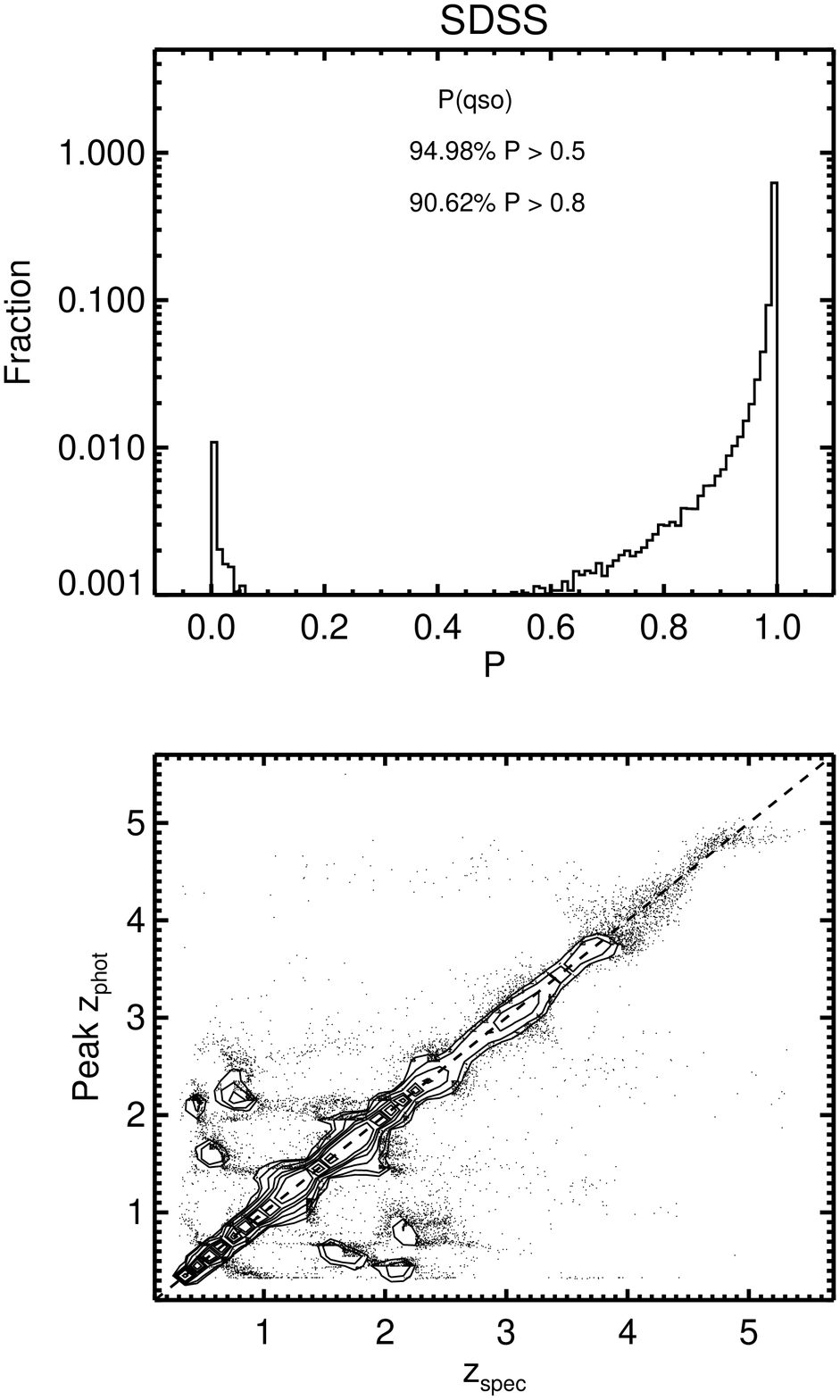}
   \includegraphics[width=5.8cm]{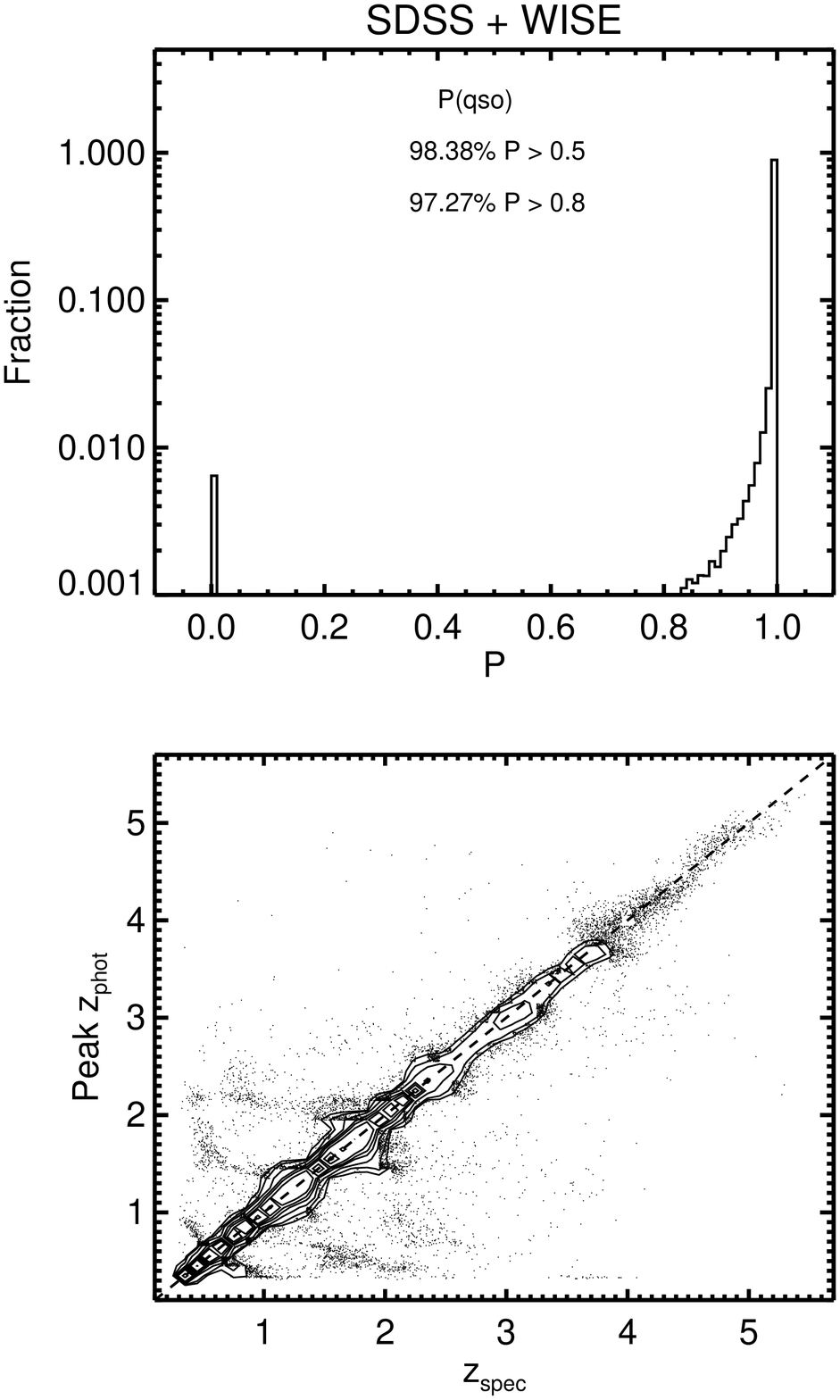}
   \includegraphics[width=5.8cm]{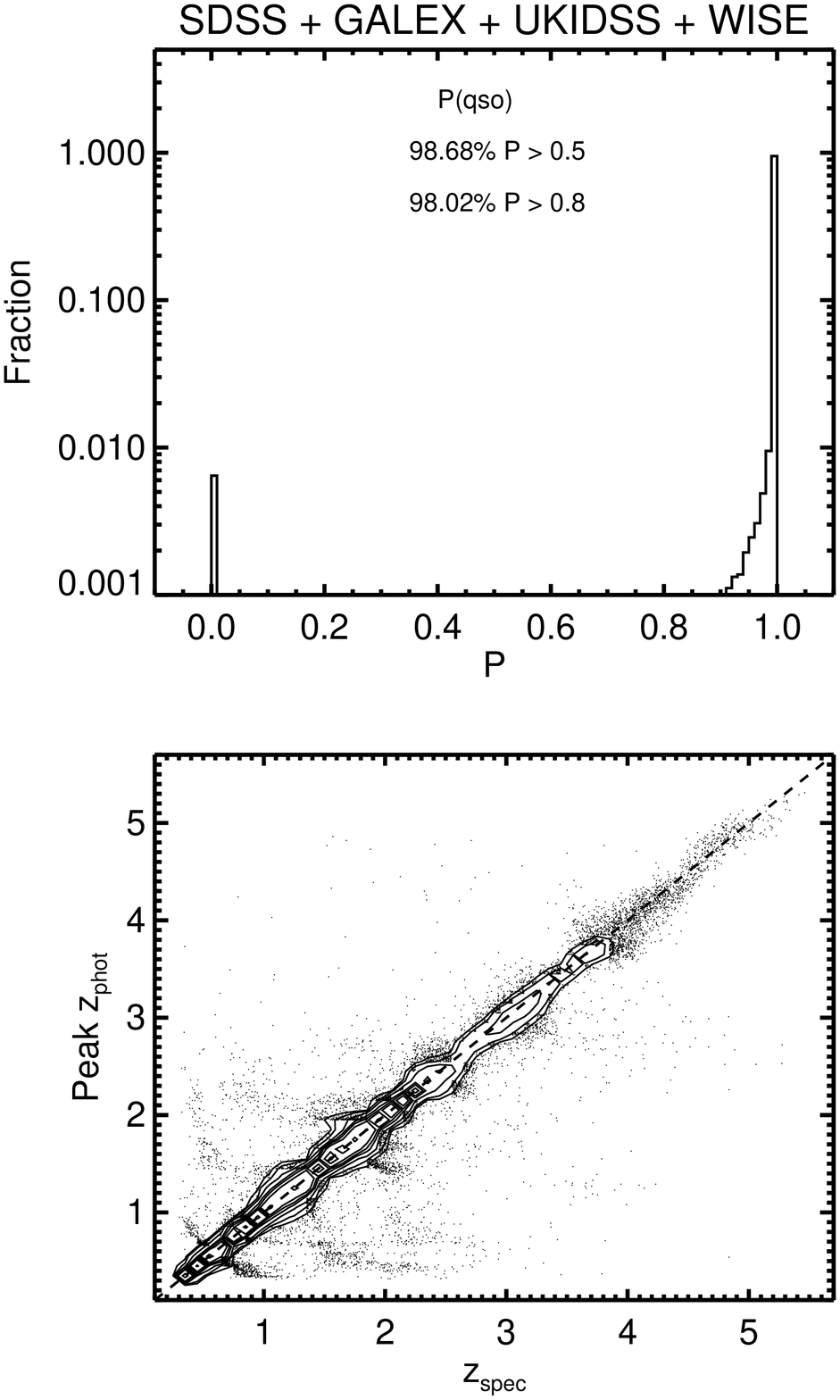}
   \vspace{0cm}
  \caption{The three panels show the distributions of $P_{\textrm{QSO}}$ over all redshifts for known spectroscopic quasars from SDSS DR7 (top panels), and a comparison of the spectroscopic redshifts and the peak of the photo-$z$ PDFs (bottom panels).  From left to right, the results using just SDSS, SDSS and \wise, and all available photometry.  Similar figures showing additional combinations of data are provided in the appendix.\label{fig:xdqsoz}}
\end{figure*}

\begin{figure}
\centering
\hspace{0cm}
   \includegraphics[width=7cm]{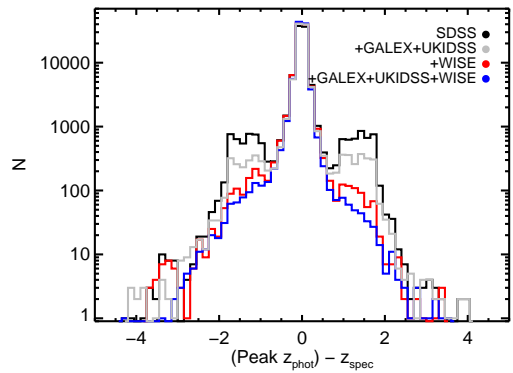}
   \vspace{0cm}
  \caption{The distribution of the difference between the peak of the photometric $z$ PDF and the spectroscopic $z$ for the DR7 quasar sample, using several different combinations of data.  Using \wise\ fluxes clearly narrows the distribution significantly, more so than using SDSS$+$\galex$+$UKIDSS.  Gains from using all of the available data are small compared to using just SDSS$+$\wise.\label{fig:z_comp}}
\end{figure}

\begin{figure}
\centering
\hspace{0cm}
   \includegraphics[width=7.5cm]{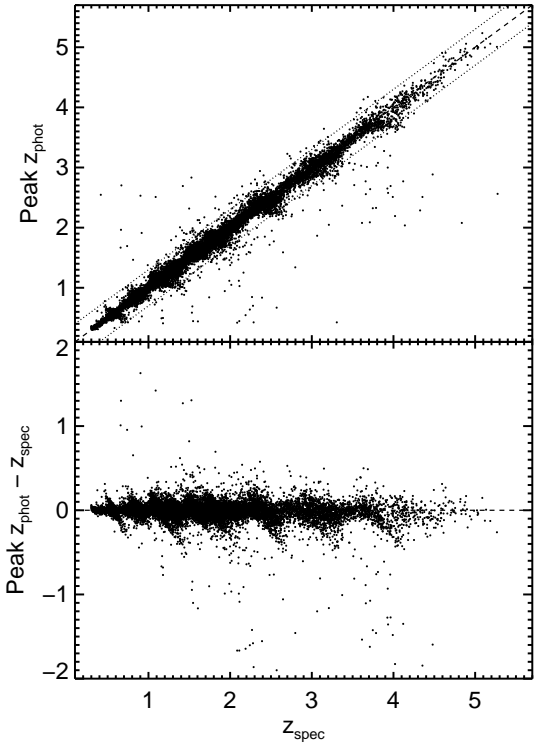}
   \vspace{0cm}
  \caption{A comparison of the photometric versus spectroscopic redshifts, limited to only objects with all of the fluxes available and with single-peaked redshift PDFs.  The dashed line shows the one-to-one line, and the dotted lines indicate $\Delta z = 0.3$.\label{fig:onepeak}}
\end{figure}

If we limit the analysis of the photometric redshifts to only objects with photometry available in all filters, and only those that have single-peaked\footnote{A peak in the redshift distribution is defined as any continuous region above the uniform distribution between $0.3 < z < 5.5$} redshift PDFs, the performance is outstanding.  This is shown in Figure~\ref{fig:onepeak}.  In this case, $\sim$98\% of the 21,929 objects have $\Delta z < 0.3$, and $\sim$84\% have $\Delta z < 0.1$.

\begin{figure*}
\centering
\hspace{0cm}
   \includegraphics[width=8.5cm]{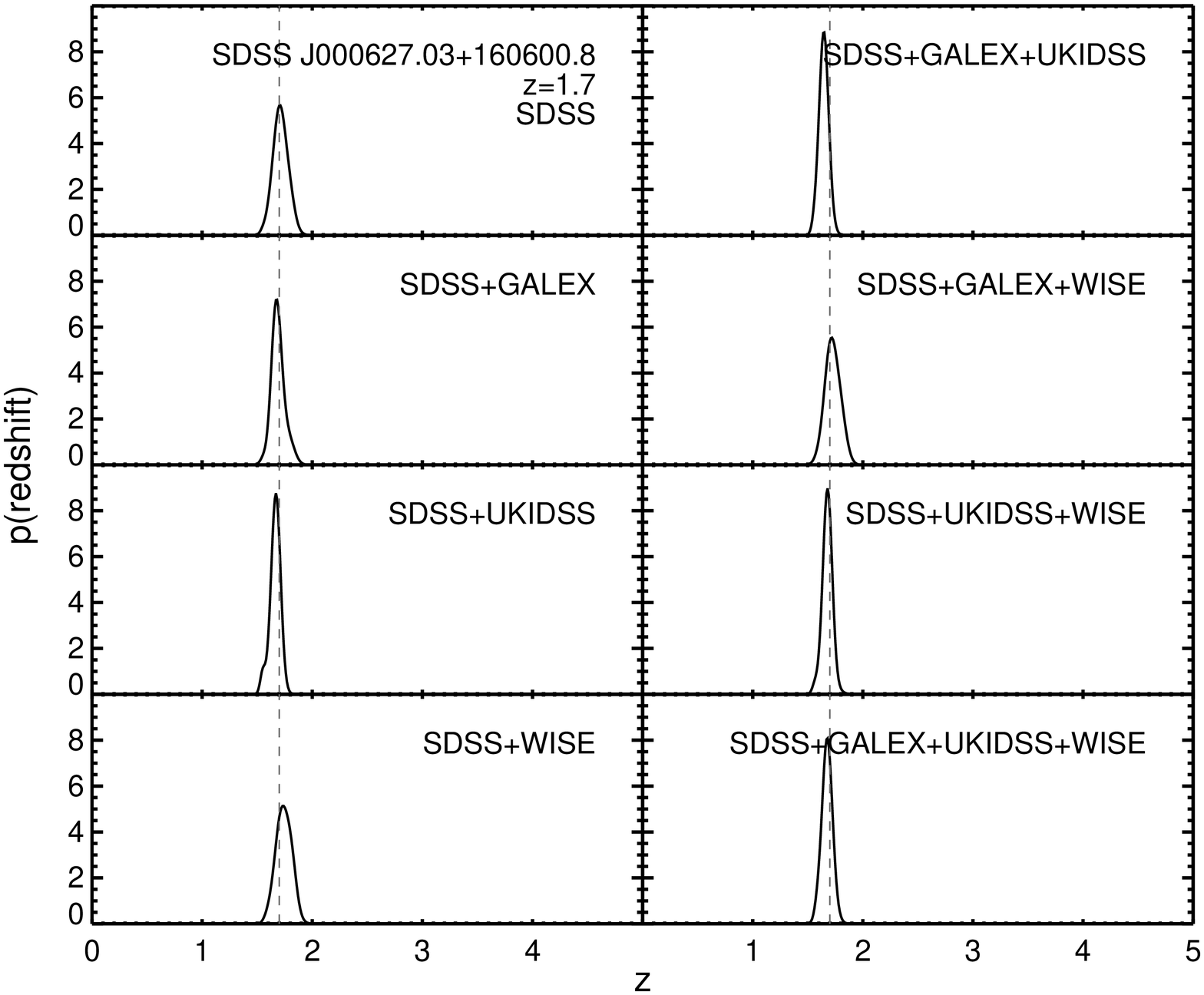}
   \includegraphics[width=8.5cm]{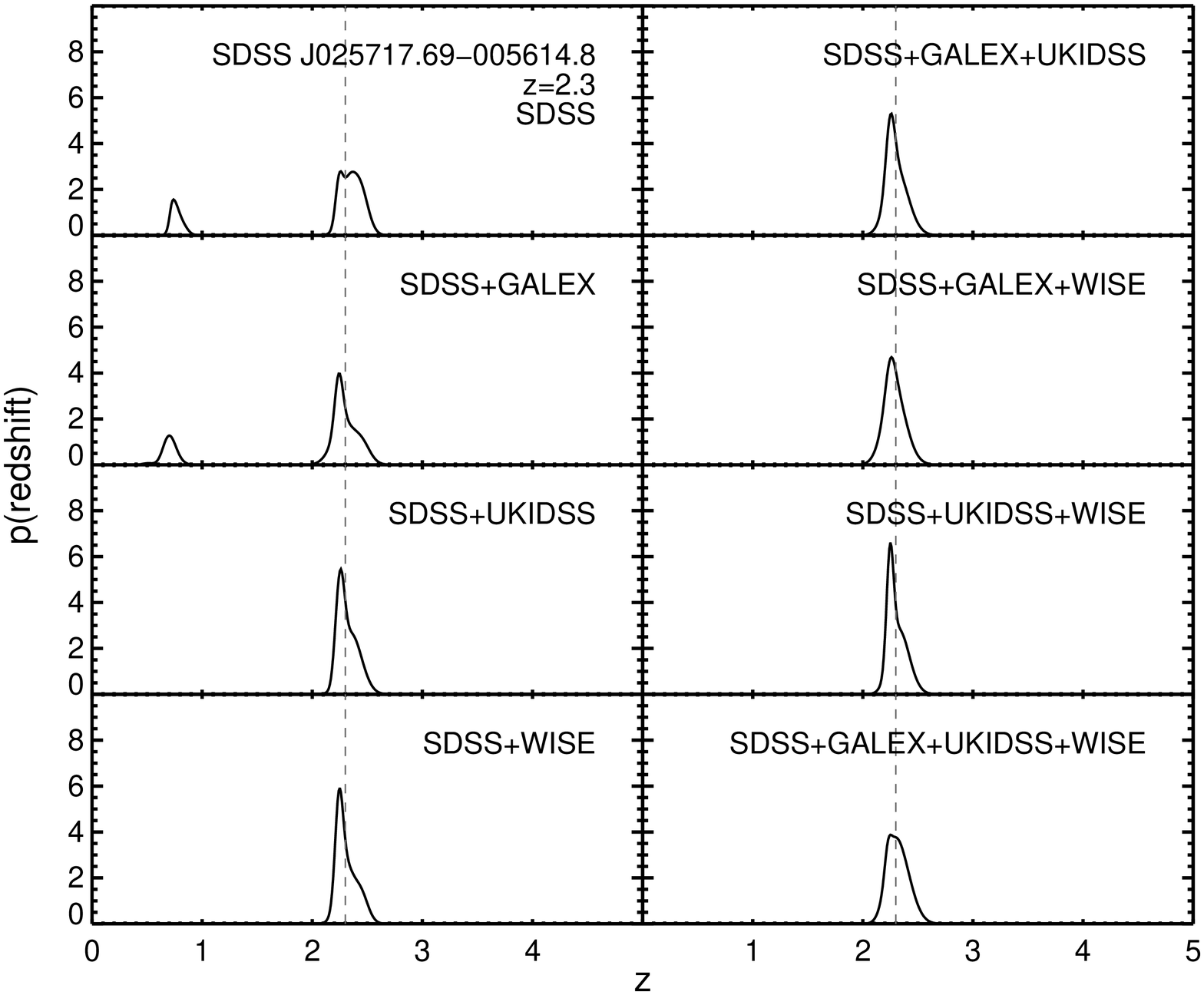}
   \includegraphics[width=8.5cm]{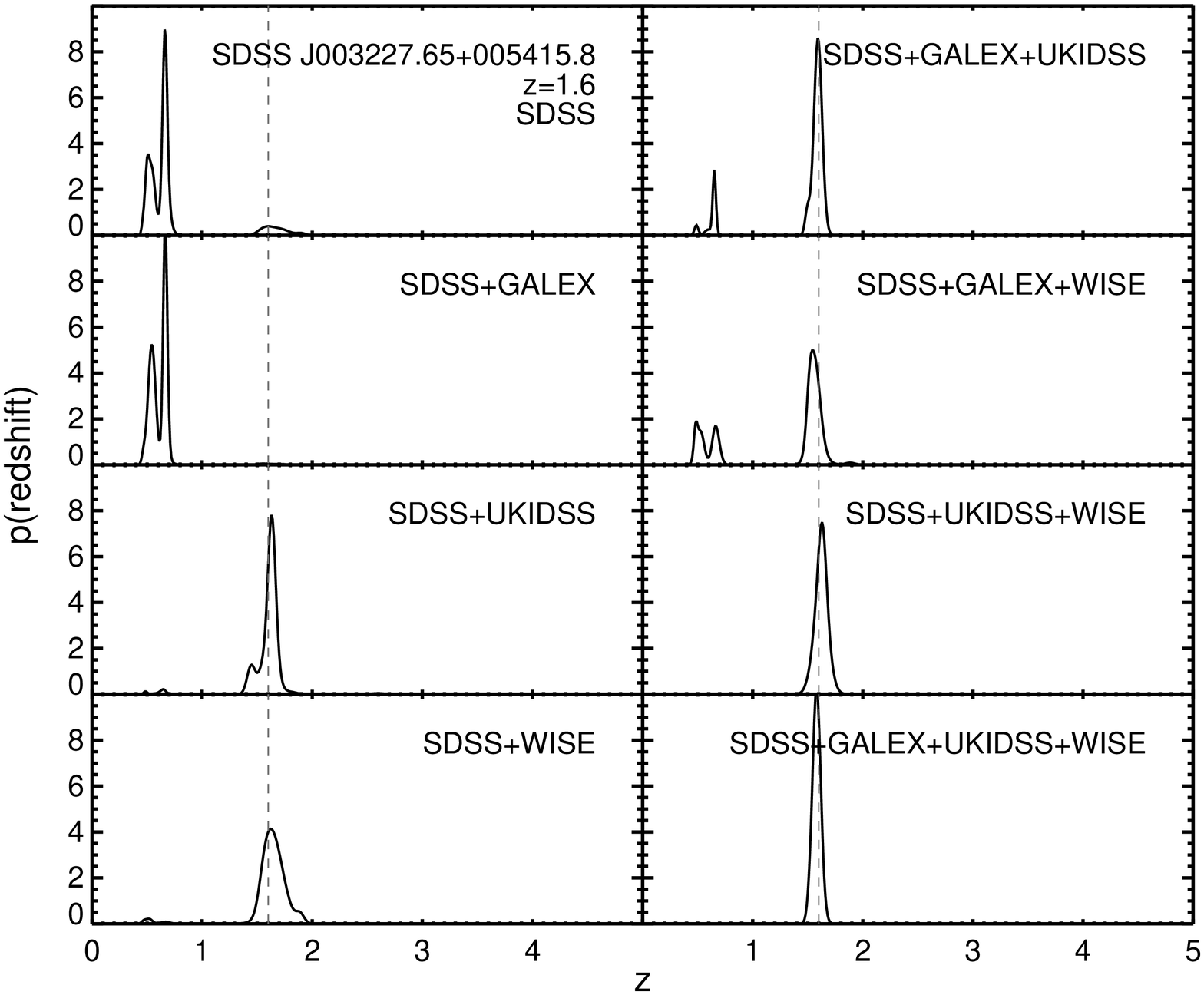}
   \includegraphics[width=8.5cm]{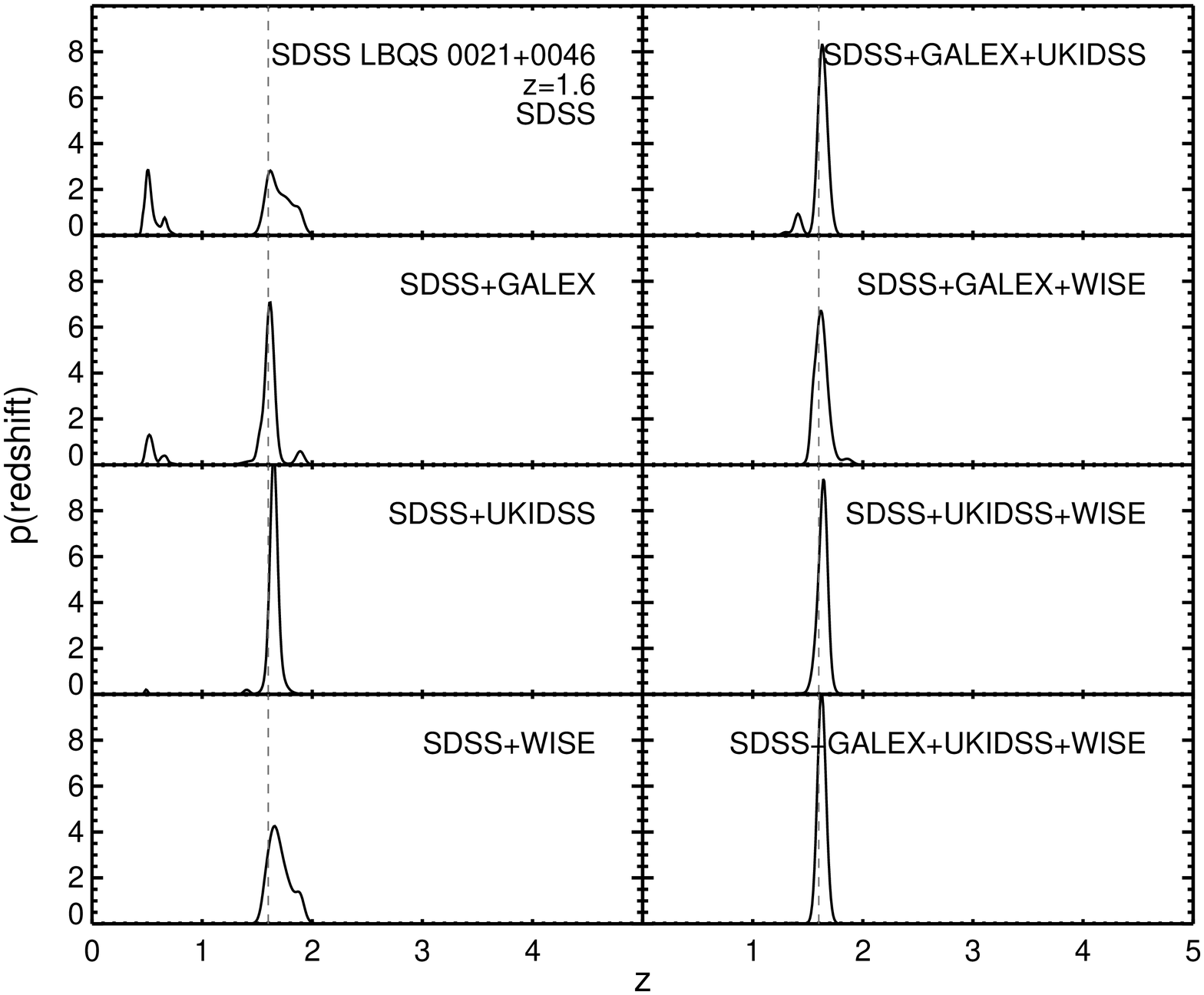}   
   \vspace{0cm}
  \caption{Examples of redshift PDFs for known spectroscopic quasars using various combinations of photometric data.\label{fig:zpdfs}}
\end{figure*}

Figure~\ref{fig:zpdfs} shows the redshift PDFs for four known quasars with various combinations of data.  These are the same objects that are shown in Figure 12 of \citet{2012ApJ...749...41B}.  In general we see that the addition of \wise\ data has the power to significantly narrow the PDF around the known spectroscopic redshifts.  However, it is not always the case that SDSS$+$\wise\ has a similar accuracy as when all of the fluxes are used.  The addition of \galex\ and UKIDSS data often has a significant effect on the photo-$z$ estimates.  The appendix includes some examples of the effect of \wise\ data on the redshift estimation of high-redshift ($z>3$) quasars.

\subsection{Comparison with \wise\ color selection}
How does the use of \wise\ photometry in the \emph{XDQSO} method compare with pure \wise-color cuts to select AGN?  Of course, with the current \emph{XDQSOz} formalism (and the use of forced-photometered \wise\ data at SDSS positions) we are limited to objects with optical SDSS detections, and so we cannot directly compare the methods for selecting obscured objects that fall below the limits of SDSS due to obscuration of the quasar.  However, it is fair to compare the two methods in identifying quasars with optical detections, as \wise\ colors are used to identify samples of unobscured quasars for comparison with obscured quasars \citep[e.g.][]{2014ApJ...789...44D, 2014MNRAS.442.3443D, 2015MNRAS.446.3492D}.  To examine this question, we start with a test sample of DR8 point sources and spectroscopically confirmed quasars from the DR7 and DR10 quasar catalogues \citep[][]{2010AJ....139.2360S, 2014A&A...563A..54P} in a circular region centered at RA=180$^{\circ}$, Dec=40$^{\circ}$ with a radius of 10$^{\circ}$.  The use of both quasar catalogues allows us to analyze the methods at high and low redshifts, as DR7 was effectively limited to $i < 20.2$ for high-redshift quasars, while the BOSS survey (included in DR10) specifically targeted $z > 2.2$ quasars with $g \leq 22$ OR $r \leq 21.85$.  After applying the flag cuts used in the construction of the catalogue presented here (section 3.3) and limiting to sources with $W1$ and $W2$ forced photometry available, there are 904,571 point sources and 8,280 spectroscopically confirmed quasars in this region.

\begin{figure*}
\centering
\hspace{0cm}
    \includegraphics[width=16cm]{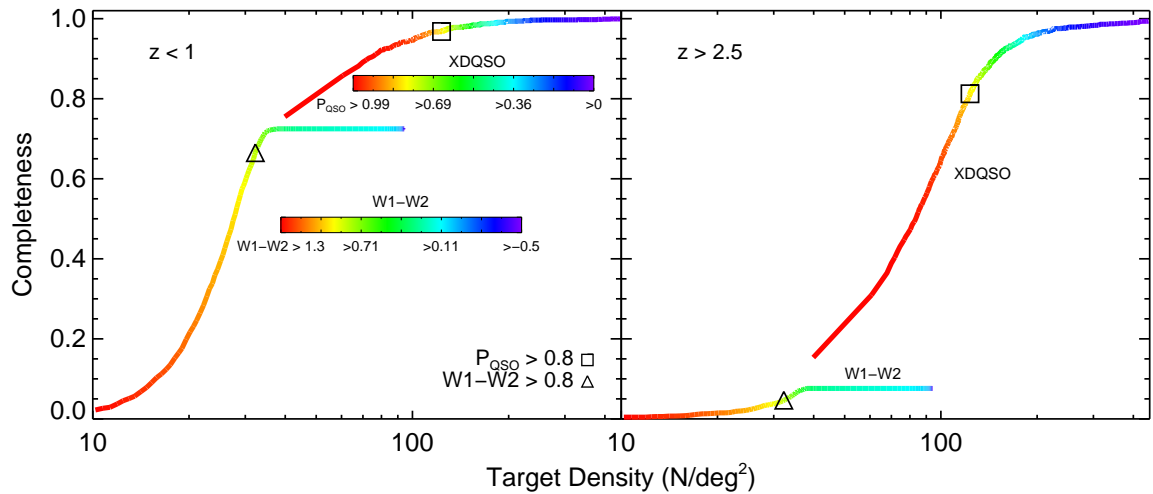}
    \vspace{0cm}
  \caption{A comparison of the completeness and efficiency in identifying quasars as a function of \emph{XDQSOz} quasar probabilities and simple \wise\ color cuts (combined with a $W2$ cut), using a test $\sim$300 deg$^2$ region, and the DR7 and DR10 spectroscopic quasar catalogues.  The left and right panels show low ($z < 1$) and high ($z > 2.5$) redshift ranges, respectively.  Colors indicate various cuts in $P_{\textrm{QSO}}$ and $W1-W2$, as shown in the color bars.  Cuts at $P_{\textrm{QSO}} = 0.8$, the inflection point in the total (all redshifts) quasar completeness for \emph{XDQSOz}, are marked with a square, and the standard $W1-W2$ cuts are indicated by a triangle.  \emph{XDQSOz} is more complete and more efficient, especially at high redshift.  This is largely because of the $W2$ restriction (Figure~\ref{fig:w2_z}). Impressively, \emph{XDQSOz} picks up where \wise\ color cuts plateau.\label{fig:comp_eff_z}}
\end{figure*}

\begin{figure}
\centering
\hspace{0cm}
   \includegraphics[width=8cm]{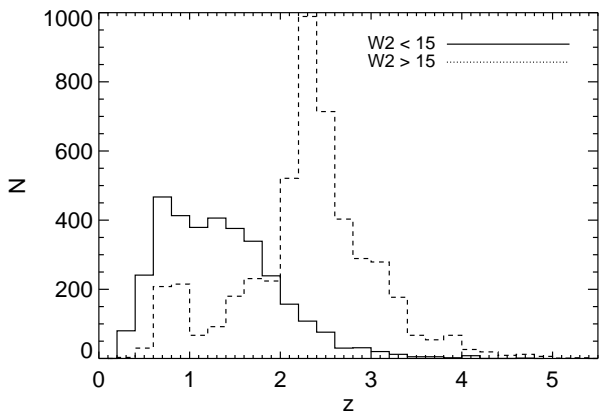}
    \vspace{0cm}
  \caption{The distributions of redshift for SDSS quasars (from both the DR7 and DR10 quasar catalogs) with $W2 < 15$ (solid) and $W2 > 15$ (dashed).  The $W2 < 15$ requirement for \wise-only quasar selection, which prevents contamination from massive, star-forming galaxies at high redshift, removes many high-redshift SDSS quasars as well.  This at least partially reflects the fact that the SDSS quasar selection uses fainter limits at high redshift.\label{fig:w2_z}}
\end{figure}

We analyze both the completeness ($N_{\textrm{QSO, sel}}/N_{\textrm{QSO}}$) and the selected target density ($N_{\textrm{point, sel}}/\textrm{area}$) for various cuts of $P_{\textrm{QSO}}$ and $W1-W2$ \citep[always applying a cut at $W2 < 15$ in the latter case;][]{2012ApJ...753...30S}.  Note again that all \wise\ magnitudes are in the Vega system.  This analysis is performed separately for two redshift ranges, $z < 1$ and $z> 2.5$.  The results are shown in Figure~\ref{fig:comp_eff_z}.  We can see that the \emph{XDQSO} method is far more complete than the simple \wise\ color cuts, especially for high redshift objects where the completeness is nearly an order of magnitude better. This is not particularly surprising, as it is well noted in the work of \citet{Stern:2005p2563}, \citet{2012ApJ...753...30S}, and \citet{2013ApJ...772...26A} that applying a simple color cut finds virtually no quasars at $z > 3$.  Figure~\ref{fig:w2_z} shows that for SDSS spectroscopic quasars, this is largely due to the $W2$ limit --- most SDSS quasars with $W2 > 15$ are at $z > 2$.  This is at least partially because the higher $z$ SDSS quasars are, on average, optically (and therefore probably bolometrically) fainter.  However, Figure~\ref{fig:color_vs_z} also shows that most high-$z$ objects have $W1-W2$ colors that are too blue to make the cut.  At lower redshift, using simple color cuts we find a completeness of $\sim$75\%, in rough agreement with \citet{2012ApJ...753...30S}.  The most striking feature in these plots is that the most conservative cuts in \emph{XDQSOz} begin where the \wise\ color cuts plateau.

For a given target density\footnote{We note that the target density at a cut of $W1 - W2 > 0.8$ is lower than what is found in \citet{2012ApJ...753...30S}.  This is because we require optical detections in the SDSS, which means that we are missing the most heavily obscured \wise-selected AGN.  These objects may make up nearly half of the quasar population \citep{2007ApJ...671.1365H, 2014MNRAS.442.3443D, 2014arXiv1408.1092A}} the \emph{XDQSO} selection is more complete, indicating that this method is also more efficient.  However, over all redshifts, to reach a completeness level above 90\% requires target densities over 100 per deg$^2$.

\subsection{Quasar catalogue and code}
We release an updated version of the quasar catalogue of \citet{2011ApJ...729..141B}, with the additional use of \wise\ data in our fits to the flux-redshift density space.  The catalogue includes quasar/star probabilities for all SDSS DR8 point sources (\verb|objc_type| $=6$) with a reasonable detection \citep[extinction-corrected magnitude in at least one band is above the completeness limit;][]{2011ApJS..193...29A}, a dereddened $i$-band magnitude in the range $17.75 \le i < 22.45$, and with $P_{\textrm{QSO}} > 0.2$.  A catalogue of quasar/star probabilities for \textit{all} of the SDSS DR8 point sources is available upon request.  The catalogue includes the entries listed in Table 2 of \citet{2011ApJ...729..141B}, along with 6 additions: \verb|galex_matched|, \verb|galex_used|, \verb|ukidss_matched|, \verb|ukidss_used|, \verb|wise_matched|, and \verb|wise_used|.  The ``matched'' tags indicate whether an object was detected in forced-photometry of each survey (true or false), and the ``used'' tags indicate if the given survey's fluxes were used in the underlying model for the probability calculation.

This catalogue contains 5,537,436 (total; 3,874,639 weighted by probability) potential quasars, and the following additional tags of redshift information: \verb|npeaks| (number of peaks in the redshift distribution, from 0 to 6), \verb|peakz| (the redshift at the highest probability of the widest continuous region above the uniform distribution), \verb|peakprob| (the probability associated with the peak redshift), \verb|peakfwhm| (the FWHM of the primary peak), \verb|otherz| (the redshifts of the peaks of up to six other regions above the uniform distribution), \verb|otherprob| (the peak probabilities of up to six secondary peaks), and \verb|otherfwhm| (the FWHM of up to six other secondary peaks).  

\begin{figure}
\centering
\hspace{0cm}
   \includegraphics[width=8cm]{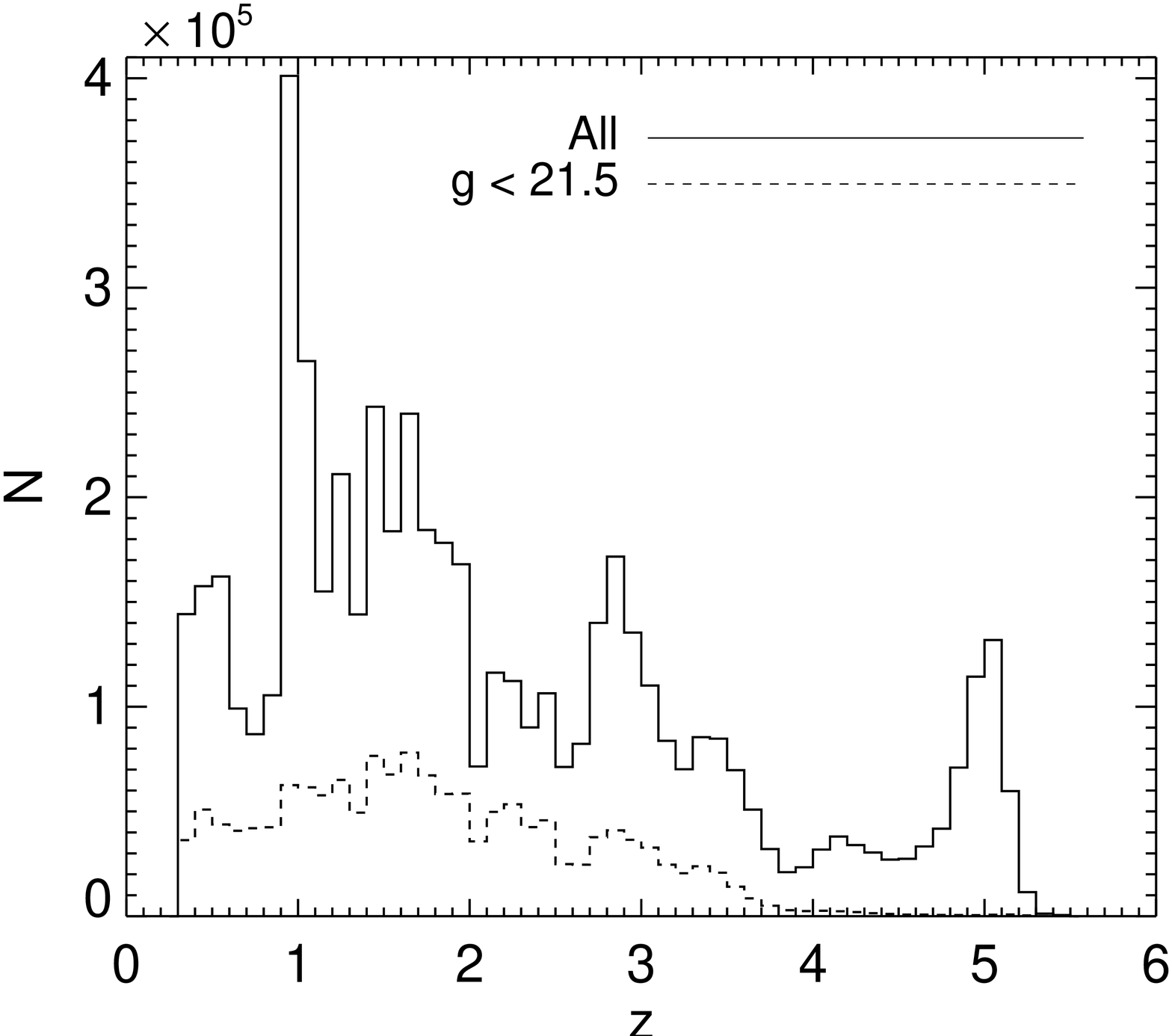}
    \vspace{0cm}
  \caption{The peak photometric redshift distribution for the new quasar catalogue (solid line).  The peak around $z =5$ can be reduced by applying additional cuts to the data to remove artifacts or regions of bad photometry (see section 3.3), but many of the sources appear real in optical images.  Additional follow-up is needed to determine if they are truly high-$z$ quasars.  Both this peak and the one at $z \sim 1$ are removed by limiting to relatively optically bright objects ($g < 21.5$; dotted line), which likely removes faint, possibly borderline quasars that fail star-galaxy separation.\label{fig:zdist}}
\end{figure}

The distribution of the peak photometric redshifts for this catalogue is shown in Figure~\ref{fig:zdist}.  There is a peak around $z \sim 1$, where star/galaxy separation becomes difficult for faint objects (see below).  Many of these sources are likely galaxies with quasar-like colors, and are potentially borderline objects.  Somewhat surprising is the additional peak at $z \sim 5$, given that high redshift quasars are relatively rare.  In visual inspection of the SDSS imaging around 100 random sources with photometric redshifts of $\sim$5, we find that approximately 20\% have nearby bright red stars that likely contaminate the photometry.  Indeed, the majority of these objects have the SDSS \verb|SUBTRACTED| flag\footnote{This flag indicates that the wings of the PSF of a bright star have been subtracted, as described in \citet{2002AJ....123..485S} and at \url{https://www.sdss3.org/dr8/algorithms/flags_detail.php}} set.  We extend the BOSS bright star mask\footnote{The original mask covering the BOSS footprint can be found at \url{http://data.sdss3.org/datamodel/files/BOSS_LSS_REDUX/reject_mask/MASK.html}.  Our extended version is provided as a \textsc{MANGLE} polygon file along with several useful \wise\ masks as presented in \citet{2014MNRAS.442.3443D}, at \url{http://faraday.uwyo.edu/~admyers/wisemask2014/wisemask.html}.  Note that there is a typo (missing ``\textasciitilde'') in the URL of \citet{2014MNRAS.442.3443D}.} \citep[as in, e.g.][]{2011ApJ...728..126W, 2012MNRAS.424..933W} across the whole SDSS region and apply it, which removes a significant number of these objects.  An additional tag has been added to the catalogue, \verb|bright_star|, to indicate objects that may suffer from contamination from nearby bright stars because they lie within the bright star mask.

There are other problematic regions in the SDSS that may cause inaccurate quasar probabilities or redshifts, including fields with poor photometry and regions in the North Galactic Cap with bad $u$-band data.  These are indicated in the catalogue in the tags \verb|bad_field| and \verb|bad_u|, respectively.  \citet{2014MNRAS.442.3443D} built a mask that removes regions around flagged/contaminated data in the \wise\ catalogue, including bright \wise\ stars (which may or may not be bright in the SDSS).  An additional tag, \verb|wise_flagged|, is included in the catalogue to indicate objects that lie in these regions.

However, even excluding objects that fall in these various masks, there is still a spike in the photometric redshift distribution of the catalogue around $z \sim 5$.  Another possibility we tested is that red galaxies that fail star-galaxy separation in the SDSS DR8 imaging (i.e.\ they are unresolved) could mimic high-$z$ quasars and cause this abundance of objects.  Because the ``stellar'' training set is from co-added Stripe 82 data, the average seeing is better than that of the full catalogue and these contaminants would likely not be included in the training.  However, if this were the case, we should see the spike at $z \sim 5$ reduced for the objects in the catalogue with the best seeing --- this is not the case.  The distribution of peak-$z$ values is not a function of the seeing. It is not immediately clear in the SDSS imaging that there are obvious problems with the vast majority of these $z \sim 5$ sources, but we recommend users exercise caution with these objects until proper follow-up is performed. 

One way to limit some of the potential contamination from e.g.\ failing star-galaxy separation is to limit the sample to relatively bright ($g < 21.5$) objects.  Indeed, as shown with the dotted histogram in Figure~\ref{fig:zdist}, doing so removes the spikes in redshift and thus likely provides a more pure quasar sample.

We stress that this catalogue is probabilistic in nature, so many of the objects are likely not quasars, and it is not intended to represent, on its own, a statistical sample.  This is largely due to the heterogeneous nature of the SDSS data it is built from, and it is possible to compile a subset of the catalogue that represents a complete statistical quasar sample.  This has been done with previous versions of the \emph{XDQSOz} catalog for an array of studies, including: probing the intergalactic 
medium with close quasar pairs \citep[e.g.][]{2013ApJ...776..136P, 2013ApJ...766...58H, 2014ApJ...796..140P, Rubin:2014uz}, photometric clustering to probe primordial non-Gaussianity \citep[][]{2013arXiv1311.2597H, 2014JCAP...02..038A, 2014MNRAS.444....2L, 2014PhRvL.113v1301L}, probing the extent of the Galactic halo \citep[e.g.][]{2014ApJ...787...30D}, and studying baryon acoustic oscillations at high redshift \citep[e.g.][]{2013JCAP...04..026S}.  The improved catalog here is sure to be a boon to future work in these areas, and is also useful for cross-matching with other catalogues/wavelengths in order to estimate quasar likelihoods or redshifts, analyses that do not require statistically complete samples, and searches for unusual objects.

The probabilistic quasar catalogue is available as a fits file (easily downloaded with a \textsc{wget} command) at \url{http://www.mpia.de/homes/joe/xdqsozcat_galex_ukidss_wise_p20.fits.gz}.  The updated \emph{XDQSO} and \emph{XDQSOz} codes for target classification and photometric redshift estimation, as well as the new flux-redshift density models including \wise\ data, are publicly available as a GitHub repository at \url{https://github.com/xdqso/xdqso/}.  The FITS files containing the model are identical to those in the previous release of \emph{XDQSOz}, with additional dimensions added to the end of each Gaussian component containing $W1$ and $W2$ information (in that order).  

\section{CONCLUSION}
We have presented an update to the \emph{XDQSOz} method of \citet{2011ApJ...729..141B, 2012ApJ...749...41B} for quasar classification and photometric redshift estimation that incorporates the two most sensitive \wise\ bands (3.6 and 4.6$\mu$m, or $W1$ and $W2$, respectively) into the relative-flux-redshift density model.  The use of \wise\ information greatly enhances the precision of the method in identifying quasars (by $\sim$5-20\%, depending on the redshift range of interest), especially at $z > 2$.  It also improves the overall accuracy of the photometric redshift estimation.  This method has better completeness compared to simple \wise\ color cuts (for quasars with lower levels of obscuration such that they are still detected and unresolved by the SDSS), again especially at high redshift, at similar efficiency.  We present a catalogue of potential quasar candidates ($P_{\textrm{QSO}} \ge 0.2$) with photometric redshift estimates.  This catalogue can be a powerful tool for identifying quasar samples or estimating redshifts for a wide variety of studies, and the improved accuracy will benefit future large quasar surveys.

\section*{Acknowledgement}
MAD and ADM were partially supported by NASA through ADAP award NNX12AE38G and EPSCoR award NNX11AM18A and by the National Science Foundation through grant numbers 1211096 and 1211112.  J.B. was supported by NASA through Hubble Fellowship grant HST-HF-51285.01 from the Space Telescope Science Institute, which is operated by the Association of Universities for Research in Astronomy, Incorporated, under NASA contract NAS5-26555.  We thank David Hogg for his feedback on the manuscript and his work on the original \emph{XDQSOz}.

\bibliography{full_library.bib}

\begin{thebibliography}{64}
\expandafter\ifx\csname natexlab\endcsname\relax\def\natexlab#1{#1}\fi

\bibitem[{Abazajian {et~al}\mbox{.}(2009)Abazajian, Adelman-McCarthy,
  Ag{\"u}eros, Allam, Allende~Prieto, An, Anderson, Anderson, Annis, Bahcall,
  Bailer-Jones, Barentine, Bassett, Becker, Beers, Bell, Belokurov, Berlind,
  Berman, Bernardi, Bickerton, Bizyaev, Blakeslee, Blanton, Bochanski, Boroski,
  Brewington, Brinchmann, Brinkmann, Brunner, Budav{\'a}ri, Carey, Carliles,
  Carr, Castander, Cinabro, Connolly, Csabai, Cunha, Czarapata, Davenport,
  de~Haas, Dilday, Doi, Eisenstein, Evans, Evans, Fan, Friedman, Frieman,
  Fukugita, G{\"a}nsicke, Gates, Gillespie, Gilmore, Gonzalez, Gonzalez,
  Grebel, Gunn, Gy{\"o}ry, Hall, Harding, Harris, Harvanek, Hawley, Hayes,
  Heckman, Hendry, Hennessy, Hindsley, Hoblitt, Hogan, Hogg, Holtzman, Hyde,
  Ichikawa, Ichikawa, Im, Ivezi{\'c}, Jester, Jiang, Johnson, Jorgensen,
  Juri{\'c}, Kent, Kessler, Kleinman, Knapp, Konishi, Kron, Krzesinski,
  Kuropatkin, Lampeitl, Lebedeva, Lee, Lee, French~Leger, L{\'e}pine, Li, Lima,
  Lin, Long, Loomis, Loveday, Lupton, Magnier, Malanushenko, Malanushenko,
  Mandelbaum, Margon, Marriner, Mart{\'\i}nez-Delgado, Matsubara, McGehee,
  McKay, Meiksin, Morrison, Mullally, Munn, Murphy, Nash, Nebot, Neilsen,
  Newberg, Newman, Nichol, Nicinski, Nieto-Santisteban, Nitta, Okamura,
  Oravetz, Ostriker, Owen, Padmanabhan, Pan, Park, Pauls, Peoples, Percival,
  Pier, Pope, Pourbaix, Price, Purger, Quinn, Raddick, Re~Fiorentin, Richards,
  Richmond, Riess, Rix, Rockosi, Sako, Schlegel, Schneider, Scholz, Schreiber,
  Schwope, Seljak, Sesar, Sheldon, Shimasaku, Sibley, Simmons, Sivarani,
  Allyn~Smith, Smith, Smolcic, Snedden, Stebbins, Steinmetz, Stoughton,
  Strauss, SubbaRao, Suto, Szalay, Szapudi, Szkody, Tanaka, Tegmark, Teodoro,
  Thakar, Tremonti, Tucker, Uomoto, Vanden~Berk, Vandenberg, Vidrih, Vogeley,
  Voges, Vogt, Wadadekar, Watters, Weinberg, West, White, Wilhite, Wonders,
  Yanny, Yocum, York, Zehavi, Zibetti, \& Zucker}]{2009ApJS..182..543A}
Abazajian K.~N. {et~al.}, 2009, ApJS, 182, 543

\bibitem[{Agarwal, Ho \& Shandera(2014)Agarwal, Ho, \&
  Shandera}]{2014JCAP...02..038A}
Agarwal N., Ho S., Shandera S., 2014, JCAP, 02, 038

\bibitem[{Ahn {et~al}\mbox{.}(2012)Ahn, Alexandroff, Allende~Prieto, Anderson,
  Anderton, Andrews, Aubourg, Bailey, Balbinot, Barnes, Bautista, Beers,
  Beifiori, Berlind, Bhardwaj, Bizyaev, Blake, Blanton, Blomqvist, Bochanski,
  Bolton, Borde, Bovy, Brandt, Brinkmann, Brown, Brownstein, Bundy, Busca,
  Carithers, Carnero, Carr, Casetti-Dinescu, Chen, Chiappini, Comparat,
  Connolly, Crepp, Cristiani, Croft, Cuesta, da~Costa, Davenport, Dawson,
  de~Putter, De~Lee, Delubac, Dhital, Ealet, Ebelke, Edmondson, Eisenstein,
  Escoffier, Esposito, Evans, Fan, Femen{\'\i}a~Castell{\'a},
  Fern{\'a}ndez~Alvar, Ferreira, Filiz~Ak, Finley, Fleming, Font-Ribera,
  Frinchaboy, Garc{\'\i}a-Hern{\'a}ndez, Garc{\'\i}a~P{\'e}rez, Ge,
  G{\'e}nova-Santos, Gillespie, Girardi, Gonz{\'a}lez~Hern{\'a}ndez, Grebel,
  Gunn, Guo, Haggard, Hamilton, Harris, Hawley, Hearty, Ho, Hogg, Holtzman,
  Honscheid, Huehnerhoff, Ivans, Ivezi{\'c}, Jacobson, Jiang, Johansson,
  Johnson, Kauffmann, Kirkby, Kirkpatrick, Klaene, Knapp, Kneib, Le~Goff,
  Leauthaud, Lee, Lee, Long, Loomis, Lucatello, Lundgren, Lupton, Ma, Ma,
  MacDonald, Mack, Mahadevan, Maia, Majewski, Makler, Malanushenko,
  Malanushenko, Manchado, Mandelbaum, Manera, Maraston, Margala, Martell,
  McBride, McGreer, McMahon, M{\'e}nard, Meszaros, Miralda-Escude,
  Montero-Dorta, Montesano, Morrison, Muna, Munn, Murayama, Myers, Neto,
  Nguyen, Nichol, Nidever, Noterdaeme, Nuza, Ogando, Olmstead, Oravetz, Owen,
  Padmanabhan, Palanque-Delabrouille, Pan, Parejko, Parihar, P{\^a}ris,
  Pattarakijwanich, Pepper, Percival, Perez-Fournon, P{\'e}rez-R{\`a}fols,
  Petitjean, Pforr, Pieri, Pinsonneault, Porto~de Mello, Prada, Price-Whelan,
  Raddick, Rebolo, Rich, Richards, Robin, Rocha-Pinto, Rockosi, Roe, Ross,
  Ross, Rossi, Rubi{\~n}o-Mart{\'\i}n, Samushia, Sanchez~Almeida, Sanchez,
  Santiago, Sayres, Schlegel, Schlesinger, Schmidt, Schneider, Schultheis,
  Schwope, Sc{\'o}ccola, Seljak, Sheldon, Shen, Shu, Simmerer, Simmons, Skibba,
  Skrutskie, Slosar, Sobreira, Sobeck, Stassun, Steele, Steinmetz, Strauss,
  Streblyanska, Suzuki, Swanson, Tal, Thakar, Thomas, Thompson, Tinker,
  Tojeiro, Tremonti, Vargas~Maga{\~n}a, Verde, Viel, Vikas, Vogt, Wake, Wang,
  Weaver, Weinberg, \& Weiner}]{2012ApJS..203...21A}
Ahn C.~P. {et~al.}, 2012, ApJS, 203, 21

\bibitem[{Aihara {et~al}\mbox{.}(2011)Aihara, Allende~Prieto, An, Anderson,
  Aubourg, Balbinot, Beers, Berlind, Bickerton, Bizyaev, Blanton, Bochanski,
  Bolton, Bovy, Brandt, Brinkmann, Brown, Brownstein, Busca, Campbell, Carr,
  Chen, Chiappini, Comparat, Connolly, Cortes, Croft, Cuesta, da~Costa,
  Davenport, Dawson, Dhital, Ealet, Ebelke, Edmondson, Eisenstein, Escoffier,
  Esposito, Evans, Fan, Femen{\'\i}a~Castell{\'a}, Font-Ribera, Frinchaboy, Ge,
  Gillespie, Gilmore, Gonz{\'a}lez~Hern{\'a}ndez, Gott, Gould, Grebel, Gunn,
  Hamilton, Harding, Harris, Hawley, Hearty, Ho, Hogg, Holtzman, Honscheid,
  Inada, Ivans, Jiang, Johnson, Jordan, Jordan, Kazin, Kirkby, Klaene, Knapp,
  Kneib, Kochanek, Koesterke, Kollmeier, Kron, Lampeitl, Lang, Le~Goff, Lee,
  Lin, Long, Loomis, Lucatello, Lundgren, Lupton, Ma, MacDonald, Mahadevan,
  Maia, Makler, Malanushenko, Malanushenko, Mandelbaum, Maraston, Margala,
  Masters, McBride, McGehee, McGreer, M{\'e}nard, Miralda-Escude, Morrison,
  Mullally, Muna, Munn, Murayama, Myers, Naugle, Neto, Nguyen, Nichol,
  O'Connell, Ogando, Olmstead, Oravetz, Padmanabhan, Palanque-Delabrouille,
  Pan, Pandey, P{\^a}ris, Percival, Petitjean, Pfaffenberger, Pforr, Phleps,
  Pichon, Pieri, Prada, Price-Whelan, Raddick, Ramos, Reyle, Rich, Richards,
  Rix, Robin, Rocha-Pinto, Rockosi, Roe, Rollinde, Ross, Ross, Rossetto,
  Sanchez, Sayres, Schlegel, Schlesinger, Schmidt, Schneider, Sheldon, Shu,
  Simmerer, Simmons, Sivarani, Snedden, Sobeck, Steinmetz, Strauss, Szalay,
  Tanaka, Thakar, Thomas, Tinker, Tofflemire, Tojeiro, Tremonti, Vandenberg,
  Vargas~Maga{\~n}a, Verde, Vogt, Wake, Wang, Weaver, Weinberg, White, White,
  Yanny, Yasuda, Yeche, \& Zehavi}]{2011ApJS..193...29A}
Aihara H. {et~al.}, 2011, ApJS, 193, 29

\bibitem[{Assef {et~al}\mbox{.}(2014)Assef, Eisenhardt, Stern, Tsai, Wu,
  Wylezalek, Blain, Bridge, Donoso, Gonzales, Griffith, \&
  Jarrett}]{2014arXiv1408.1092A}
Assef R.~J. {et~al.}, 2014, arXiv, 1092

\bibitem[{Assef {et~al}\mbox{.}(2013)Assef, Stern, Kochanek, Blain, Brodwin,
  Brown, Donoso, Eisenhardt, Jannuzi, Jarrett, Stanford, Tsai, Wu, \&
  Yan}]{2013ApJ...772...26A}
Assef R.~J. {et~al.}, 2013, ApJ, 772, 26

\bibitem[{Baldwin(1977)}]{1977ApJ...214..679B}
Baldwin J.~A., 1977, ApJ, 214, 679

\bibitem[{Ball {et~al}\mbox{.}(2008)Ball, Brunner, Myers, Strand, Alberts, \&
  Tcheng}]{2008ApJ...683...12B}
Ball N.~M., Brunner R.~J., Myers A.~D., Strand N.~E., Alberts S.~L., Tcheng D.,
  2008, ApJ, 683, 12

\bibitem[{Ball {et~al}\mbox{.}(2007)Ball, Brunner, Myers, Strand, Alberts,
  Tcheng, \& Llor{\`a}}]{2007ApJ...663..774B}
Ball N.~M., Brunner R.~J., Myers A.~D., Strand N.~E., Alberts S.~L., Tcheng D.,
  Llor{\`a} X., 2007, ApJ, 663, 774

\bibitem[{Bovy {et~al}\mbox{.}(2011)Bovy, Hennawi, Hogg, Myers, Kirkpatrick,
  Schlegel, Ross, Sheldon, McGreer, Schneider, \& Weaver}]{2011ApJ...729..141B}
Bovy J. {et~al.}, 2011, ApJ, 729, 141

\bibitem[{Bovy, Hogg \& Roweis(2011)Bovy, Hogg, \& Roweis}]{Bovy:2011bt}
Bovy J., Hogg D.~W., Roweis S.~T., 2011, Ann. App. Stat., 5, 1657

\bibitem[{Bovy {et~al}\mbox{.}(2012)Bovy, Myers, Hennawi, Hogg, McMahon,
  Schiminovich, Sheldon, Brinkmann, Schneider, \& Weaver}]{2012ApJ...749...41B}
Bovy J. {et~al.}, 2012, ApJ, 749, 41

\bibitem[{Budav{\'a}ri {et~al}\mbox{.}(2001)Budav{\'a}ri, Csabai, Szalay,
  Connolly, Szokoly, Vanden~Berk, Richards, Weinstein, Schneider, Ben{\'\i}tez,
  Brinkmann, Brunner, Hall, Hennessy, Ivezi{\'c}, Kunszt, Munn, Nichol, Pier,
  \& York}]{2001AJ....122.1163B}
Budav{\'a}ri T. {et~al.}, 2001, AJ, 122, 1163

\bibitem[{Dawson {et~al}\mbox{.}(2013)Dawson, Schlegel, Ahn, Anderson, Aubourg,
  Bailey, Barkhouser, Bautista, Beifiori, Berlind, Bhardwaj, Bizyaev, Blake,
  Blanton, Blomqvist, Bolton, Borde, Bovy, Brandt, Brewington, Brinkmann,
  Brown, Brownstein, Bundy, Busca, Carithers, Carnero, Carr, Chen, Comparat,
  Connolly, Cope, Croft, Cuesta, da~Costa, Davenport, Delubac, de~Putter,
  Dhital, Ealet, Ebelke, Eisenstein, Escoffier, Fan, Filiz~Ak, Finley,
  Font-Ribera, G{\'e}nova-Santos, Gunn, Guo, Haggard, Hall, Hamilton, Harris,
  Harris, Ho, Hogg, Holder, Honscheid, Huehnerhoff, Jordan, Jordan, Kauffmann,
  Kazin, Kirkby, Klaene, Kneib, Le~Goff, Lee, Long, Loomis, Lundgren, Lupton,
  Maia, Makler, Malanushenko, Malanushenko, Mandelbaum, Manera, Maraston,
  Margala, Masters, McBride, McDonald, McGreer, McMahon, Mena, Miralda-Escude,
  Montero-Dorta, Montesano, Muna, Myers, Naugle, Nichol, Noterdaeme, Nuza,
  Olmstead, Oravetz, Oravetz, Owen, Padmanabhan, Palanque-Delabrouille, Pan,
  Parejko, P{\^a}ris, Percival, Perez-Fournon, P{\'e}rez-R{\`a}fols, Petitjean,
  Pfaffenberger, Pforr, Pieri, Prada, Price-Whelan, Raddick, Rebolo, Rich,
  Richards, Rockosi, Roe, Ross, Ross, Rossi, Rubi{\~n}o-Mart{\'\i}n, Samushia,
  Sanchez, Sayres, Schmidt, Schneider, Sc{\'o}ccola, Seo, Shelden, Sheldon,
  Shen, Shu, Slosar, Smee, Snedden, Stauffer, Steele, Strauss, Streblyanska,
  Suzuki, Swanson, Tal, Tanaka, Thomas, Tinker, Tojeiro, Tremonti,
  Vargas~Maga{\~n}a, Verde, Viel, Wake, Watson, Weaver, Weinberg, Weiner, West,
  White, Wood-Vasey, Yeche, Zehavi, Zhao, \& Zheng}]{2013AJ....145...10D}
Dawson K.~S. {et~al.}, 2013, AJ, 145, 10

\bibitem[{Deason {et~al}\mbox{.}(2014)Deason, Belokurov, Koposov, \&
  Rockosi}]{2014ApJ...787...30D}
Deason A.~J., Belokurov V., Koposov S.~E., Rockosi C.~M., 2014, ApJ, 787, 30

\bibitem[{DiPompeo {et~al}\mbox{.}(2014)DiPompeo, Myers, Hickox, Geach, \&
  Hainline}]{2014MNRAS.442.3443D}
DiPompeo M.~A., Myers A.~D., Hickox R.~C., Geach J.~E., Hainline K.~N., 2014,
  MNRAS, 442, 3443

\bibitem[{DiPompeo {et~al}\mbox{.}(2015)DiPompeo, Myers, Hickox, Geach, Holder,
  Hainline, \& Hall}]{2015MNRAS.446.3492D}
DiPompeo M.~A., Myers A.~D., Hickox R.~C., Geach J.~E., Holder G., Hainline
  K.~N., Hall S.~W., 2015, MNRAS, 446, 3492

\bibitem[{Donoso {et~al}\mbox{.}(2014)Donoso, Yan, Stern, \&
  Assef}]{2014ApJ...789...44D}
Donoso E., Yan L., Stern D., Assef R.~J., 2014, ApJ, 789, 44

\bibitem[{Eisenstein {et~al}\mbox{.}(2011)Eisenstein, Weinberg, Agol, Aihara,
  Allende~Prieto, Anderson, Arns, Aubourg, Bailey, Balbinot, Barkhouser, Beers,
  Berlind, Bickerton, Bizyaev, Blanton, Bochanski, Bolton, Bosman, Bovy,
  Brandt, Breslauer, Brewington, Brinkmann, Brown, Brownstein, Burger, Busca,
  Campbell, Cargile, Carithers, Carlberg, Carr, Chang, Chen, Chiappini,
  Comparat, Connolly, Cortes, Croft, Cunha, da~Costa, Davenport, Dawson,
  De~Lee, Porto~de Mello, de~Simoni, Dean, Dhital, Ealet, Ebelke, Edmondson,
  Eiting, Escoffier, Esposito, Evans, Fan, Femen{\'\i}a~Castell{\'a},
  Dutra~Ferreira, Fitzgerald, Fleming, Font-Ribera, Ford, Frinchaboy, Elia
  Garc{\'\i}a~P{\'e}rez, Gaudi, Ge, Ghezzi, Gillespie, Gilmore, Girardi, Gott,
  Gould, Grebel, Gunn, Hamilton, Harding, Harris, Hawley, Hearty, Hennawi,
  Gonz{\'a}lez~Hern{\'a}ndez, Ho, Hogg, Holtzman, Honscheid, Inada, Ivans,
  Jiang, Jiang, Johnson, Jordan, Jordan, Kauffmann, Kazin, Kirkby, Klaene,
  Knapp, Kneib, Kochanek, Koesterke, Kollmeier, Kron, Lampeitl, Lang, Lawler,
  Le~Goff, Lee, Lee, Leisenring, Lin, Liu, Long, Loomis, Lucatello, Lundgren,
  Lupton, Ma, Ma, MacDonald, Mack, Mahadevan, Maia, Majewski, Makler,
  Malanushenko, Malanushenko, Mandelbaum, Maraston, Margala, Maseman, Masters,
  McBride, McDonald, McGreer, McMahon, Mena~Requejo, M{\'e}nard,
  Miralda-Escude, Morrison, Mullally, Muna, Murayama, Myers, Naugle, Neto,
  Nguyen, Nichol, Nidever, O'Connell, Ogando, Olmstead, Oravetz, Padmanabhan,
  Paegert, Palanque-Delabrouille, Pan, Pandey, Parejko, P{\^a}ris, Pellegrini,
  Pepper, Percival, Petitjean, Pfaffenberger, Pforr, Phleps, Pichon, Pieri,
  Prada, Price-Whelan, Raddick, Ramos, Reid, Reyle, Rich, Richards, Rieke,
  Rieke, Rix, Robin, Rocha-Pinto, Rockosi, Roe, Rollinde, Ross, Ross, Rossetto,
  Sanchez, Santiago, Sayres, Schiavon, Schlegel, Schlesinger, Schmidt,
  Schneider, Sellgren, Shelden, Sheldon, Shetrone, Shu, Silverman, Simmerer,
  Simmons, Sivarani, Skrutskie, Slosar, Smee, Smith, Snedden, Stassun, Steele,
  Steinmetz, Stockett, Stollberg, Strauss, Szalay, Tanaka, Thakar, Thomas,
  Tinker, Tofflemire, Tojeiro, Tremonti, \&
  Vargas~Maga{\~n}a}]{2011AJ....142...72E}
Eisenstein D.~J. {et~al.}, 2011, AJ, 142, 72

\bibitem[{Fukugita {et~al}\mbox{.}(1996)Fukugita, Ichikawa, Gunn, Doi,
  Shimasaku, \& Schneider}]{1996AJ....111.1748F}
Fukugita M., Ichikawa T., Gunn J.~E., Doi M., Shimasaku K., Schneider D.~P.,
  1996, Astronomical Journal v.111, 111, 1748

\bibitem[{Giannantonio {et~al}\mbox{.}(2006)Giannantonio, Crittenden, Nichol,
  Scranton, Richards, Myers, Brunner, Gray, Connolly, \&
  Schneider}]{2006PhRvD..74f3520G}
Giannantonio T. {et~al.}, 2006, PhRvD, 74, 63520

\bibitem[{Giannantonio {et~al}\mbox{.}(2008)Giannantonio, Scranton, Crittenden,
  Nichol, Boughn, Myers, \& Richards}]{2008PhRvD..77l3520G}
Giannantonio T., Scranton R., Crittenden R.~G., Nichol R.~C., Boughn S.~P.,
  Myers A.~D., Richards G.~T., 2008, PhRvD, 77, 123520

\bibitem[{Hennawi \& Prochaska(2013)}]{2013ApJ...766...58H}
Hennawi J.~F., Prochaska J.~X., 2013, ApJ, 766, 58

\bibitem[{Hickox {et~al}\mbox{.}(2007)Hickox, Jones, Forman, Murray, Brodwin,
  Brown, Eisenhardt, Stern, Kochanek, Eisenstein, Cool, Jannuzi, Dey, Brand,
  Gorjian, \& Caldwell}]{2007ApJ...671.1365H}
Hickox R.~C. {et~al.}, 2007, ApJ, 671, 1365

\bibitem[{Hickox {et~al}\mbox{.}(2011)Hickox, Myers, Brodwin, Alexander,
  Forman, Jones, Murray, Brown, Cool, Kochanek, Dey, Jannuzi, Eisenstein,
  Assef, Eisenhardt, Gorjian, Stern, Le~Floc'h, Caldwell, Goulding, \&
  Mullaney}]{2011ApJ...731..117H}
Hickox R.~C. {et~al.}, 2011, ApJ, 731, 117

\bibitem[{Ho {et~al}\mbox{.}(2013)Ho, Agarwal, Myers, Lyons, Disbrow, Seo,
  Ross, Hirata, Padmanabhan, O'Connell, Huff, Schlegel, Slosar, Weinberg,
  Strauss, Ross, Schneider, Bahcall, Brinkmann, Palanque-Delabrouille, \&
  Yeche}]{2013arXiv1311.2597H}
Ho S. {et~al.}, 2013, arXiv, 2597

\bibitem[{Hook {et~al}\mbox{.}(1994)Hook, McMahon, Boyle, \&
  Irwin}]{1994MNRAS.268..305H}
Hook I.~M., McMahon R.~G., Boyle B.~J., Irwin M.~J., 1994, MNRAS, 268, 305

\bibitem[{Hopkins, Richards \& Hernquist(2007)Hopkins, Richards, \&
  Hernquist}]{2007ApJ...654..731H}
Hopkins P.~F., Richards G.~T., Hernquist L., 2007, ApJ, 654, 731

\bibitem[{Ivezi{\'c} {et~al}\mbox{.}(2003)Ivezi{\'c}, Lupton, Anderson, Eyer,
  Gunn, Juri{\'c}, Knapp, Miknaitis, Rockosi, Schlegel, Strauss, Stubbs, \&
  Vanden~Berk}]{2003MmSAI..74..978I}
Ivezi{\'c} {\v Z}. {et~al.}, 2003, Mem. Soc. Astron. Ital., 74, 978

\bibitem[{Ivezi{\'c} {et~al}\mbox{.}(2007)Ivezi{\'c}, Smith, Miknaitis, Lin,
  Tucker, Lupton, Gunn, Knapp, Strauss, Sesar, Doi, Tanaka, Fukugita, Holtzman,
  Kent, Yanny, Schlegel, Finkbeiner, Padmanabhan, Rockosi, Juri{\'c}, Bond,
  Lee, Stoughton, Jester, Harris, Harding, Morrison, Brinkmann, Schneider, \&
  York}]{2007AJ....134..973I}
Ivezi{\'c} {\v Z}. {et~al.}, 2007, AJ, 134, 973

\bibitem[{Kirkpatrick {et~al}\mbox{.}(2011)Kirkpatrick, Schlegel, Ross, Myers,
  Hennawi, Sheldon, Schneider, \& Weaver}]{2011ApJ...743..125K}
Kirkpatrick J.~A., Schlegel D.~J., Ross N.~P., Myers A.~D., Hennawi J.~F.,
  Sheldon E.~S., Schneider D.~P., Weaver B.~A., 2011, ApJ, 743, 125

\bibitem[{Koz{\l}owski {et~al}\mbox{.}(2010)Koz{\l}owski, Kochanek, Stern,
  Ashby, Assef, Bock, Borys, Brand, Brodwin, Brown, Cool, Cooray, Croft, Dey,
  Eisenhardt, Gonzalez, Gorjian, Griffith, Grogin, Ivison, Jacob, Jannuzi,
  Mainzer, Moustakas, R{\"o}ttgering, Seymour, Smith, Stanford, Stauffer,
  Sullivan, van Breugel, Willner, \& Wright}]{2010ApJ...716..530K}
Koz{\l}owski S. {et~al.}, 2010, ApJ, 716, 530

\bibitem[{Lang(2014)}]{2014AJ....147..108L}
Lang D., 2014, AJ, 147, 108

\bibitem[{Lang, Hogg \& Schlegel(2014)Lang, Hogg, \&
  Schlegel}]{2014arXiv1410.7397L}
Lang D., Hogg D.~W., Schlegel D.~J., 2014, arXiv, 1410.7397

\bibitem[{Leistedt \& Peiris(2014)}]{2014MNRAS.444....2L}
Leistedt B., Peiris H.~V., 2014, MNRAS, 444, 2

\bibitem[{Leistedt, Peiris \& Roth(2014)Leistedt, Peiris, \&
  Roth}]{2014PhRvL.113v1301L}
Leistedt B., Peiris H.~V., Roth N., 2014, Physical Review Letters, 113, 221301

\bibitem[{Martin {et~al}\mbox{.}(2005)Martin, Fanson, Schiminovich, Morrissey,
  Friedman, Barlow, Conrow, Grange, Jelinsky, Milliard, Siegmund, Bianchi,
  Byun, Donas, Forster, Heckman, Lee, Madore, Malina, Neff, Rich, Small,
  Surber, Szalay, Welsh, \& Wyder}]{2005ApJ...619L...1M}
Martin D.~C. {et~al.}, 2005, ApJ, 619, L1

\bibitem[{Mateos {et~al}\mbox{.}(2013)Mateos, Alonso-Herrero, Carrera, Blain,
  Severgnini, Caccianiga, \& Ruiz}]{2013MNRAS.434..941M}
Mateos S., Alonso-Herrero A., Carrera F.~J., Blain A., Severgnini P.,
  Caccianiga A., Ruiz A., 2013, MNRAS, 434, 941

\bibitem[{Mateos {et~al}\mbox{.}(2012)Mateos, Alonso-Herrero, Carrera, Blain,
  Watson, Barcons, Braito, Severgnini, Donley, \& Stern}]{2012MNRAS.426.3271M}
Mateos S. {et~al.}, 2012, MNRAS, 426, 3271

\bibitem[{Myers {et~al}\mbox{.}(2007{\natexlab{a}})Myers, Brunner, Nichol,
  Richards, Schneider, \& Bahcall}]{2007ApJ...658...85M}
Myers A.~D., Brunner R.~J., Nichol R.~C., Richards G.~T., Schneider D.~P.,
  Bahcall N.~A., 2007{\natexlab{a}}, ApJ, 658, 85

\bibitem[{Myers {et~al}\mbox{.}(2007{\natexlab{b}})Myers, Brunner, Richards,
  Nichol, Schneider, \& Bahcall}]{2007ApJ...658...99M}
Myers A.~D., Brunner R.~J., Richards G.~T., Nichol R.~C., Schneider D.~P.,
  Bahcall N.~A., 2007{\natexlab{b}}, ApJ, 658, 99

\bibitem[{Myers {et~al}\mbox{.}(2006)Myers, Brunner, Richards, Nichol,
  Schneider, Vanden~Berk, Scranton, Gray, \& Brinkmann}]{2006ApJ...638..622M}
Myers A.~D. {et~al.}, 2006, ApJ, 638, 622

\bibitem[{P{\^a}ris {et~al}\mbox{.}(2014)P{\^a}ris, Petitjean, Aubourg, Ross,
  Myers, Streblyanska, Bailey, Hall, Strauss, Anderson, Bizyaev, Borde,
  Brinkmann, Bovy, Brandt, Brewington, Brownstein, Cook, Ebelke, Fan, Filiz~Ak,
  Finley, Font-Ribera, Ge, Hamann, Ho, Jiang, Kinemuchi, Malanushenko,
  Malanushenko, Marchante, McGreer, McMahon, Miralda-Escude, Muna, Noterdaeme,
  Oravetz, Palanque-Delabrouille, Pan, Perez-Fournon, Pieri, Riffel, Schlegel,
  Schneider, Simmons, Viel, Weaver, Wood-Vasey, Yeche, \&
  York}]{2014A&A...563A..54P}
P{\^a}ris I. {et~al.}, 2014, A{\&}A, 563, 54

\bibitem[{Prochaska {et~al}\mbox{.}(2013)Prochaska, Hennawi, Lee, Cantalupo,
  Bovy, Djorgovski, Ellison, Lau, Martin, Myers, Rubin, \&
  Simcoe}]{2013ApJ...776..136P}
Prochaska J.~X. {et~al.}, 2013, ApJ, 776, 136

\bibitem[{Prochaska, Lau \& Hennawi(2014)Prochaska, Lau, \&
  Hennawi}]{2014ApJ...796..140P}
Prochaska J.~X., Lau M.~W., Hennawi J.~F., 2014, ApJ, 796, 140

\bibitem[{Richards {et~al}\mbox{.}(2009{\natexlab{a}})Richards, Deo, Lacy,
  Myers, Nichol, Zakamska, Brunner, Brandt, Gray, Parejko, Ptak, Schneider,
  Storrie-Lombardi, \& Szalay}]{2009AJ....137.3884R}
Richards G.~T. {et~al.}, 2009{\natexlab{a}}, AJ, 137, 3884

\bibitem[{Richards {et~al}\mbox{.}(2001{\natexlab{a}})Richards, Fan, Schneider,
  Vanden~Berk, Strauss, York, Anderson, Anderson, Annis, Bahcall, Bernardi,
  Briggs, Brinkmann, Brunner, Burles, Carey, Castander, Connolly, Crocker,
  Csabai, Doi, Finkbeiner, Friedman, Frieman, Fukugita, Gunn, Hindsley,
  Ivezi{\'c}, Kent, Knapp, Lamb, Leger, Long, Loveday, Lupton, McKay, Meiksin,
  Merrelli, Munn, Newberg, Newcomb, Nichol, Owen, Pier, Pope, Richmond,
  Rockosi, Schlegel, Siegmund, Smee, Snir, Stoughton, Stubbs, SubbaRao, Szalay,
  Szokoly, Tremonti, Uomoto, Waddell, Yanny, \& Zheng}]{2001AJ....121.2308R}
Richards G.~T. {et~al.}, 2001{\natexlab{a}}, AJ, 121, 2308

\bibitem[{Richards {et~al}\mbox{.}(2009{\natexlab{b}})Richards, Myers, Gray,
  Riegel, Nichol, Brunner, Szalay, Schneider, \&
  Anderson}]{2009ApJS..180...67R}
Richards G.~T. {et~al.}, 2009{\natexlab{b}}, ApJS, 180, 67

\bibitem[{Richards {et~al}\mbox{.}(2004)Richards, Nichol, Gray, Brunner,
  Lupton, Vanden~Berk, Chong, Weinstein, Schneider, Anderson, Munn, Harris,
  Strauss, Fan, Gunn, Ivezi{\'c}, York, Brinkmann, \&
  Moore}]{2004ApJS..155..257R}
Richards G.~T. {et~al.}, 2004, ApJS, 155, 257

\bibitem[{Richards {et~al}\mbox{.}(2001{\natexlab{b}})Richards, Weinstein,
  Schneider, Fan, Strauss, Vanden~Berk, Annis, Burles, Laubacher, York,
  Frieman, Johnston, Scranton, Gunn, Ivezi{\'c}, Nichol, Budav{\'a}ri, Csabai,
  Szalay, Connolly, Szokoly, Bahcall, Ben{\'\i}tez, Brinkmann, Brunner,
  Fukugita, Hall, Hennessy, Knapp, Kunszt, Lamb, Munn, Newberg, \&
  Stoughton}]{2001AJ....122.1151R}
Richards G.~T. {et~al.}, 2001{\natexlab{b}}, AJ, 122, 1151

\bibitem[{Ross {et~al}\mbox{.}(2012)Ross, Myers, Sheldon, Yeche, Strauss, Bovy,
  Kirkpatrick, Richards, Aubourg, Blanton, Brandt, Carithers, Croft, da~Silva,
  Dawson, Eisenstein, Hennawi, Ho, Hogg, Lee, Lundgren, McMahon,
  Miralda-Escude, Palanque-Delabrouille, P{\^a}ris, Petitjean, Pieri, Rich,
  Roe, Schiminovich, Schlegel, Schneider, Slosar, Suzuki, Tinker, Weinberg,
  Weyant, White, \& Wood-Vasey}]{2012ApJS..199....3R}
Ross N.~P. {et~al.}, 2012, ApJS, 199, 3

\bibitem[{Rubin {et~al}\mbox{.}(2014)Rubin, Hennawi, Prochaska, Simcoe, Myers,
  \& Lau}]{Rubin:2014uz}
Rubin K. H.~R., Hennawi J.~F., Prochaska J.~X., Simcoe R.~A., Myers A., Lau
  M.~W., 2014, arXiv

\bibitem[{Schneider {et~al}\mbox{.}(2010)Schneider, Richards, Hall, Strauss,
  Anderson, Boroson, Ross, Shen, Brandt, Fan, Inada, Jester, Knapp, Krawczyk,
  Thakar, Vanden~Berk, Voges, Yanny, York, Bahcall, Bizyaev, Blanton,
  Brewington, Brinkmann, Eisenstein, Frieman, Fukugita, Gray, Gunn, Hibon,
  Ivezi{\'c}, Kent, Kron, Lee, Lupton, Malanushenko, Malanushenko, Oravetz,
  Pan, Pier, Price, Saxe, Schlegel, Simmons, Snedden, SubbaRao, Szalay, \&
  Weinberg}]{2010AJ....139.2360S}
Schneider D.~P. {et~al.}, 2010, AJ, 139, 2360

\bibitem[{Scranton {et~al}\mbox{.}(2005)Scranton, M{\'e}nard, Richards, Nichol,
  Myers, Jain, Gray, Bartelmann, Brunner, Connolly, Gunn, Sheth, Bahcall,
  Brinkman, Loveday, Schneider, Thakar, \& York}]{2005ApJ...633..589S}
Scranton R. {et~al.}, 2005, ApJ, 633, 589

\bibitem[{Slosar {et~al}\mbox{.}(2013)Slosar, Ir{\v s}i{\v c}, Kirkby, Bailey,
  Busca, Delubac, Rich, Aubourg, Bautista, Bhardwaj, Blomqvist, Bolton, Bovy,
  Brownstein, Carithers, Croft, Dawson, Font-Ribera, Le~Goff, Ho, Honscheid,
  Lee, Margala, McDonald, Medolin, Miralda-Escude, Myers, Nichol, Noterdaeme,
  Palanque-Delabrouille, P{\^a}ris, Petitjean, Pieri, Pi{\v s}kur, Roe, Ross,
  Rossi, Schlegel, Schneider, Suzuki, Sheldon, Seljak, Viel, Weinberg, \&
  Yeche}]{2013JCAP...04..026S}
Slosar A. {et~al.}, 2013, JCAP, 04, 026

\bibitem[{Stern {et~al}\mbox{.}(2012)Stern, Assef, Benford, Blain, Cutri, Dey,
  Eisenhardt, Griffith, Jarrett, Lake, Masci, Petty, Stanford, Tsai, Wright,
  Yan, Harrison, \& Madsen}]{2012ApJ...753...30S}
Stern D. {et~al.}, 2012, ApJ, 753, 30

\bibitem[{Stern {et~al}\mbox{.}(2005)Stern, Eisenhardt, Gorjian, Kochanek,
  Caldwell, Eisenstein, Brodwin, Brown, Cool, Dey, Green, Jannuzi, Murray,
  Pahre, \& Willner}]{Stern:2005p2563}
Stern D. {et~al.}, 2005, ApJ, 631, 163

\bibitem[{Stoughton {et~al}\mbox{.}(2002)Stoughton, Lupton, Bernardi, Blanton,
  Burles, Castander, Connolly, Eisenstein, Frieman, Hennessy, Hindsley,
  Ivezi{\'c}, Kent, Kunszt, Lee, Meiksin, Munn, Newberg, Nichol, Nicinski,
  Pier, Richards, Richmond, Schlegel, Smith, Strauss, SubbaRao, Szalay, Thakar,
  Tucker, Vanden~Berk, Yanny, Adelman, Anderson, Anderson, Annis, Bahcall,
  Bakken, Bartelmann, Bastian, Bauer, Berman, B{\"o}hringer, Boroski, Bracker,
  Briegel, Briggs, Brinkmann, Brunner, Carey, Carr, Chen, Christian, Colestock,
  Crocker, Csabai, Czarapata, Dalcanton, Davidsen, Davis, Dehnen, Dodelson,
  Doi, Dombeck, Donahue, Ellman, Elms, Evans, Eyer, Fan, Federwitz, Friedman,
  Fukugita, Gal, Gillespie, Glazebrook, Gray, Grebel, Greenawalt, Greene, Gunn,
  de~Haas, Haiman, Haldeman, Hall, Hamabe, Hansen, Harris, Harris, Harvanek,
  Hawley, Hayes, Heckman, Helmi, Henden, Hogan, Hogg, Holmgren, Holtzman,
  Huang, Hull, Ichikawa, Ichikawa, Johnston, Kauffmann, Kim, Kimball, Kinney,
  Klaene, Kleinman, Klypin, Knapp, Korienek, Krolik, Kron, Krzesinski, Lamb,
  Leger, Limmongkol, Lindenmeyer, Long, Loomis, Loveday, MacKinnon, Mannery,
  Mantsch, Margon, McGehee, McKay, McLean, Menou, Merelli, Mo, Monet, Nakamura,
  Narayanan, Nash, Neilsen, Newman, Nitta, Odenkirchen, Okada, Okamura,
  Ostriker, Owen, Pauls, Peoples, Peterson, Petravick, Pope, Pordes, Postman,
  Prosapio, Quinn, Rechenmacher, Rivetta, Rix, Rockosi, Rosner, Ruthmansdorfer,
  Sandford, Schneider, Scranton, Sekiguchi, Sergey, Sheth, Shimasaku, Smee,
  Snedden, Stebbins, Stubbs, Szapudi, Szkody, Szokoly, Tabachnik, Tsvetanov,
  Uomoto, Vogeley, Voges, Waddell, Walterbos, Wang, Watanabe, Weinberg, White,
  White, Wilhite, Wolfe, Yasuda, York, Zehavi, \& Zheng}]{2002AJ....123..485S}
Stoughton C. {et~al.}, 2002, AJ, 123, 485

\bibitem[{Suchkov, Hanisch \& Margon(2005)Suchkov, Hanisch, \&
  Margon}]{2005AJ....130.2439S}
Suchkov A.~A., Hanisch R.~J., Margon B., 2005, AJ, 130, 2439

\bibitem[{White {et~al}\mbox{.}(2011)White, Blanton, Bolton, Schlegel, Tinker,
  Berlind, da~Costa, Kazin, Lin, Maia, McBride, Padmanabhan, Parejko, Percival,
  Prada, Ramos, Sheldon, de~Simoni, Skibba, Thomas, Wake, Zehavi, Zheng,
  Nichol, Schneider, Strauss, Weaver, \& Weinberg}]{2011ApJ...728..126W}
White M. {et~al.}, 2011, ApJ, 728, 126

\bibitem[{White {et~al}\mbox{.}(2012)White, Myers, Ross, Schlegel, Hennawi,
  Shen, McGreer, Strauss, Bolton, Bovy, Fan, Miralda-Escude,
  Palanque-Delabrouille, Paris, Petitjean, Schneider, Viel, Weinberg, Yeche,
  Zehavi, Pan, Snedden, Bizyaev, Brewington, Brinkmann, Malanushenko,
  Malanushenko, Oravetz, Simmons, Sheldon, \& Weaver}]{2012MNRAS.424..933W}
White M. {et~al.}, 2012, MNRAS, 424, 933

\bibitem[{Wright {et~al}\mbox{.}(2010)Wright, Eisenhardt, Mainzer, Ressler,
  Cutri, Jarrett, Kirkpatrick, Padgett, McMillan, Skrutskie, Stanford, Cohen,
  Walker, Mather, Leisawitz, Gautier, McLean, Benford, Lonsdale, Blain, Mendez,
  Irace, Duval, Liu, Royer, Heinrichsen, Howard, Shannon, Kendall, Walsh,
  Larsen, Cardon, Schick, Schwalm, Abid, Fabinsky, Naes, \&
  Tsai}]{2010AJ....140.1868W}
Wright E.~L. {et~al.}, 2010, AJ, 140, 1868

\bibitem[{Yip {et~al}\mbox{.}(2004)Yip, Connolly, Vanden~Berk, Ma, Frieman,
  SubbaRao, Szalay, Richards, Hall, Schneider, Hopkins, Trump, \&
  Brinkmann}]{2004AJ....128.2603Y}
Yip C.~W. {et~al.}, 2004, AJ, 128, 2603

\bibitem[{York {et~al}\mbox{.}(2000)York, Adelman, Anderson, Anderson, Annis,
  Bahcall, Bakken, Barkhouser, Bastian, Berman, Boroski, Bracker, Briegel,
  Briggs, Brinkmann, Brunner, Burles, Carey, Carr, Castander, Chen, Colestock,
  Connolly, Crocker, Csabai, Czarapata, Davis, Doi, Dombeck, Eisenstein,
  Ellman, Elms, Evans, Fan, Federwitz, Fiscelli, Friedman, Frieman, Fukugita,
  Gillespie, Gunn, Gurbani, de~Haas, Haldeman, Harris, Hayes, Heckman,
  Hennessy, Hindsley, Holm, Holmgren, Huang, Hull, Husby, Ichikawa, Ichikawa,
  Ivezi{\'c}, Kent, Kim, Kinney, Klaene, Kleinman, Kleinman, Knapp, Korienek,
  Kron, Kunszt, Lamb, Lee, Leger, Limmongkol, Lindenmeyer, Long, Loomis,
  Loveday, Lucinio, Lupton, MacKinnon, Mannery, Mantsch, Margon, McGehee,
  McKay, Meiksin, Merelli, Monet, Munn, Narayanan, Nash, Neilsen, Neswold,
  Newberg, Nichol, Nicinski, Nonino, Okada, Okamura, Ostriker, Owen, Pauls,
  Peoples, Peterson, Petravick, Pier, Pope, Pordes, Prosapio, Rechenmacher,
  Quinn, Richards, Richmond, Rivetta, Rockosi, Ruthmansdorfer, Sandford,
  Schlegel, Schneider, Sekiguchi, Sergey, Shimasaku, Siegmund, Smee, Smith,
  Snedden, Stone, Stoughton, Strauss, Stubbs, SubbaRao, Szalay, Szapudi,
  Szokoly, Thakar, Tremonti, Tucker, Uomoto, Vanden~Berk, Vogeley, Waddell,
  Wang, Watanabe, Weinberg, Yanny, Yasuda, \&
  collaboration}]{2000AJ....120.1579Y}
York D.~G. {et~al.}, 2000, AJ, 120, 1579

\end{thebibliography}

\appendix

\section{Additional combinations of data}

\begin{figure*}
\centering
\hspace{0cm}
   \includegraphics[width=5.8cm]{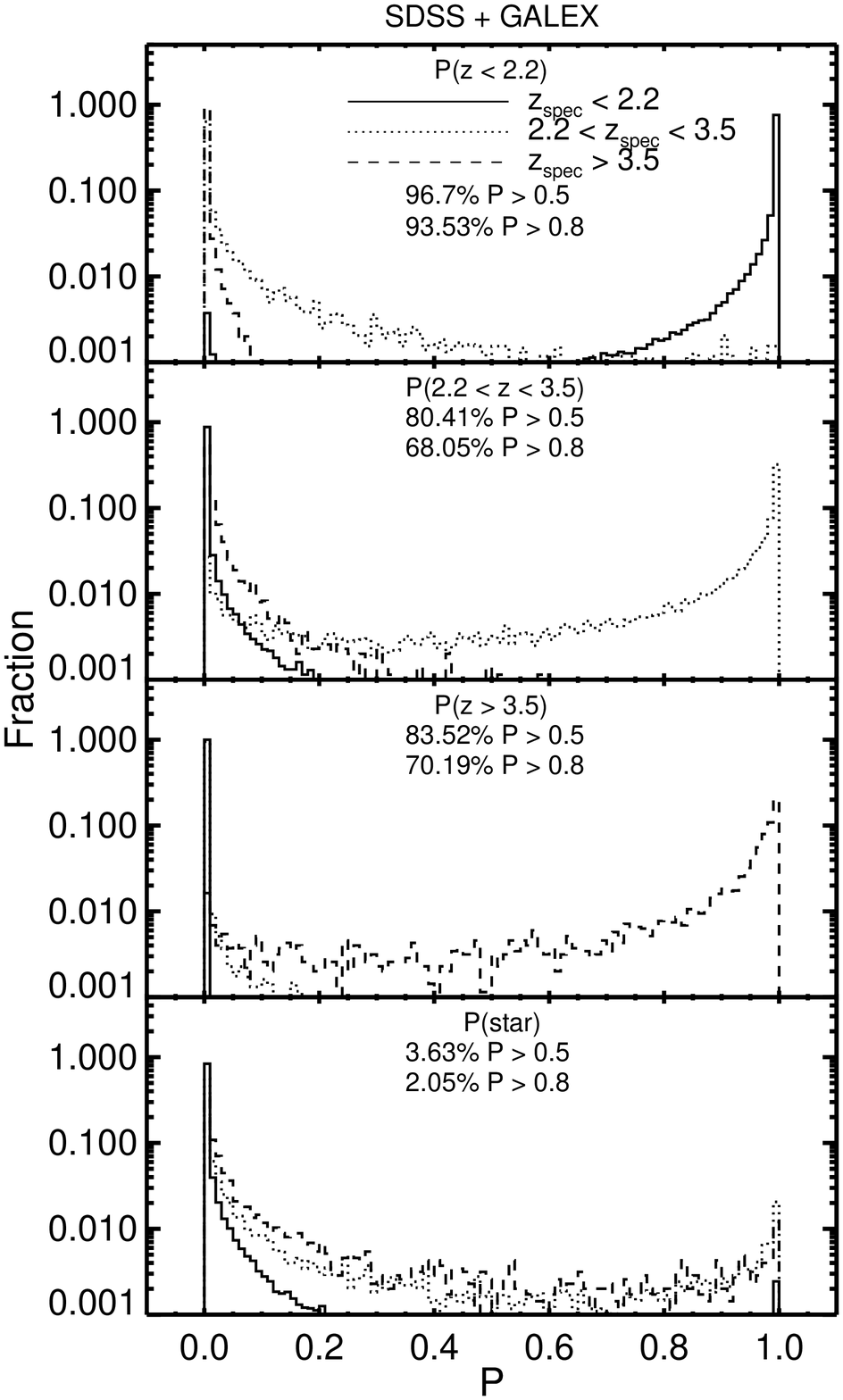}
   \includegraphics[width=5.8cm]{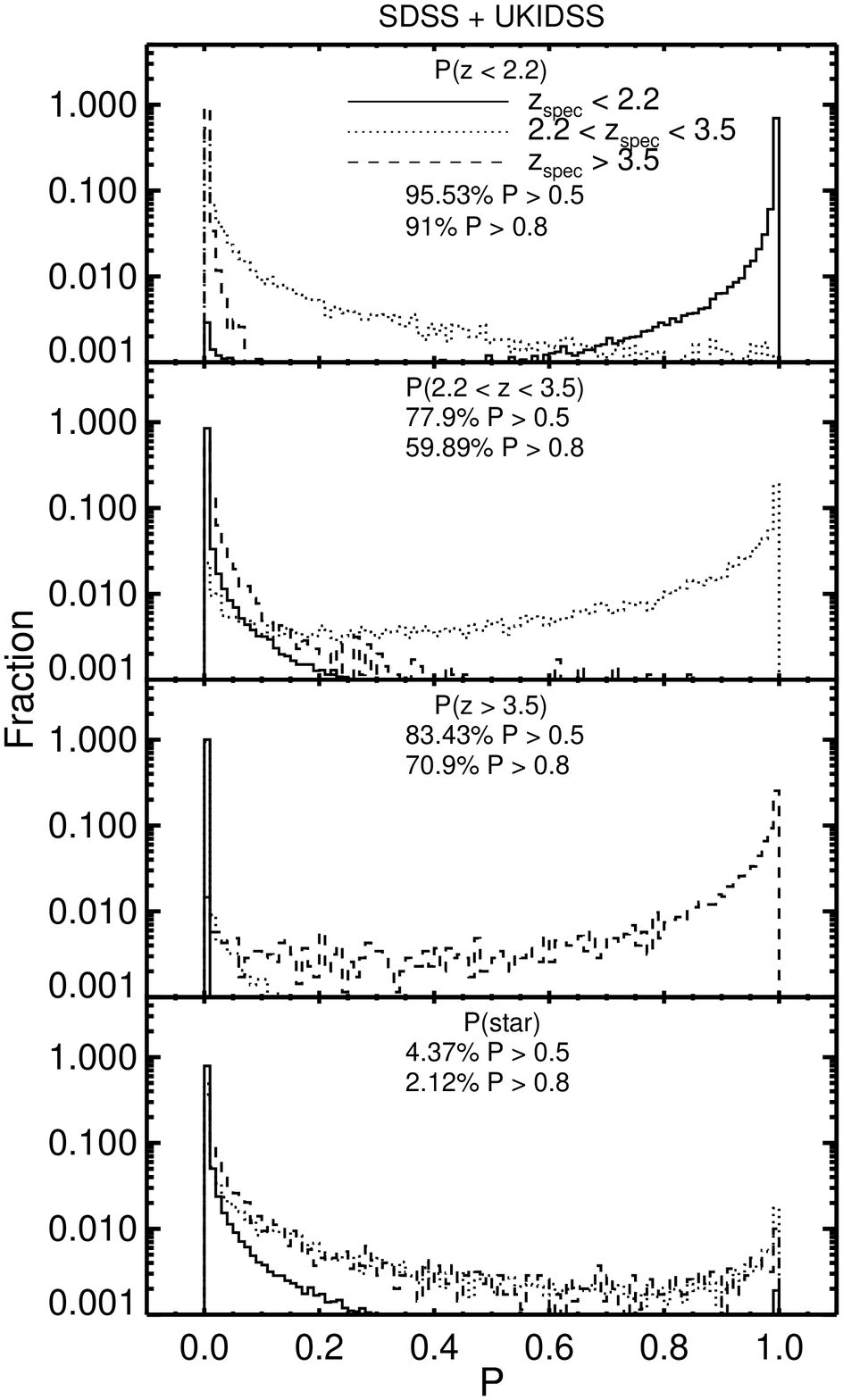}
   \includegraphics[width=5.8cm]{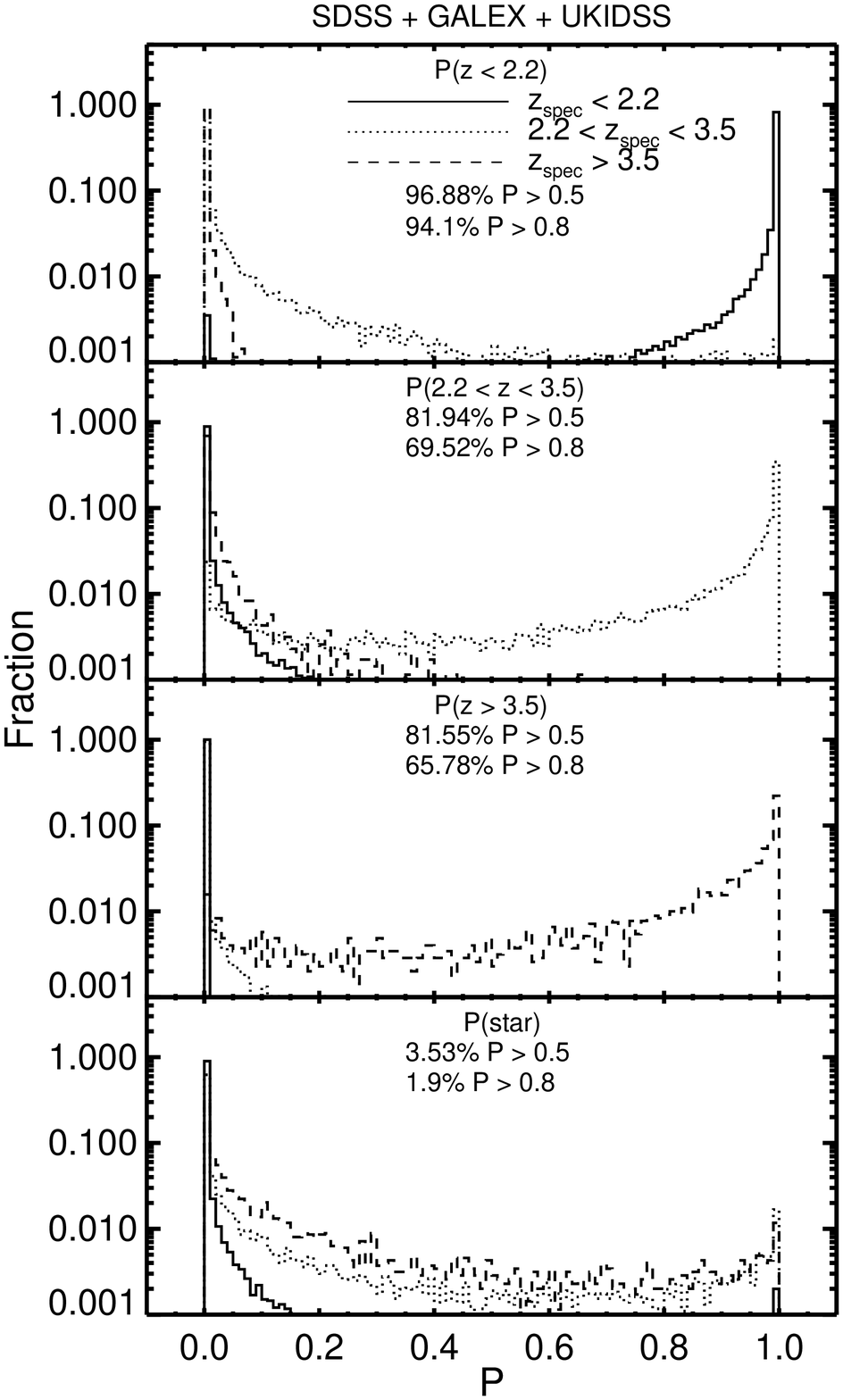}
   \includegraphics[width=5.8cm]{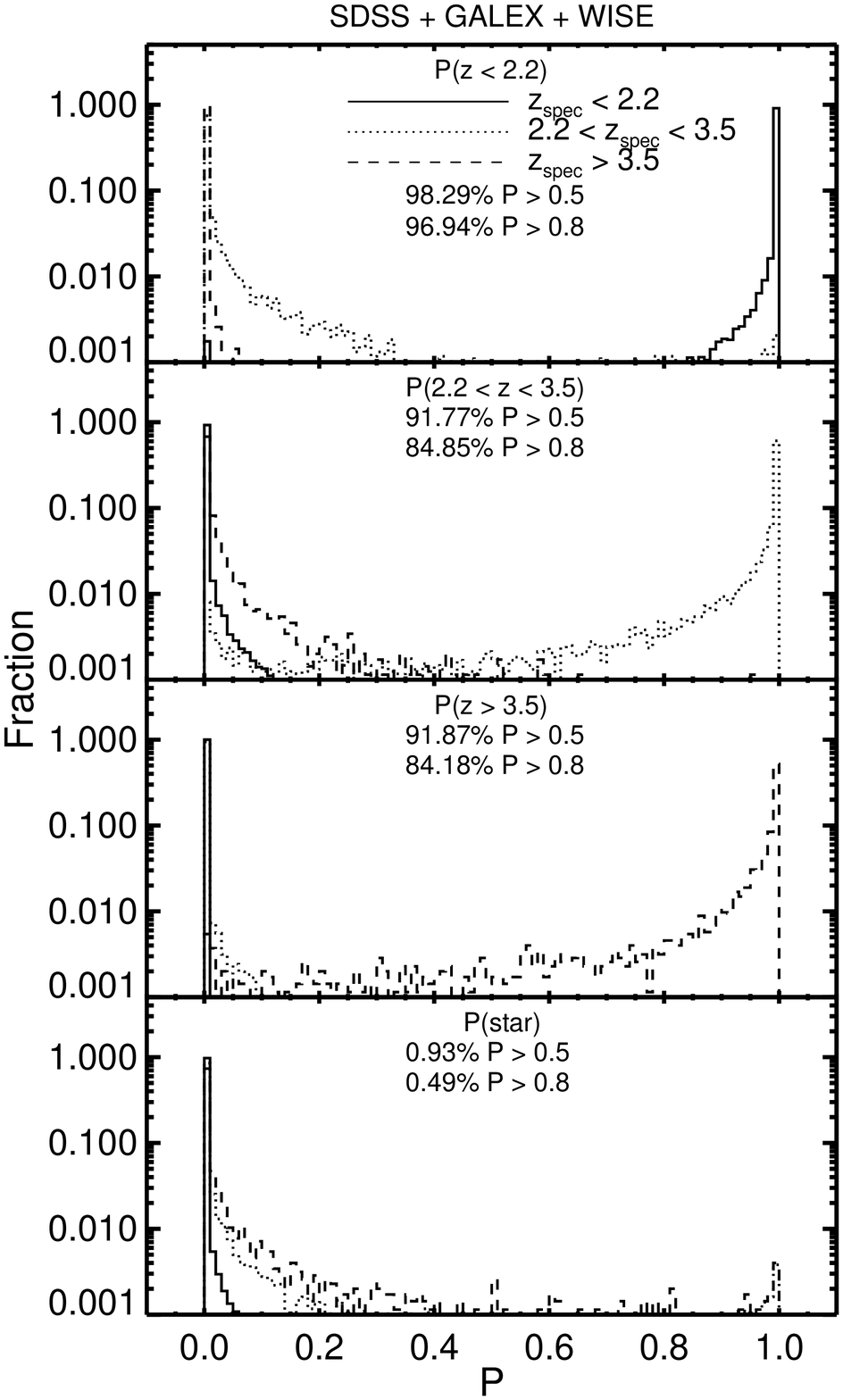}
   \includegraphics[width=5.8cm]{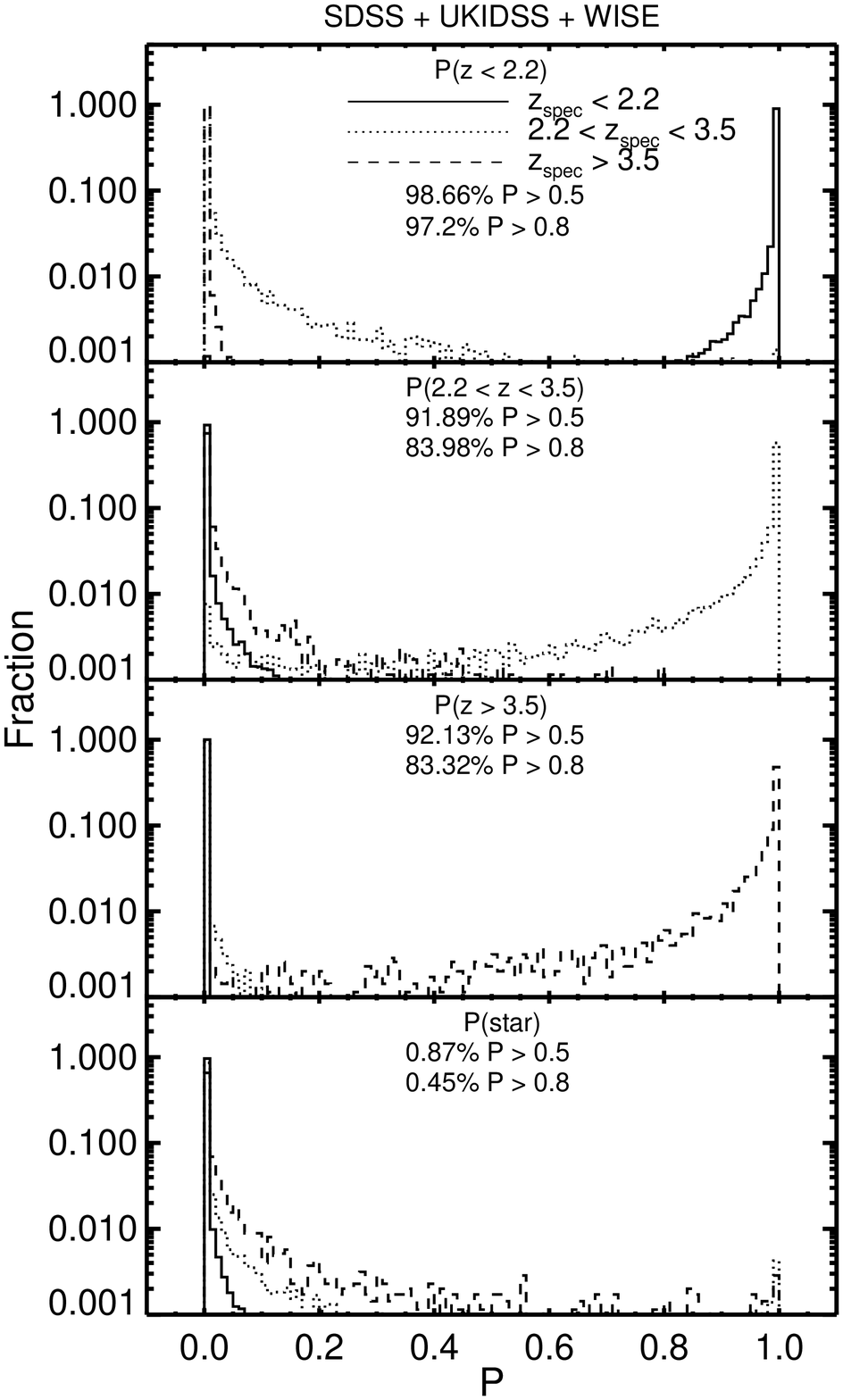}
    \vspace{0cm}
  \caption{The same as Figure~\ref{fig:quasars}, illustrating the performance of \emph{XDQSOz} on known quasars in broad redshift bins but for other combinations of photometric data.\label{fig:quasars_extra}}
\end{figure*}

\begin{figure*}
\centering
\hspace{0cm}
    \includegraphics[width=5.8cm]{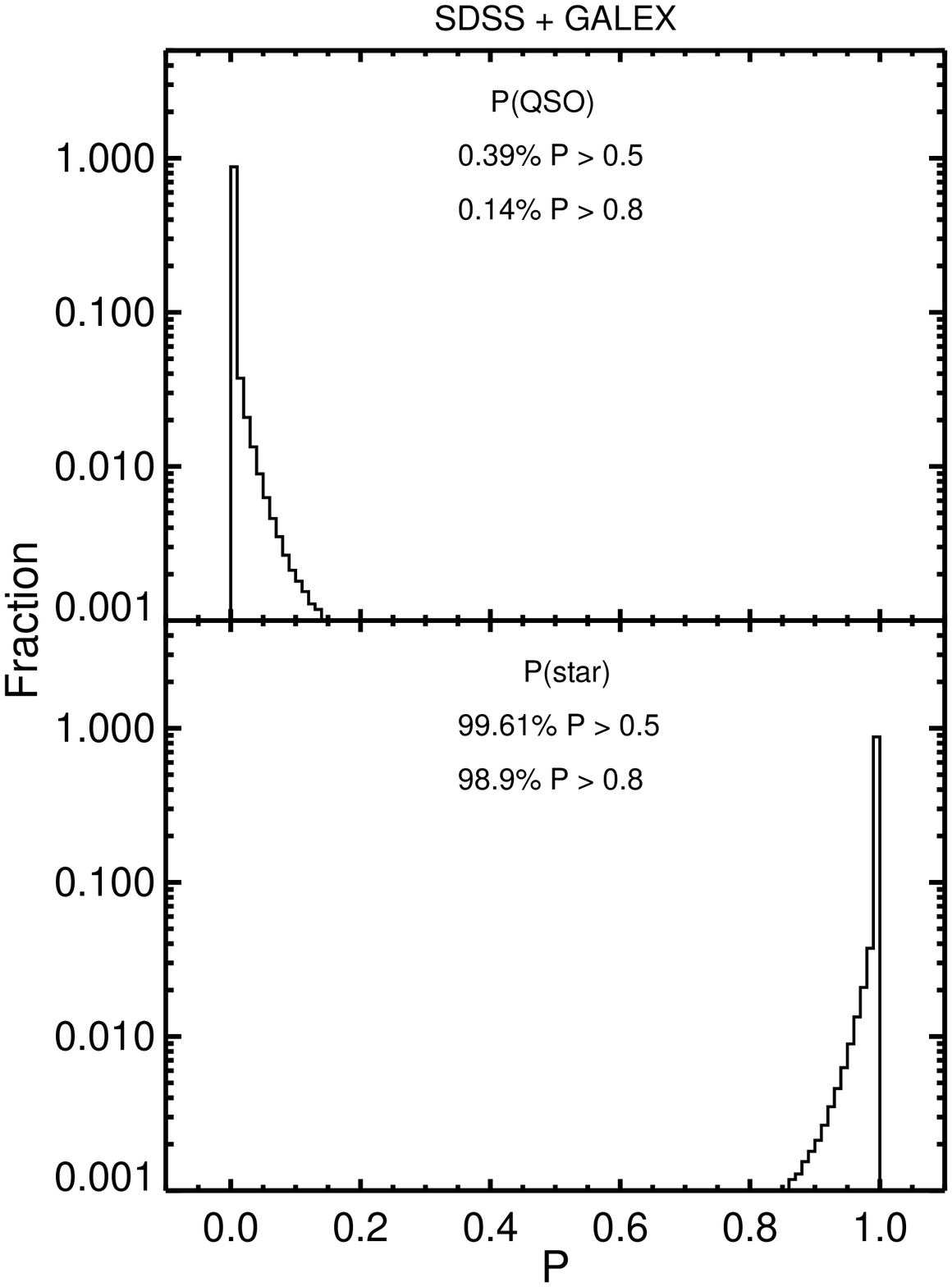}
    \includegraphics[width=5.8cm]{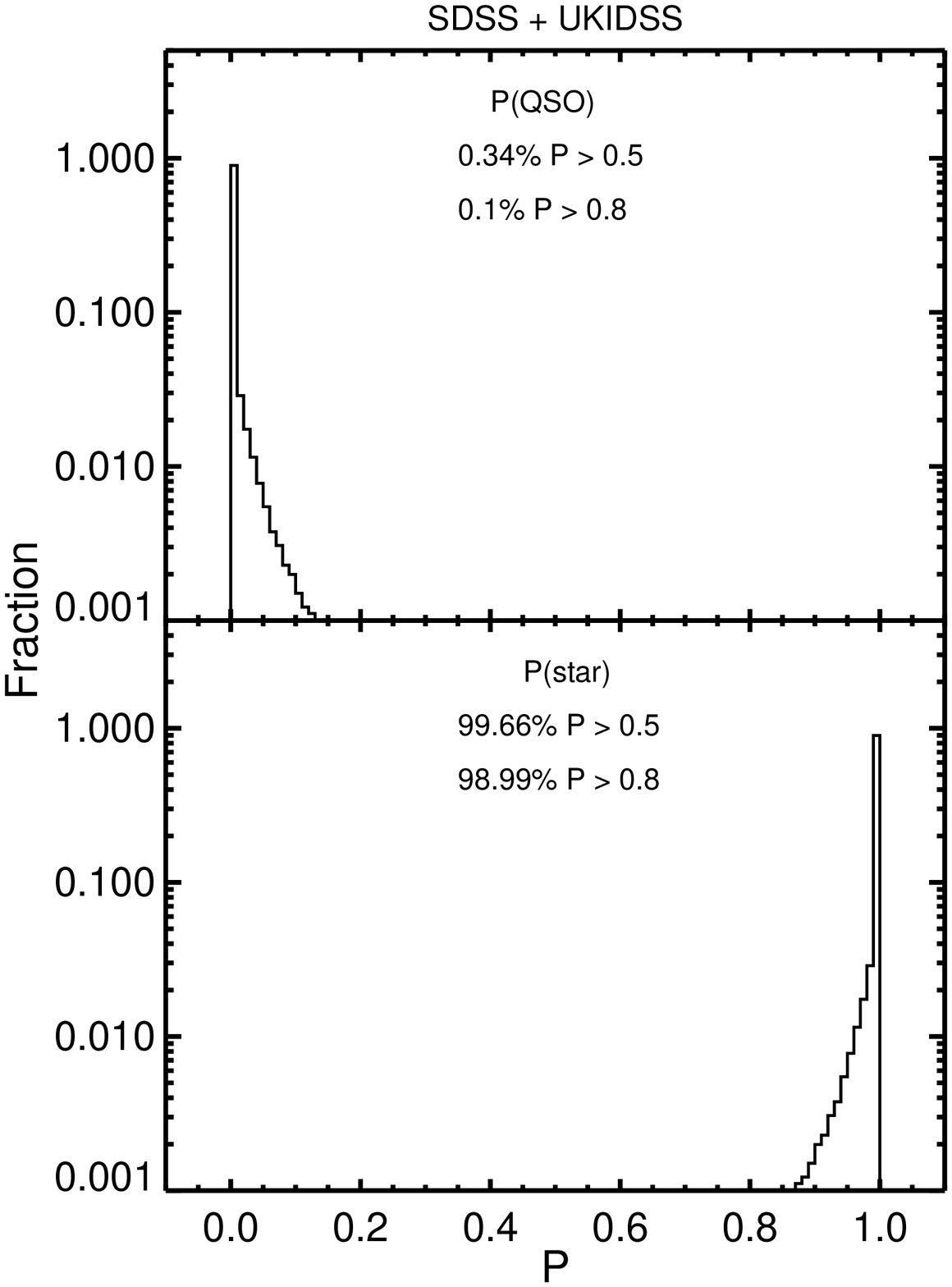}
    \includegraphics[width=5.8cm]{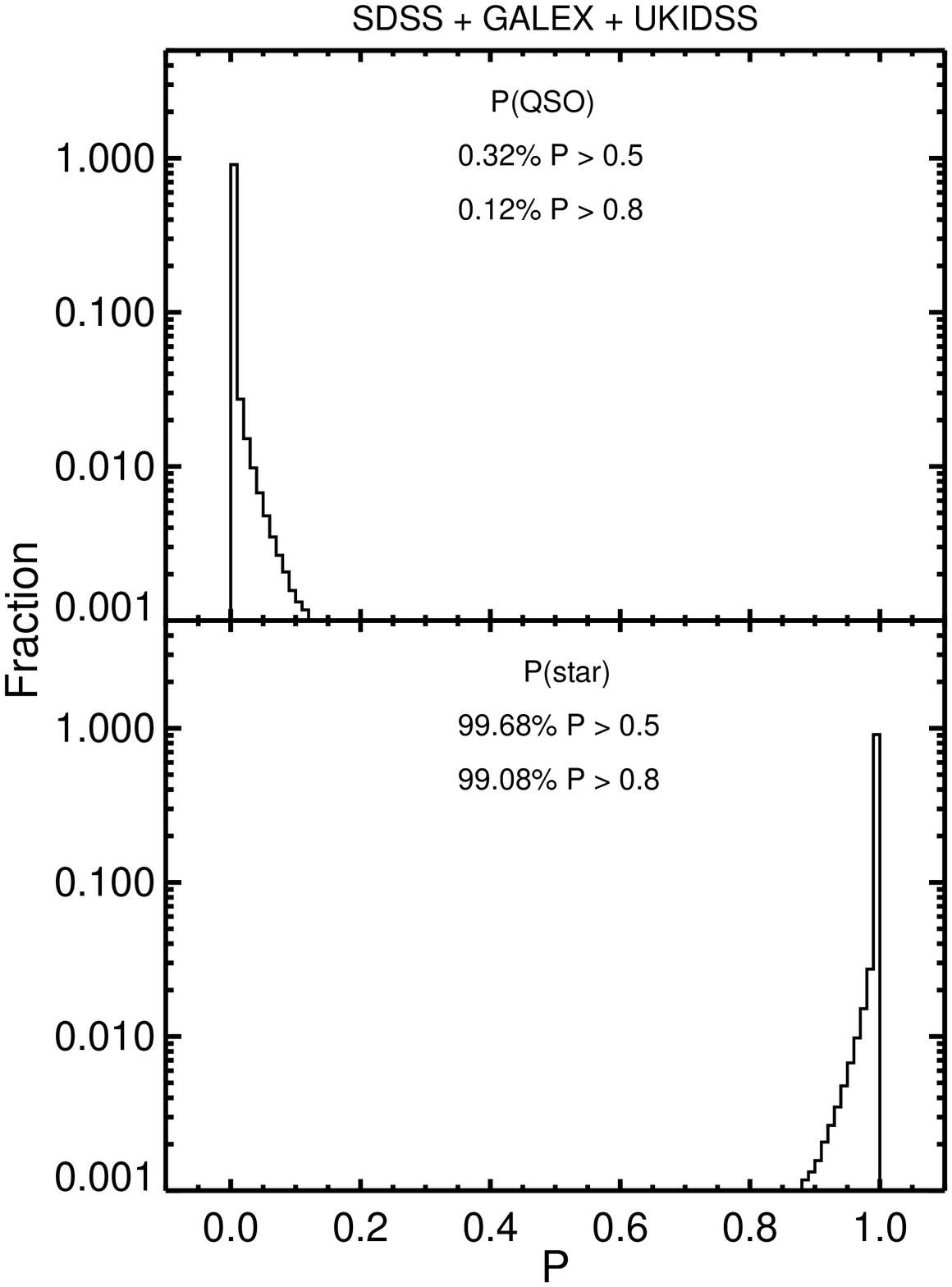}
    \includegraphics[width=5.8cm]{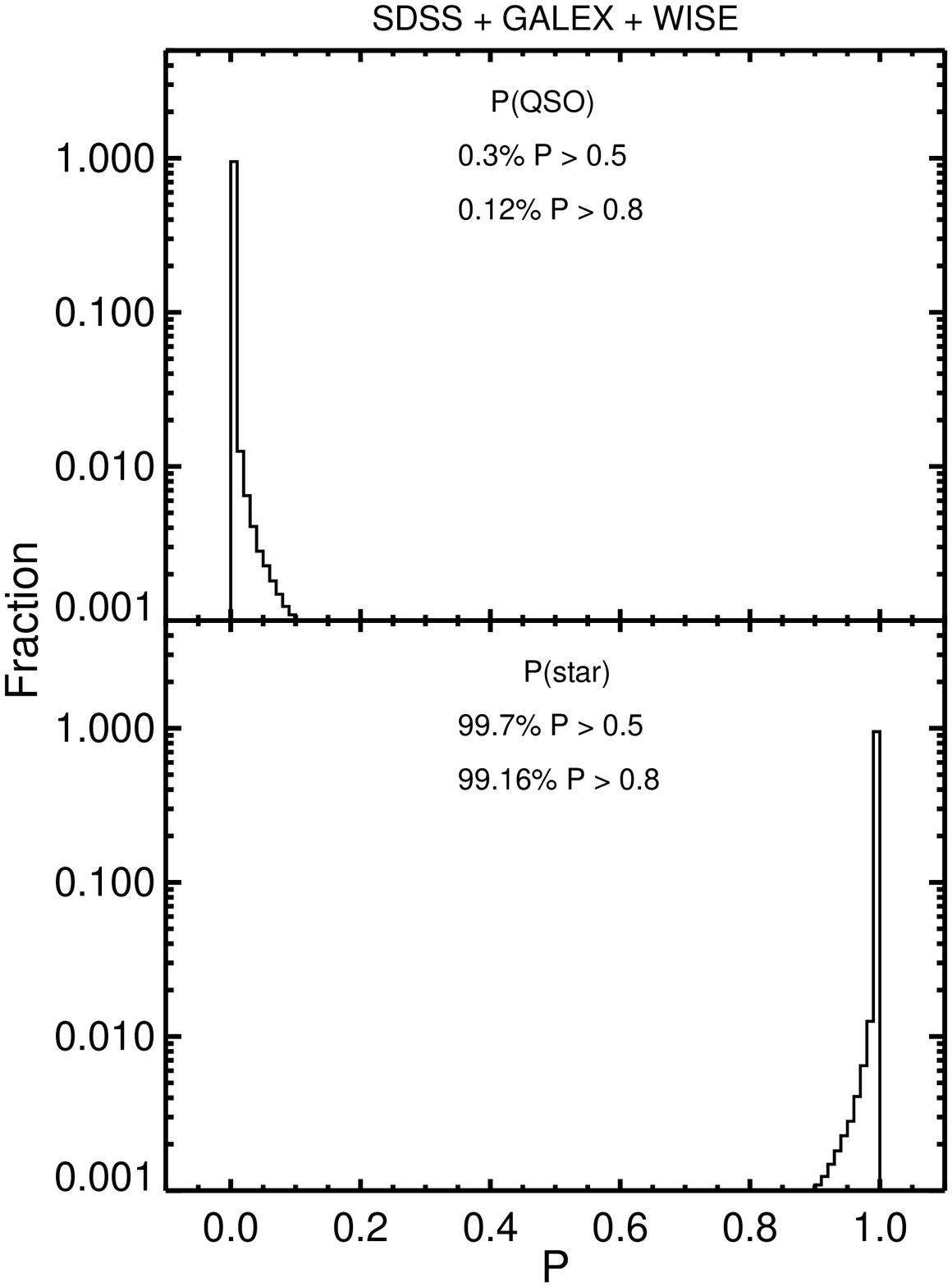}
    \includegraphics[width=5.8cm]{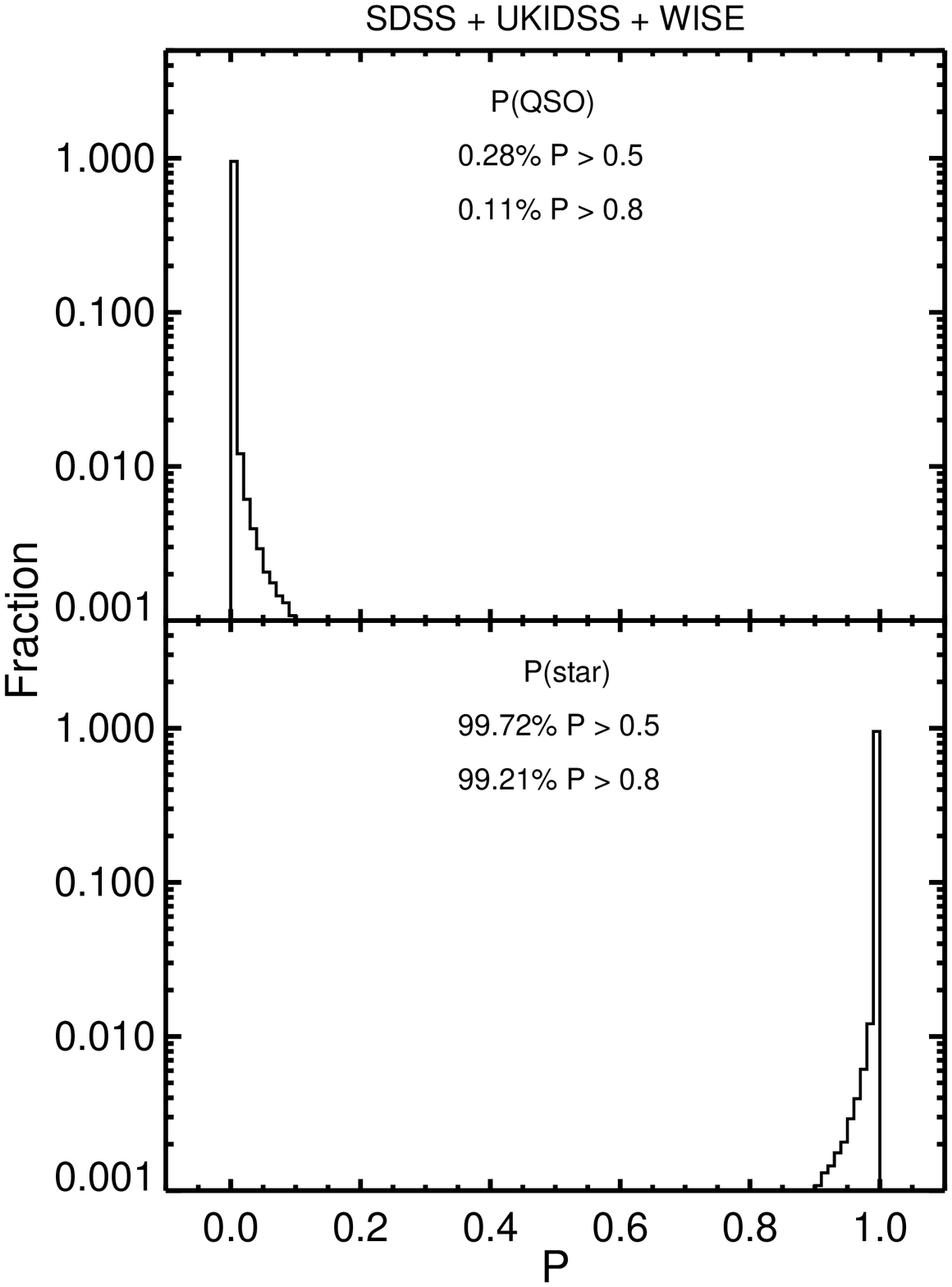}
    \vspace{0cm}
  \caption{The same as Figure~\ref{fig:stars}, illustrating the performance of \emph{XDQSOz} on known stars but for other combinations of photometric data.\label{fig:stars_extra}}
\end{figure*}

\begin{figure*}
\centering
\hspace{0cm}
   \includegraphics[width=5.8cm]{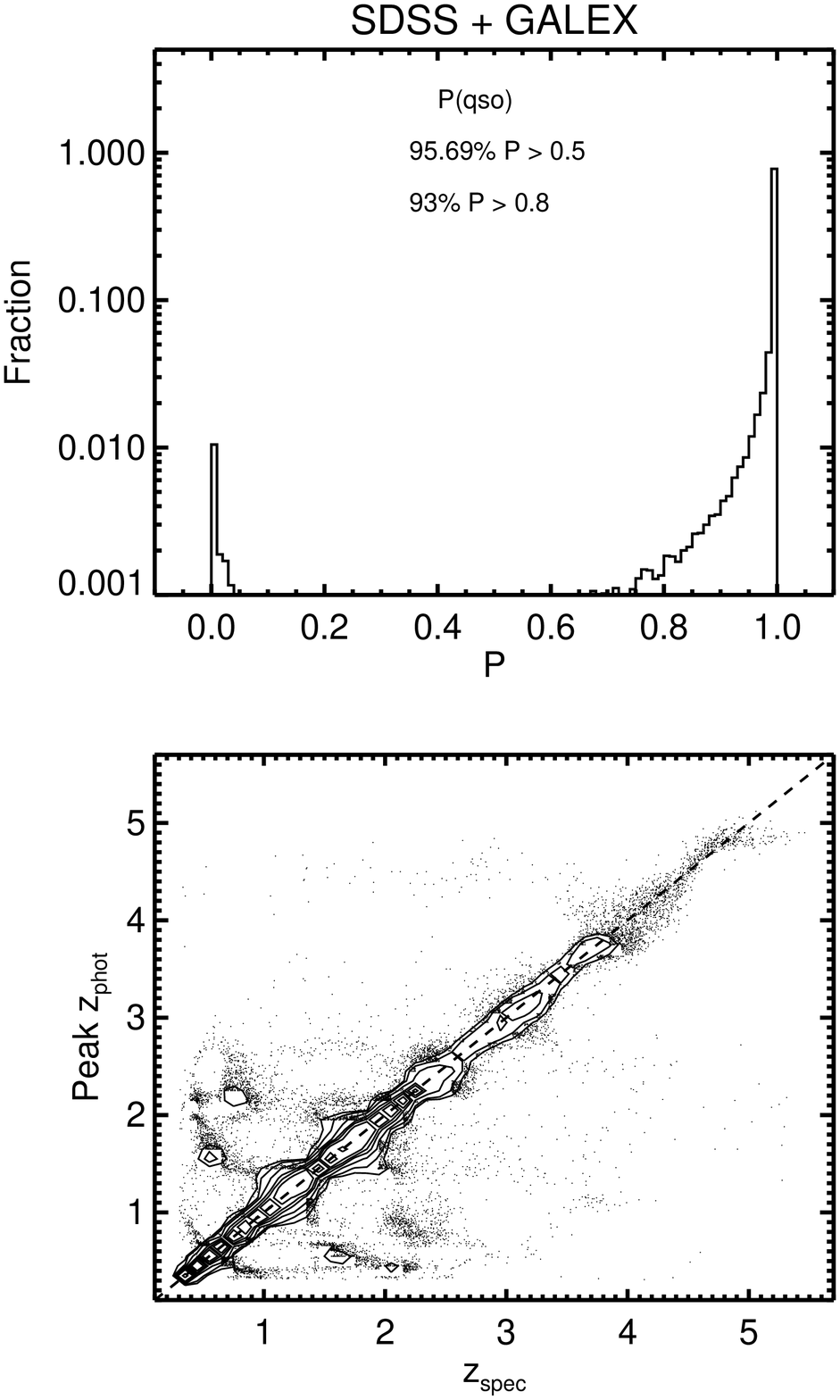}
   \includegraphics[width=5.8cm]{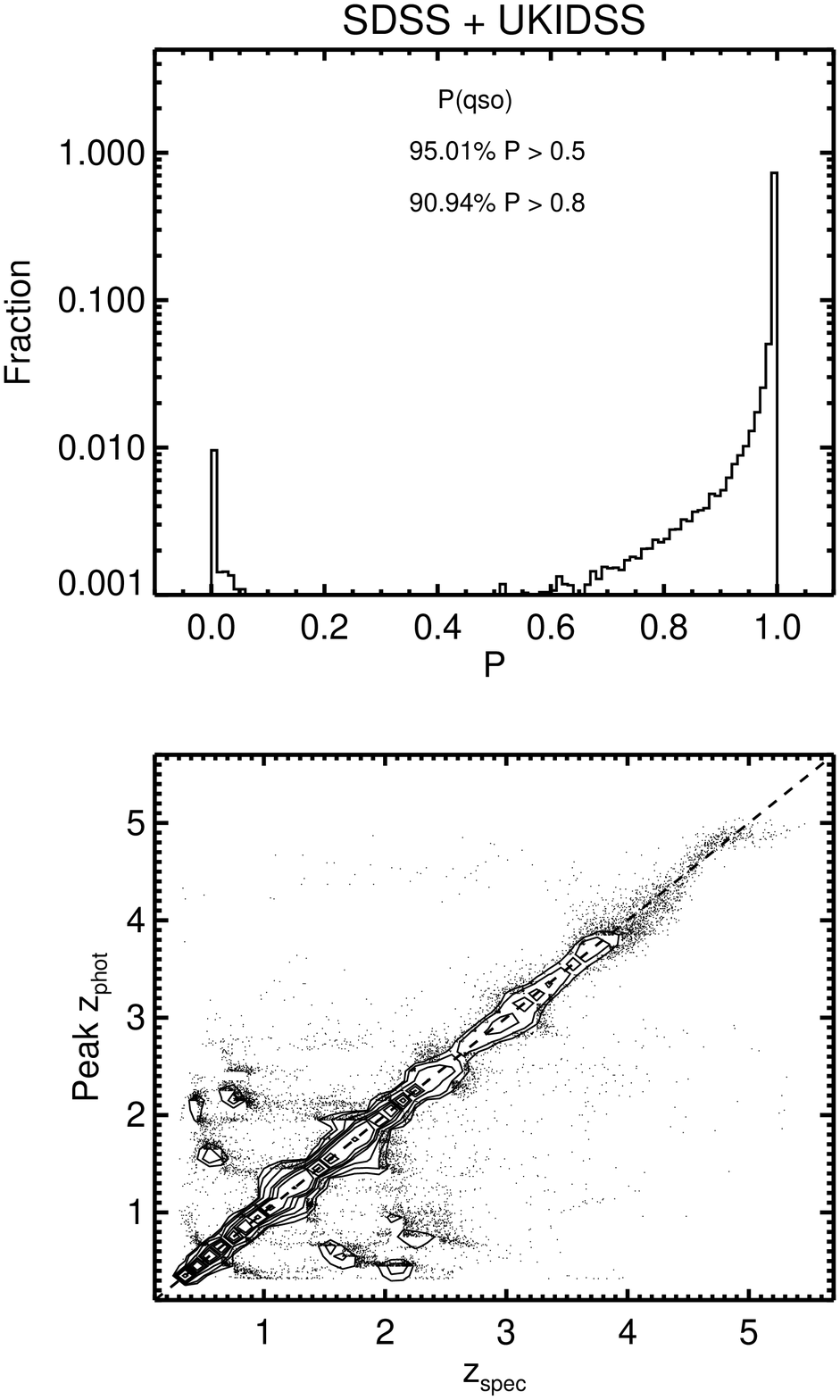}
   \includegraphics[width=5.8cm]{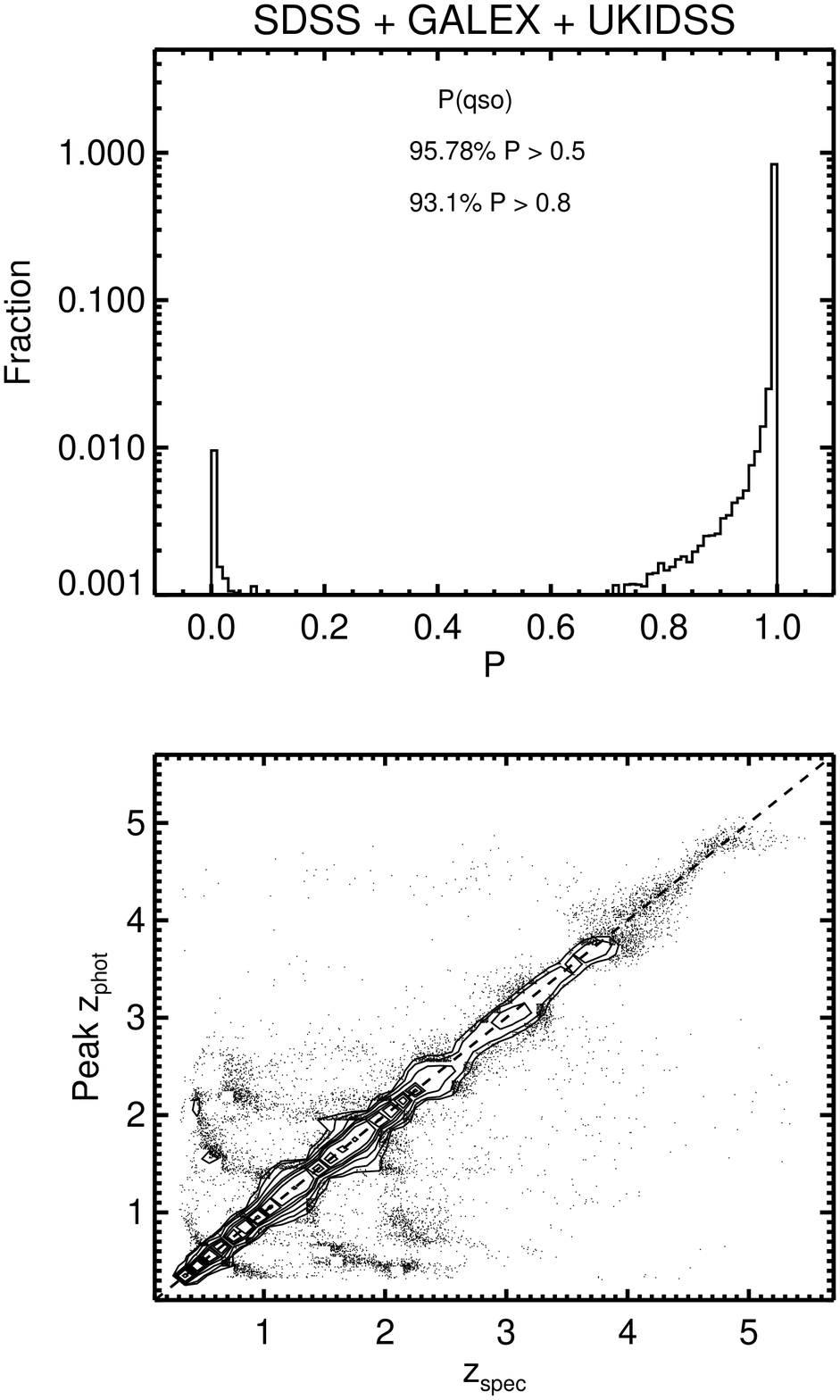}
   \includegraphics[width=5.8cm]{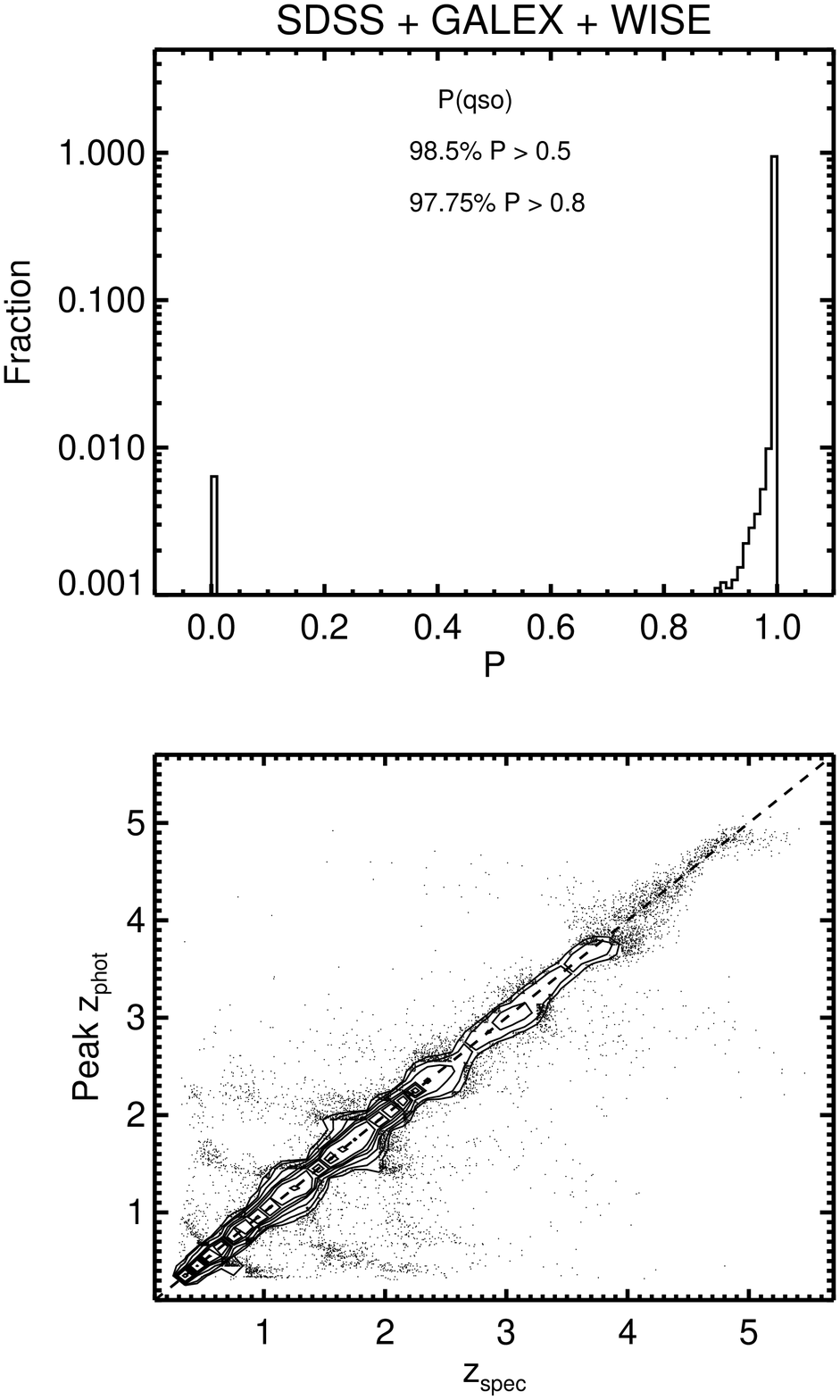}
   \includegraphics[width=5.8cm]{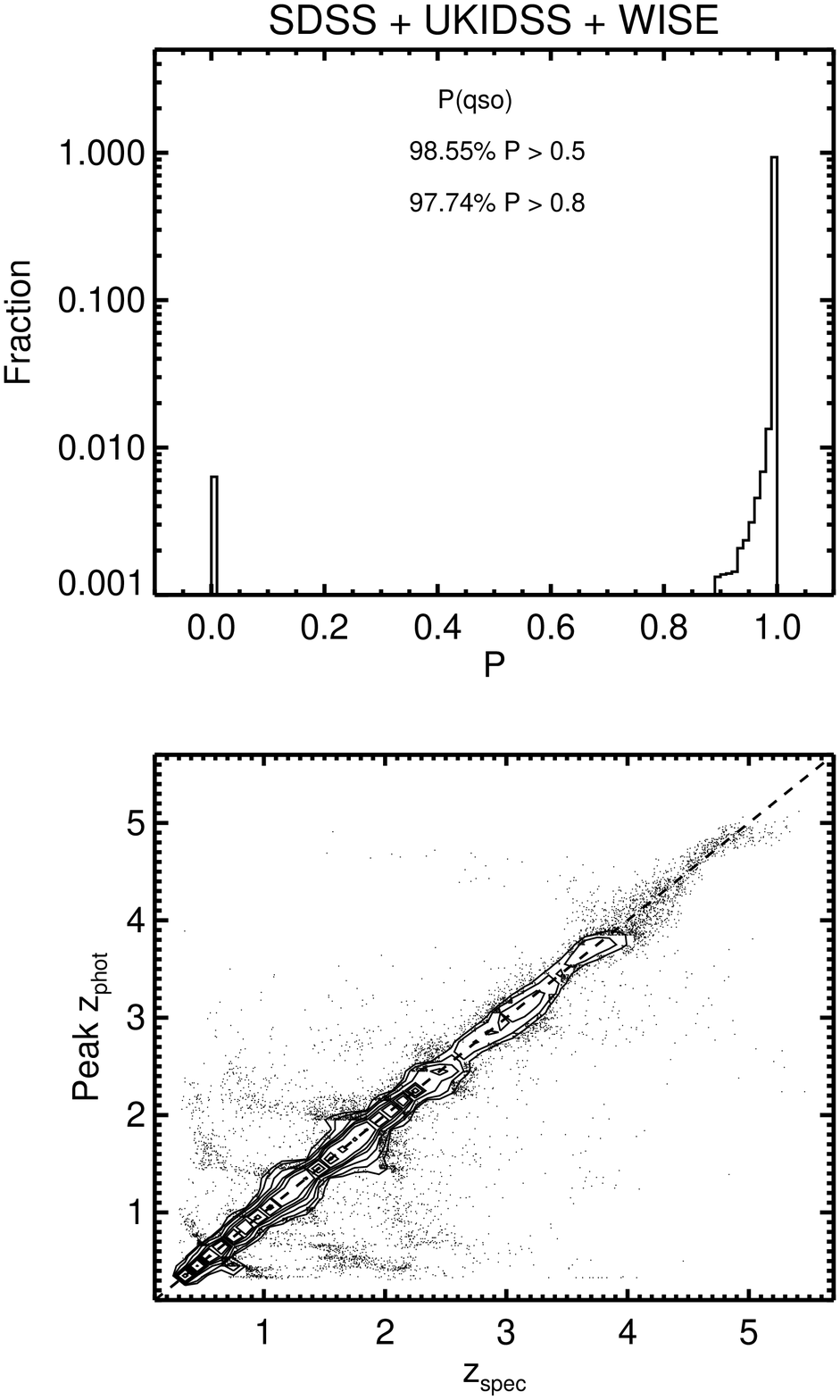}
   \vspace{0cm}
  \caption{The same as Figure~\ref{fig:xdqsoz}, illustrating the performance of \emph{XDQSOz} on known quasars at any redshift but for other combinations of photometric data.\label{fig:xdqsoz_extra}}
\end{figure*}

Given that this work uses updated catalogues of \galex\ and UKIDSS forced-photometry, in addition to the new \wise\ data, we have performed fits to the relative-flux-redshift space using all possible combinations of data so that the appropriate model can be used for any combination of available data.  These are included in the new release of the \emph{XDQSOz} code. Figures~\ref{fig:quasars_extra}, \ref{fig:stars_extra}, and~\ref{fig:xdqsoz_extra} show the distributions of probabilities returned for the known training sets and the photometric versus spectroscopic redshifts using these various combinations of data, for comparison with Figures~\ref{fig:quasars}, \ref{fig:stars}, and~\ref{fig:xdqsoz}.

\section{High-redshift quasars}
One of the motivations for incorporating \wise\ data into the \emph{XDQSO} training is because of the utility of the mid-IR in finding highly reddened and/or high-redshift quasars that are missed using pure optical selection.  Here we illustrate the improvement in constraining the redshifts of high-$z$ quasars when \wise\ fluxes are incorporated.  Because these are examples using known spectroscopic quasars, they are fairly bright in the optical --- the use of \wise\ data will likely show an even more dramatic improvement in identifying these objects when the optical data is fainter.

\begin{figure*}
\centering
\hspace{0cm}
   \includegraphics[width=8.5cm]{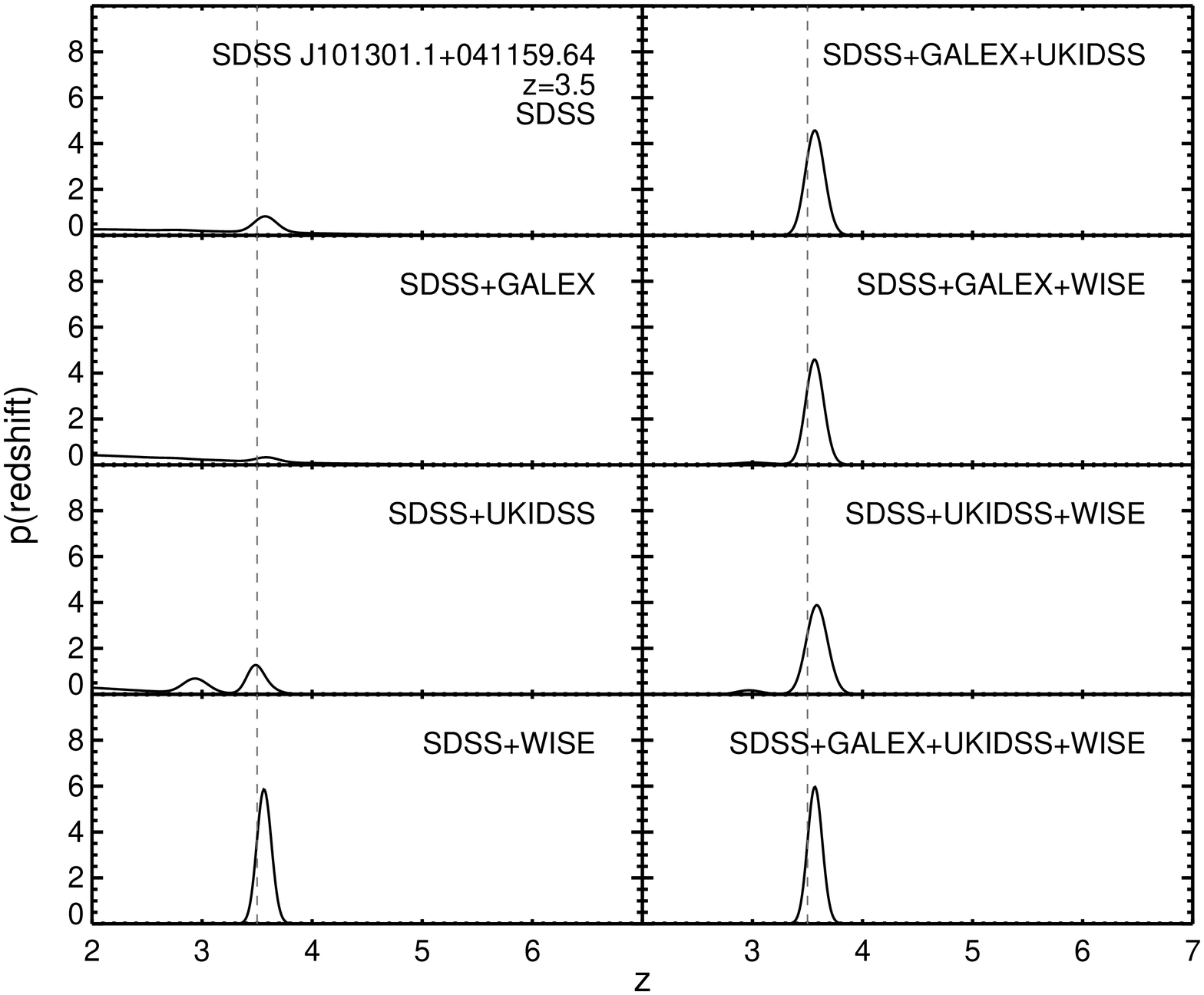}
   \includegraphics[width=8.5cm]{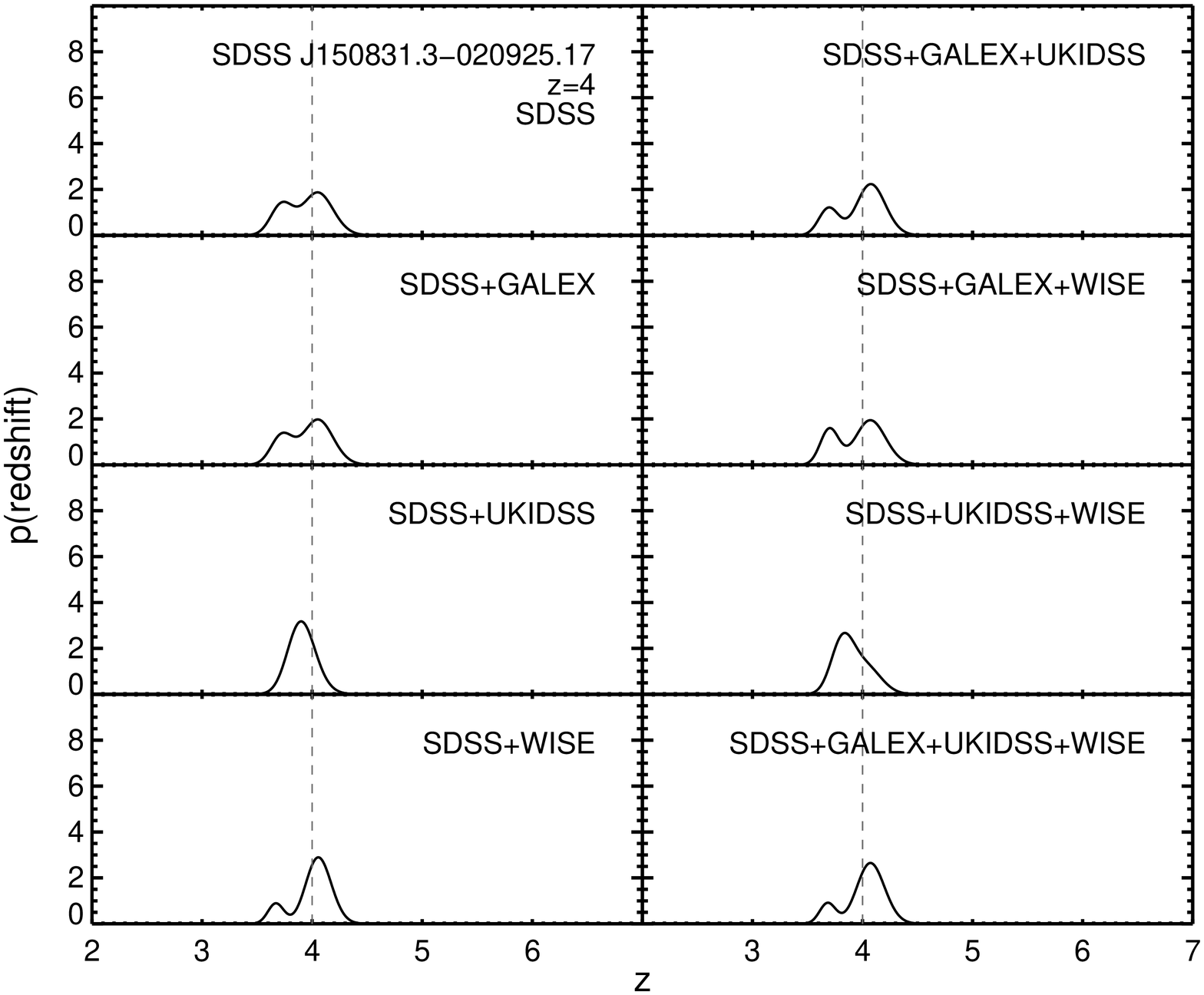}
   \includegraphics[width=8.5cm]{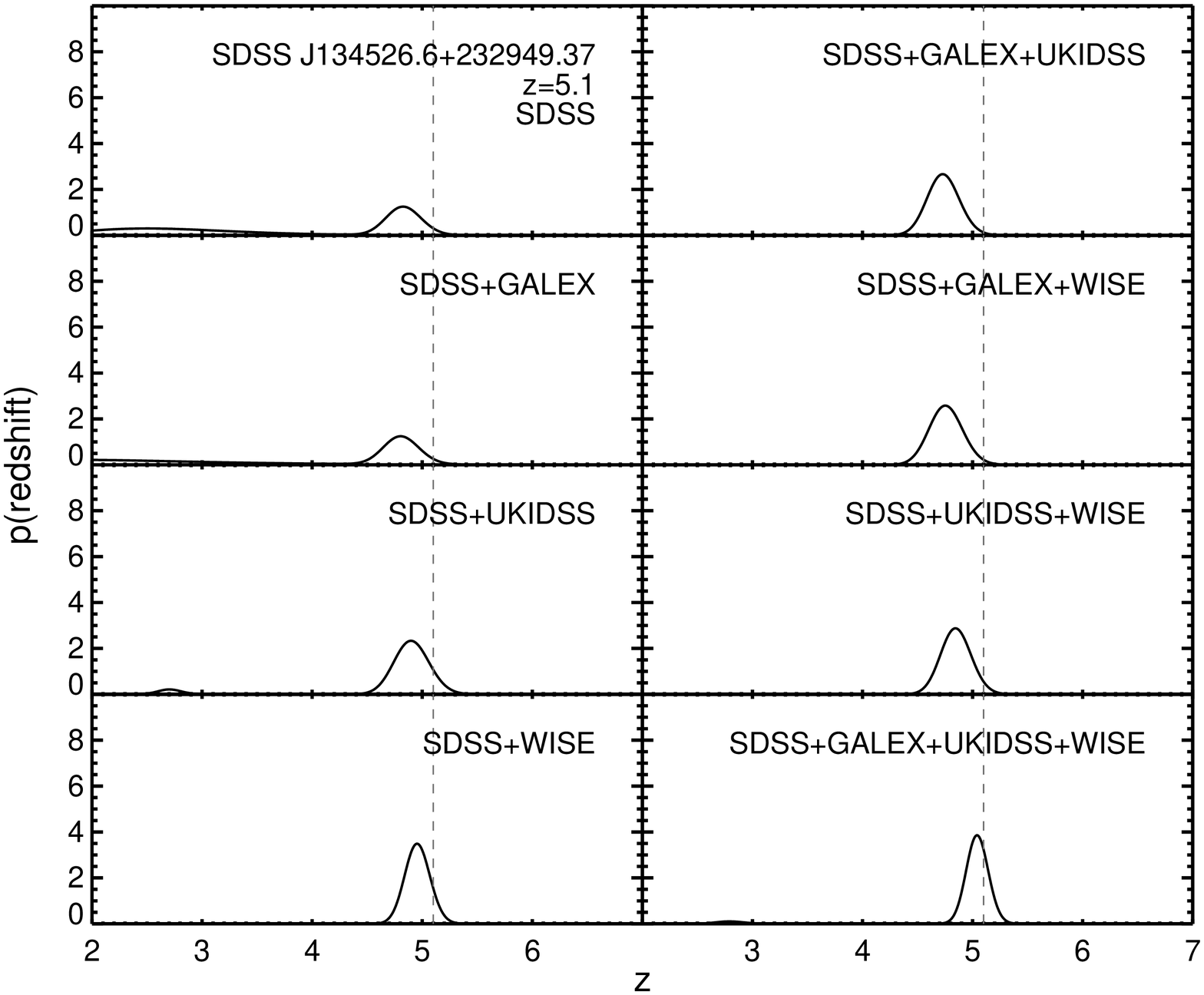}   
   \vspace{0cm}
  \caption{Examples of the effect of \wise\ information on the photometric redshift estimation for quasars at high redshift ($z=3.5$, top left; $z=4$, top right; $z=5$, bottom).\label{fig:hiz}}
\end{figure*}

\label{lastpage}

\clearpage

\end{document}